\documentclass{eurovisDefinitions/egpubl}
\usepackage{eurovisDefinitions/eurovis2023}

\SpecialIssuePaper

\usepackage[T1]{fontenc}
\usepackage{eurovisDefinitions/dfadobe}  

\usepackage{cite}  
\BibtexOrBiblatex
\electronicVersion
\PrintedOrElectronic
\ifpdf \usepackage[pdftex]{graphicx} \pdfcompresslevel=9
\else \usepackage[dvips]{graphicx} \fi

\usepackage{eurovisDefinitions/egweblnk}

\usepackage{microtype}                 
\usepackage{textcomp}                  
\usepackage{mathptmx}                  
\usepackage{times}                     
\usepackage{cite}                      
\usepackage{tabu}                      
\usepackage{booktabs}                  
\usepackage{url}
\usepackage{tabularx}
\usepackage{multirow}

\usepackage{xcolor}

\newcommand{\squishlist}{
	\begin{list}{$\bullet$}
		{ \setlength{\itemsep}{0pt}      \setlength{\parsep}{3pt}
			\setlength{\topsep}{3pt}       \setlength{\partopsep}{0pt}
			\setlength{\leftmargin}{1.5em} \setlength{\labelwidth}{1em}
			\setlength{\labelsep}{0.5em} } }
	
	\newcommand{\squishlisttwo}{
		\begin{list}{$\bullet$}
			{ \setlength{\itemsep}{0pt}    \setlength{\parsep}{0pt}
				\setlength{\topsep}{0pt}     \setlength{\partopsep}{0pt}
				\setlength{\leftmargin}{2em} \setlength{\labelwidth}{1.5em}
				\setlength{\labelsep}{0.5em} } }
		
		\newcommand{\squishend}{
		\end{list}  }

\documentclass{eurovisDefinitions/egpubl}
\usepackage{eurovisDefinitions/eurovis2023}

\SpecialIssuePaper         

\usepackage[T1]{fontenc}
\usepackage{eurovisDefinitions/dfadobe}  

\usepackage{cite}  
\BibtexOrBiblatex
\electronicVersion
\PrintedOrElectronic
\ifpdf \usepackage[pdftex]{graphicx} \pdfcompresslevel=9
\else \usepackage[dvips]{graphicx} \fi

\usepackage{eurovisDefinitions/egweblnk}

\graphicspath{{figures/}{pictures/}{images/}{./}} 

\usepackage{microtype}                 
\PassOptionsToPackage{warn}{textcomp}  
\usepackage{textcomp}                  
\usepackage{mathptmx}                  
\usepackage{times}                     
\usepackage{cite}                      
\usepackage{tabu}                      
\usepackage{booktabs}                  

\usepackage{url}
\usepackage{tabularx}
\usepackage{multirow}
\usepackage{amsmath}
\usepackage{amssymb}
\usepackage{nicematrix}
\usepackage{tikz}
\usepackage{array}
\usepackage{subfigure}

\newcommand{\squishlist}{
	\begin{list}{$\bullet$}
		{ \setlength{\itemsep}{0pt}      \setlength{\parsep}{3pt}
			\setlength{\topsep}{3pt}       \setlength{\partopsep}{0pt}
			\setlength{\leftmargin}{1.5em} \setlength{\labelwidth}{1em}
			\setlength{\labelsep}{0.5em} } }
	
	\newcommand{\squishlisttwo}{
		\begin{list}{$\bullet$}
			{ \setlength{\itemsep}{0pt}    \setlength{\parsep}{0pt}
				\setlength{\topsep}{0pt}     \setlength{\partopsep}{0pt}
				\setlength{\leftmargin}{2em} \setlength{\labelwidth}{1.5em}
				\setlength{\labelsep}{0.5em} } }
		
		\newcommand{\squishend}{
		\end{list}  }

\makeatletter
\renewcommand{\ps@plain}{%
\renewcommand{\@oddfoot}{\hfil\textrm{\thepage}\hfil}%
\renewcommand{\@evenfoot}{\@oddfoot}%
}
\renewcommand{\ps@empty}{%
\renewcommand{\@oddfoot}{\hfil\textrm{\thepage}\hfil}%
\renewcommand{\@evenfoot}{\@oddfoot}%
}
\makeatother

\title[Knowledge-Decks]%
      {Knowledge-Decks: Automatically Generating Presentation Slide
Decks of Visual Analytics Knowledge Discovery Applications}

\author[L. Christino \& T. Hill \& F. V. Paulovich]
{\parbox{\textwidth}{\centering L. Christino$^{1}$\orcid{0000-0002-8754-8460}, T. Hill$^{1}$\orcid{0000-0003-0048-7712} and F. V. Paulovich$^{2}$\orcid{0000-0002-2316-760X} 
        }
        \\
{\parbox{\textwidth}{\centering $^1$Dalhousie University, Canada\\
         $^2$Eindhoven University of Technology (TU/e), Netherlands
       } 
}
}

\begin{document}


\maketitle
\begin{abstract}
Visual Analytics (VA) tools provide ways for users to harness insights and knowledge from datasets. Recalling and retelling user experiences while utilizing VA tools has attracted significant interest. Nevertheless, each user sessions are unique. Even when different users have the same intention when using a VA tool, they may follow different paths and uncover different insights. Current methods of manually processing such data to recall and retell users' knowledge discovery paths may also be time-consuming, especially when there is the need to present users' findings to third parties. This paper presents a novel system that collects user intentions, behavior, and insights during knowledge discovery sessions, automatically structure the data, and extracts narrations of knowledge discovery as PowerPoint slide decks. The system is powered by a Knowledge Graph designed based on a formal and reproducible modeling process. To evaluate our system, we have attached it to two existing VA tools where users were asked to perform pre-defined tasks. Several slide decks and other analysis metrics were extracted from the generated Knowledge Graph. Experts scrutinized and confirmed the usefulness of our automated process for using the slide decks to disclose knowledge discovery paths to others and to verify whether the VA tools themselves were effective.


\end{abstract}


\section{Introduction}
Visual Analytics (VA) tools provide ways for users to harness insights and knowledge from datasets~\cite{sacha2014knowledge}. In general, through VA tools, users wish to answer an existing question, execute a pre-planned task, or aim to explore the information contained in the data. In all cases, the user performs actions and interactions considering an initial intention until they reach their goal, such as uncovering new insights. This process of user-guided knowledge discovery, also called VA's knowledge-gathering workflow, is extensively studied within VA. Theoretical knowledge models~\cite{sacha2014knowledge, federico2017role}, behavior collection techniques~\cite{mathisen2019insideinsights, guan2022event}, behavior analysis~\cite{heer2008graphical, battle2019characterizing, christino2022data} and storytelling through slide decks~\cite{park2022storyfacets} are recent works which attempt to study how to understand and analyze user-guided knowledge discovery in VA, or how to recall and retell the story of user intentions, behaviors, and insights, putting VA in a unique position of narrating knowledge discovery through storytelling~\cite{knaflic2015storytelling, xu2015analytic}.

However, typically each user-guided knowledge-gathering session is unique and non-linear. Even when different users have the same intention when using a VA tool, they may uncover different insights due to their differences in tacit knowledge~\cite{federico2017role}. Likewise, even when the same insight is found, users may have used different ways to arrive at it~\cite{heer2008graphical}. These inconsistencies and the subsequent manual labor of transforming the collected user data, such as video/audio/logs recordings of surveys~\cite{xu2015analytic} into a cohesive data structure wherein storytelling~\cite{knaflic2015storytelling, lipford2010helping} can be derived, is not just time-consuming~\cite{park2022storyfacets, schoeneborn2013pervasive}, but also it is often not reproducible with other users or VA tools~\cite{xu2015analytic, christino2022data, federico2017role}.

Yet, albeit these limitations, researchers still use these data to recall users' knowledge discovery process~\cite{ragan2015characterizing} for the user's benefit or to a third party by using a process called provenance~\cite{ragan2015characterizing}. Provenance can be used to recall the story of what and how users found a new piece of knowledge~\cite{xu2015analytic}, which can then be structured as a linear sequence of events, snapshots, or discoveries. This resulting knowledge is then usually manually formatted into info-graphs~\cite{knaflic2015storytelling, zhu2020survey}, slide decks~\cite{park2022storyfacets, schoeneborn2013pervasive}, or included in visualizations and tools as explicit knowledge~\cite{xu2020survey}. For instance, if a third party (e.g., the user's manager) needs to use the collected knowledge discovery (e.g., the slide decks) to retell the story to an external actor (e.g., presentation to external partners or board of directors), any difference of expertise level may cause issues when manually building the presentation. Though manual means have proven valuable for the user to recall their past experiences~\cite{lipford2010helping}, a process that could automatically generate a presentation to disclose knowledge discovery to others may be advantageous~\cite{xu2015analytic}. However, this process has received little attention. 
%
%
%
Instead, the methods of collection and retelling of users' behavior when using a VA tool~\cite{da2009towards, ragan2015characterizing} have been largely decoupled in the VA literature from how one would publish or present a user's findings to third parties from when they had used a VA tool~\cite{knaflic2015storytelling, xu2015analytic}.


In this paper, we discuss, design, and implement \textit{Knowledge-Decks}\footnote{https://youtu.be/ghe7zYPLUWs}, a novel approach that collects user intentions, behaviors, and insights during knowledge discovery sessions and provides an interface (API) where one can extract narrations of knowledge discovery as PowerPoint slide decks. \textit{Knowledge-Decks} has two components: the front-end library and the back-end web-service. The front-end can be attached to any VA tool that follows specific pre-defined design requirements to simultaneously collect user behavior as temporal events and user intentions and insights as visual annotations. The back-end is powered by a \textit{Knowledge Graph (KG)} modeled to match the same pre-defined design requirements. With these two components, we collect user intentions, behaviors, and insights and structure them into a specifically designed KG. \textit{Knowledge-Decks} also provides automatic slide deck generation and other analytical features through a well-known KG query language without the need for extra pre-processing manual labor. To evaluate our approach, we have attached it to a VA tool and asked users to perform pre-defined tasks. From the resulting KG with the collective knowledge discovery of all users, we used pre-defined queries to provide a results overview of the tool's usage and to generate different knowledge discovery slide decks, such as a description of all insights found during a task or the process a user should take to reach a given insight. An expert in the tool and its domain was interviewed to discuss the utilization of the slide decks as the starting point for the presentation of the tool itself or to discuss the insights found through it.
%

In summary, our main contributions are: 1) automatic collection of user behavior as event sequences from web-based VA tools and user intentions and insights through user input; 2) modeling and association of user's behaviors, intentions, and insights through user-guided state matching and structuring the collected data as a knowledge graph; 3) automatic generation of knowledge discovery slide decks from the relationship paths of the knowledge graph through pre-defined queries; and 4) a blueprint to extend our system to other VA tools. In the next section, we discuss the techniques and strategies we used to implement our goals and the existing related work.

\section{Background and Related Work}\label{sec:related}

The modeling and design of \textit{Knowledge-Decks} are based on existing knowledge model research. We have implemented ways to collect and transform user data into a retelling of the users' insights and knowledge discovery process through slide decks. Therefore, this section discusses VA knowledge models and their practical usage when applied to knowledge discovery. Then, we discuss several provenance~\footnote{Provenance: tracking and using data collected from a tool's usage~\cite{ragan2015characterizing}} approaches, including data and analysis provenance, which enables one to collect users' insights and knowledge discovery process. Finally, we discuss existing approaches to recall and retell users' knowledge discovery as storytelling presentations, such as slide decks.

\subsection{Knowledge Modeling and Provenance in VA}

VA literature in knowledge modeling has shown that users' interactivity with VA tools can be modeled as an iterative workflow of user intentions, behaviors, and insights~\cite{sacha2014knowledge, sacha2016analytic, federico2017role}. For instance, Federico et al. ~\cite{federico2017role} describes how VA tools and frameworks can be modeled following knowledge modeling methodology and explain how to interpret VA research in light of their model. This methodology of describing VA as an iterative workflow of events between Human and Machine actors shows that it is possible to understand a VA problem as a sequence of events. This model also includes automatic processes, such as data mining or machine learning, as part of the sequence of events, even if a Human interaction did not trigger it. Sacha et.al.~\cite{sacha2018vis4ml}, among others, show through their detailed machine-learning ontology that these Computer-generated events permeate much of VA as a whole. 

%
%


Usually, VA developers and researchers use more practical means to model, collect, store, and utilize their tools or users' experiences. For instance, some researchers collect and analyze user behavior using provenance~\cite{da2009towards, ragan2015characterizing}. Landesberger et.al.~\cite{von2014interaction} perform behavior provenance by modeling user behavior as a graph network, which is then used for analysis. Other works also use graph networks as a way to model knowledge~\cite{fensel2020introduction}, data provenance~\cite{simmhan2005survey, da2009towards}, insight provenance~\cite{gomez2014insight}, and analytic provenance~\cite{xu2015analytic, madanagopal2019analytic}. Each of these tools and techniques aim to collect some part of the user's knowledge discovery process and model it into some structure.

Among the literature, graph networks, or more specifically \textit{Knowledge Graphs (KGs)}~\cite{fensel2020introduction, guan2022event}, are uniquely positioned to model the user's knowledge discovery process due to their specific applicability in collecting and analyzing two kinds of data: temporally-based event data~\cite{guan2022event} and relationship-centric data~\cite{auer2007dbpedia}. As VA is modeled based on events~\cite{sacha2014knowledge, federico2017role} and our goal is to find and analyze the \textit{relationship} between VA events~\cite{fujiwara2018concise}, we decided to use KGs as one of the cornerstones of our approach, \textit{Knowledge-Decks}.

In order to perform provenance, however, one must first collect information from the user. Existing works have collected changes in datasets~\cite{da2009towards}, updates in visualizations~\cite{battle2019characterizing, xu2020survey}, and other similar events in order to recreate and recall user behavior. On the other hand, tracking user's \textit{Tacit Knowledge} as defined by Federico et al.~\cite{federico2017role} is either done by manual feedback systems~\cite{bernard2017comparing, mathisen2019insideinsights}, by manual annotations over visualizations~\cite{soares2019vista}, or by inference methods that attempt to extract users' insights by recording the users' screens, video or logs and extract from them interactivity patterns as a post-mortem task~\cite{battle2019characterizing, guo2015case}. Among these VA systems, InsideInsights~\cite{mathisen2019insideinsights} is an approach to recording insights through annotations during the user's analytical process. It has demonstrated that collecting user annotations is a legitimate way to extract and store user insights. Similar to InsideInsights, Knowledge-Decks also requests user annotation through text inputs, but different from other works, Knowledge-Decks calculates the similarity of the text input to inputs from other users in order to coalesce the collective knowledge discovery into a single structure.
%




After collecting data from users' knowledge discovery process, we must impose a structure to it. The structural design through knowledge ontologies has been a central goal of KGs~\cite{fensel2020introduction}. KGs~\cite{chen2020review, guan2022event} is a widely used technique to store and structure knowledge and has shown to be a successful instrument to interpret, store and query explicit knowledge, such as the information within Wikipedia~\cite{auer2007dbpedia}. KG methodology defines that the relationships between nodes of a graph network, and not just the nodes themselves, can represent knowledge~\cite{guan2022event}. Among similar structures, Event KGs~\cite{guan2022event} expands on this by including the concept of event-based data by describing that paths of the graph's relationships depict sequences of events over time.

In order to store and analyze a KG, graph databases (e.g. neo4j~\cite{noauthororeditorneo4j}) have emerged, exhibiting a wide array of techniques. Such specialized databases are needed since the structure of KGs focuses less on the usual transactional or row-based structure of typical relational databases and their transactional structure~\cite{cashman2020cava}. Applications of KGs are also part of the overall graph network research, which means that graph network techniques such as Graph Neural Networks (GNNs)~\cite{jin2020gnnvis}, graph visualizations~\cite{chang2016appgrouper, he2019aloha} and graph operations~\cite{ilievski2020kgtk}, such as page rank and traveling salesman, can be applied to KGs.

Among works that utilize KGs to model and store the user's knowledge discovery process, VAKG~\cite{christino2022data} connects the VA knowledge modeling theory with KGs to structure and store knowledge provenance. Indeed, as is later discussed, our solution is built around the VAKG methodology. We have modeled two VA tools, one of which is described here, in light of existing knowledge models. From this methodology, we developed a KG ontology (or schema) to store users' knowledge discovery process. With VAKG we can identify what should be collected from the user. However, it by itself does not aid in performing data collection, nor aids in defining how to analyze or how to extract KG event paths (e.g. slide decks).
%

Certain limitations from these works show that tracking automatically-generated insights~\cite{spinner2019explainer} or accounting for automatic computer processes~\cite{federico2017role} is still a challenging and largely unsolved problem. Additionally, since few existing frameworks or systems attempt to recall and retell a user's knowledge discovery process, no existing work solves how to link user-related and computer-related provenance for retelling the collective knowledge discovery of all users as far as the authors are aware. To solve that, we have opted to follow the argumentation of Xu et al.~\cite{xu2015analytic} and focus on manual means to collect user insights and intentions while automatically collecting computer-processes~\cite{federico2017role, battle2019characterizing}, such as user interaction. We also automatically link all user-related to their respective computer-related events as described by VAKG~\cite{christino2022data}.


Outside the realm of provenance, other researchers have attempted to analyze KGs to extract information. CAVA~\cite{cashman2020cava}, for instance, enables the exploration and analysis of KGs through visual interaction. KG4Vis~\cite{li2021kg4vis}, on the other hand, uses the advantages of the KG structure to provide visualization recommendations. ExeKG~\cite{zheng2022exekg} uses KGs to record paths of execution within data mining and machine learning pipelines. Indeed, there is an ever-increasing number of works using KGs in VA, all of which agree that KGs are suitable for knowledge analysis and extraction. Our novelty among such works is the usage of KGs to structure the knowledge discovery process as paths within the KG, which are then reformatted as slide decks for presentation purposes.

Hardly any research has attempted to model user behavior or the associated knowledge discovery as we aim to do. Info-graphs are one of the most used to encode storytelling visually~\cite{knaflic2015storytelling}. Indeed, Zhu et al.~\cite{zhu2020survey} details how info-graphs can be automatically generated through machine learning~\cite{chen2019towards} or pre-defined rules~\cite{gao2014newsviews}. When applied to storytelling, displaying insights as visual or textual annotations on top of visualizations has been the aim of several works~\cite{gao2014newsviews, chen2020augmenting, chen2019towards}. Although their results are very relevant for storytelling, they only tell what insights were found, not the process of how a user might have reached them. Instead, slide decks are better suited to narrate the events which lead to the insight~\cite{schoeneborn2013pervasive}.

Regarding slide decks, StoryFacets~\cite{park2022storyfacets} is a system that displays visualizations both in dashboard and slide-deck formats. Although slide decks have well-researched limitations~\cite{knaflic2015storytelling}, StoryFacets also argue about how and why slide decks are advantageous when one wishes to narrate a sequence of events. Indeed, slide decks are yet to be dethroned as the best way to give presentations~\cite{schoeneborn2013pervasive}. Of course, other works have succeeded in recalling and retelling the knowledge discovery process in other means~\cite{xu2015analytic}. For instance, Ragan et al. ~\cite{ragan2015characterizing} list many tools that collect and structure user behavior and insights in a queriable format. Nevertheless, they have not proposed a means to extract slide decks from user behavior to narrate the knowledge discovery process to third parties.

\section{Methodology}

This section describes the goals that motivated \textit{Knowledge-Decks}, its design, and its implementation.

\subsection{Research Question and Goals}\label{sec:goals}

In this paper, our research question is: \textit{\textbf{how to automate the process of creating a draft slide deck presentation that tells the story of which and how insight(s) were achieved using a VA tool?}} In other words, how to automatically create linear storytelling based on VA's non-linear knowledge discovery process? To answer this, other sub-questions emerged, such as: what would cause a user to reach such an insight? Which questions have the users had to ask throughout their interaction with the tool to reach such an insight? Which other insights were found along the way? And how to automatically generate a draft slide deck presentation that can be used to show these answers to others. As discussed in Sec~\ref{sec:related}, current solutions do not generically solve this, so that the same process could be applied to different VA tools. A generalized solution that addresses these primary and sub-questions should not just retell a user's insights, but also create or propose new optimized knowledge discovery paths extracted from the experiences of many users.
%

This led us to develop an approach that attaches itself to existing VA tools to (1) collect users' knowledge discovery processes and, from it, (2) extract knowledge discovery paths to (3) retell the underlying story using a slide deck. Based on existing methods of modeling, collecting, and analyzing user behavior and knowledge discovery, we identified our goals as: 

\squishlist
\item[G1:] Collect user's intentions and insights while using a VA tool;
\item[G2:] Collect the user behavior and assign which intentions/insights were related to them;
\item[G3:] Provide a way to extract knowledge discovery paths as narrations of user's experience;
\item[G4:] Format the knowledge discovery paths as slide deck presentations;
\squishend


To extract knowledge discovery paths from a collection of user experiences, we followed existing works by utilizing \textit{Knowledge Graphs (KGs)}~\cite{guan2022event} to structure and store user intentions, behavior, and insights. We also utilized the methodology of VAKG~\cite{christino2022data} to model the expected user experience based on the tools they would be using to derive what pieces of data should be stored within the KG given our goals. In order to apply VAKG, however, we first need to describe what kind of VA tools are targeted by our system.

After analyzing existing VA tools, we decided to focus on a specific subset of tools that provides visualizations to better understand and explore the available data through exploratory data analysis (EDA). More specifically:

\squishlist
\item The tool must be web-based and be comprised of a set of visualizations that displays data but does not allow any modification or addition of new data;
\item The tool may also have means to filter the data through visual selection, aggregations, or other manual means;
\item The web-based tool must allow our system to recall any previously visited state through its URL;
\squishend

We can see from the tool's requirements that there are no domain-specific or data-type requirements. However, the tools should aim to use visualizations to allow users to explore and better understand the available data. Indeed, this common point provides us with the means by which we can apply the same VAKG modeling to the tools to extract and retell users' knowledge discovery paths of any such tool, promoting reproducibility and generalization of \textit{Knowledge-Decks}.


\subsection{Modeling and Schema Design}\label{sec:modeling}

\begin{figure}
    \centering
    \includegraphics[width=\columnwidth]{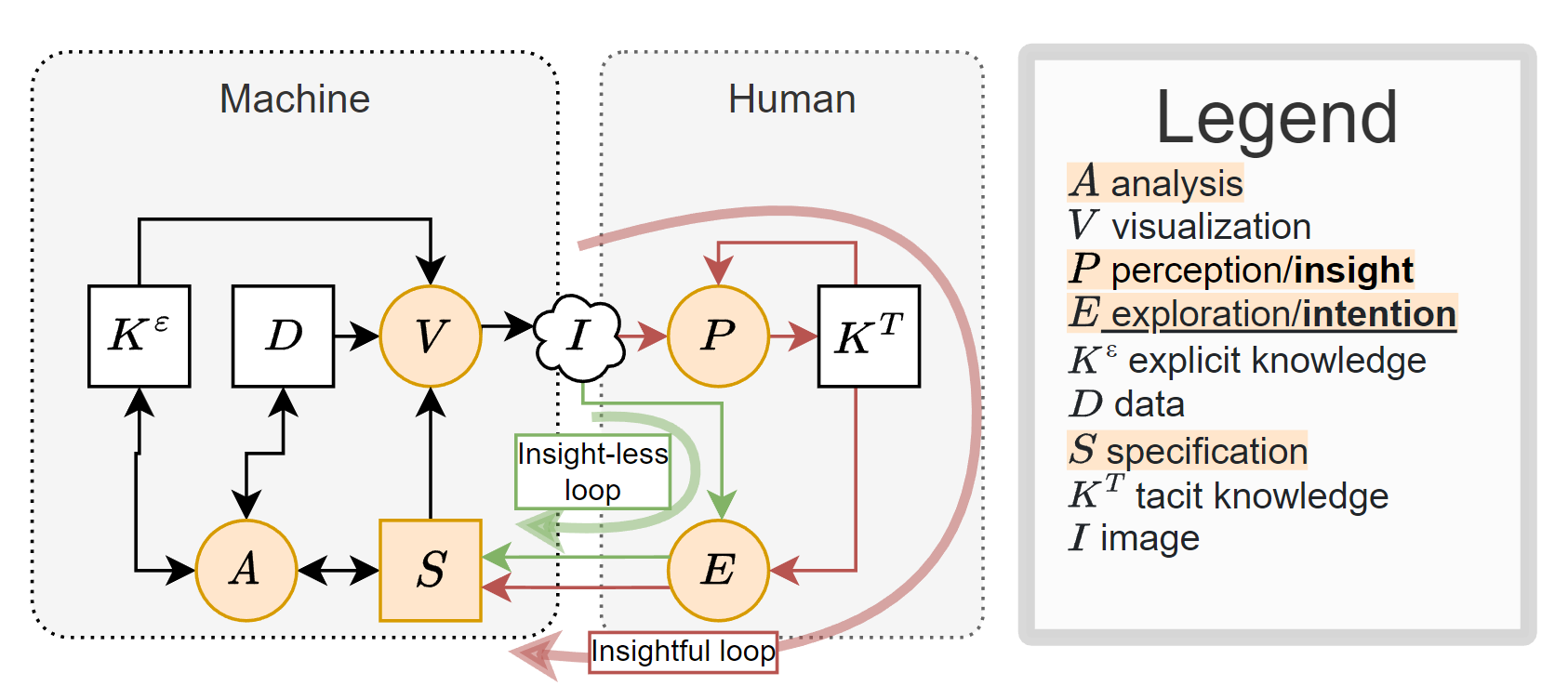}
    \caption{Simplified knowledge-assisted VA model. In red from $V$ to $E$ is the \textbf{insightful flow} where the user has a new insight, while in green is the \textbf{insight-less flow} when the user performs an action without any new insight. The two parallel red and green links between $E$ and $S$ represent the user's intent when a new specification, such as an interaction with some visualization, is made. This causes the execution of some algorithm $A$ and/or the generation of a new visualization $V$. Nodes in orange identify which data will be tracked in our system. }
    \label{fig:flow}
\end{figure}

The first step of VAKG~\cite{christino2022data} is to formally define the expected knowledge model as a flow graph. Thus, we took the existing flow graph example from VAKG and removed the aspects of VA not present within the proposed target tools. For instance, one aspect of the general model not present in the target tools is the ability for users to modify the data. The resulting model is presented in Fig.~\ref{fig:flow}.

This flow graph shows four connections between the \textsc{Machine} and \textsc{Human} sides, two in each way. Within the \textsc{Human} side, we see the possibility that the user gained new tacit knowledge due to a new perception ($P$) and interacted with the tool ($E$). That said, the user may also interact with the tool without any new perception, usually when multiple interactions are done without any new knowledge being gained. On the \textsc{Machine} side, we see how explicit knowledge $K^\epsilon$, data $D$, and schema $S$ generate new visualizations $V$ for the user. As further detailed by others~\cite{federico2017role, christino2022data}, this flow graph indeed models the knowledge flow while users are using a VA tool. Next, VAKG requires the creation of the Set-Theory equations. From these equations, we can formally define the schema of our KG and its property maps, which is the definition of the expected content of each of the KG's node types. The equations are listed below:

\begin{align}
U^h_t =& \{P_t, E_t\}, & U^h_t \equiv U_t &\cap H_t \\
U^m_t =& \{V_t, A_t\}, & U^m_t \equiv U_t &\cap M_t \\
C^h_{t+1} =& \{K^T_{t+1}\} \equiv U^h_t(C^m_t), & C^h_t \equiv C_t &\cap H_t \\
C^m_{t+1} =& \{D_{t+1}, S_{t+1}, K^{\epsilon}_{t+1}, I^{t+1}\} \equiv& \\
& U^m_t(C^m_t) + U^h_t(C^h_t), & C^m_t \equiv C_t &\cap M_t  \nonumber
\label{eq:VAKG}
\end{align}

VAKG ontology refers to classes of events and their relationships within a KG. By following the same definition of~\cite{christino2022data}, we have four classes: the \textsc{Human} temporal sequence, which holds the user's temporally-dependent events, such as insights, perceptions, and explorations; \textsc{Human} state-space, which holds the state-space of all user events; \textsc{Machine} temporal sequence, which holds all computer-side events like changes to the dataset, filters applied to the data, automated processes, and changes to the visualization; and \textsc{Machine} state-space, which holds the state-space of all computer-side events. Note that the temporal aspect is the difference between a ``temporal sequence'' and a ``state space''. If many users perform the same exact interaction, each user's interaction will create a new ``temporal sequence'' event. However, only one state-space event would be created. Our model also uses the same $9$ relationships of VAKG~\cite{christino2022data}.

The property maps define the content of each event class and relationship. For that, we follow a design based on the equations above, where the relationships hold no content, and each event class holds information collected from the tool, as will be discussed later. Note that this design choice aims to be simple to explain and reproduce but could be constructed differently depending on other goals or requirements. The full ontology schematic of our KG is included as supplemental material.

From the description of the VA tools being targeted, the tool has one or multiple interlinked visualizations where the raw data is visualized. As per the requirement, the only interaction events available in the tool are the filters, selections, and aggregations being applied at any given time and the state of each visualization, such as map position, zoom, or the selection of which part of the data is to be visualized. Therefore, the data $D$ and explicit knowledge $K^\epsilon$ in Fig.~\ref{fig:flow} do not change over time. Only filtering ($A$), aggregation operations ($A$), and selections ($S$) are expected. As we discussed previously, new \textsc{Machine} temporal sequence events are created at every new user interaction. Therefore we consider the above-described information as our KG's property map definition, which also includes a timestamp and an anonymous id of who made any given interaction.

On the \textsc{Human} side, the property map is required to contain the events happening on or within the user while using the VA tools. In order to keep the design simple while also being able to answer the hypothesis of Sec.~\ref{sec:goals}, we deviate from related works which capture user interactivity and only process it into a specific format after the fact~\cite{battle2019characterizing, guo2015case}. In other words, we do not use recording software such as cameras or screen capture to populate said property map. Instead, we use a more active method to extract user events: by requesting user inputs~\cite{mathisen2019insideinsights, bernard2017comparing}. Namely, by pairing the flow of Fig.~\ref{fig:flow} and the hypothesis of Sec.~\ref{sec:goals}, we identified two main user events to track: user intentions and user insights. In short, user intentions relate to anything the user wishes to do, which is represented by all arrows from $E$ to $S$. User insights relate to any new information perceived by the user, which is represented by all arrow sequences between $I$ to $E$. These two events represent the possible property maps of the \textsc{Human} space-state and \textsc{Human} temporal sequence classes. Also, similar to before, the \textsc{Human} temporal sequence property map has a timestamp of the event and an id of the user.

Finally, after performing preemptive tests with the KG defined so far, we identified one issue: when two users were to type the same insight or intention, it was very uncommon for the text to match exactly. In order to better allow for matching texts of similar intentions or insights, we modified the property map of the \textsc{Human} space-state class. Instead of holding the text inputted by the user, the property map holds a set of keywords that represents said text. The exact process of calculating the keywords, how all property maps fit within a KG, and how the user interaction can be collected and stored for analysis will be discussed in the next section.

\subsection{Knowledge-Decks}
\label{sec:system}

With the tool modeled following VAKG, we now discuss how we implemented and attached them to the VA tools. Our implementation of the VAKG method has two components: a python web server with a graph database for storage and a library with which tools can use to send the expected events to the web server.
%


\noindent\textbf{Knowledge Graph Database.} First, to collect and store the VAKG data, our implementation must receive and store events in a database while following the design of Sec.~\ref{sec:modeling}. We chose Python as the language of choice for its wide usage within the community and Neo4J~\cite{noauthororeditorneo4j} as the graph database. We chose to use an API-based design where our code creates a server to which any tool can send data through the web~\cite{leonardo_christino_2021_5750019}. With this, any number of VA tools following the same knowledge model can connect and send their events to the same VAKG implementation. It is important to note that the implementation provided is specific to the modeling process above but can be modified to satisfy other models.


\noindent\textbf{Collecting \textsc{Machine} Events. } Next, to populate the machine-side events, we are required to keep track of the VA tool's state at all times. For this, we developed a library with which the VA tools can inform of any new events. The VA tools being used are required to be web-based tools implemented in JavaScript. Hence, the library to collect and send all events from the VA tool to the web-server are implemented to be used as a JavaScript library. The library is also publically available~\cite{leonardo_christino_2021_5750019}.

In short, the library implements two methods with which the VA tool informs the web-server of any new interaction or change within itself. This requires the tool to inform its current state as a hashable dictionary (or hashmap), which is transmitted to the web server as JSON and then stored within our KG. This information will also be used later to recall the state of the VA tool during the analysis of the resulting KG. Since the most common manner to recall a state in a website is through its URL, we extract the current state of the VA tool in this way. Therefore, this implementation leaves the VA tool to control what is included in VAKG's property maps.

\noindent\textbf{Collecting \textsc{Human} Events. } Similar to the \textsc{Machine} events, the library also provides ways for the VA tool to send \textsc{Human} events to the web-server. One core difference, however, is that we expect users to type their insights or intentions as a sentence. We extract keywords from the text, and the resulting combination of keywords plus text is sent to the web-server. In order to generate such keywords, we use the natural language processing library \textit{Spacy}~\cite{vasiliev2020natural} to retrieve the list of lemmas from a given natural language text.

From preliminary tests, we found that a more reliable way to compare previous texts (e.g. insights or intentions) and the new text being typed by a user is to request confirmation from the user. We also noticed that short, concise texts are more reliable than larger ones. Therefore, we limited the number of characters allowed per text to $75$. Once the text is typed, we gather similar texts and display them to the user asking whether any existing text is similar to theirs. To calculate this text similarity, we use the text's keywords (lemmas) and scan the database for similar keywords among the \textsc{Human} state-space nodes. The similarity measure used is a cosine similarity measure calculated from a word2vec representation of the keyword list and is also implemented by \textit{Spacy}~\cite{vasiliev2020natural}. If there is any match with a score greater than $0.5$, we collect the texts that generated these keywords, which can be accessed from its neighboring \textsc{Human} temporal sequence node, and display up to $5$ of these texts to the user ranked by similarity score. The user can say whether their intention or insight is equivalent to any of the existing ones suggested or if it is new. This user-feedback step provided a reasonably reliable way for our system to better match new intentions or insights with previous ones compared to purely automatic processes.

Finally, the \textit{Knowledge-Decks} asks whether there was an existing element in the VA tool, such as visualization, text, or color, which caused the user's intention or insight. The user is given the ability to draw circles or arrows on the screen in this step. Users can also say that nothing in the interface caused the new intention or insight. The drawn elements and the current URL of the website, similar to Collecting \textsc{Machine} Events, are also saved as part of the user's \textsc{Human} temporal sequence to recall the elements of interest.

With this, the flow of a new insight or intention from the perspective of the user is as follows: the user types a new entry within a text field of the VA tool and selects whether it is a new intention or new insight, the VA tool searches and displays similar intentions or insights from previous users and asks if the intention or insight is new or if it is equivalent as a previous one. The user is also asked to indicate what element in the VA tool, if any, caused this new intention or insight. After this flow, the KG will store the user's typed text, keywords, the website URL, the drawn elements, and whether this was a new intention or insight by following the structure defined in Sec.~\ref{sec:modeling}.

\noindent\textbf{Knowledge Discovery Recall as Narrative Slide Decks.} 
%
%
Once the VA tools and our approach can communicate the required data and structure it as a VAKG knowledge graph, we can analyze the extracted data. For that, \textit{Knowledge-Decks} accepts Neo4J's cypher queries~\cite{francis2018cypher} as the means to extract information from its KG. 

Cypher queries are one of the main ways to query for raw or aggregated data from graph databases~\cite{francis2018cypher}. Its format follows strict specifications, where nodes and/or relationships of interest are defined within the query to specify either aggregation values or graph paths of interest. For instance, we can verify how many users have used the system by aggregating all \textsc{Computer} temporal sequence nodes with different users and retrieving its count. We can also extract the path within the graph a particular user took from a particular intention to a specific insight, which returns us an ordered list of \textsc{Human} temporal sequence nodes the user took from beginning to end. This flexibility provides endless ways to analyze and extract data from the KG. However, we will focus on one goal: extracting a narration of the user's knowledge discovery process as a slide deck.

Given a KG collected from several users utilizing a VA tool, we aim to construct a set of cypher queries that narrates the tool's knowledge discovery process. The first question is: how to extract the user's insights that began with a known intention? Given an $\$intention$, the cypher query below solves this question, discovering all insights \textit{the same user had} which follows the intention in question.
\begin{verbatim}
MATCH ((n:H_UPDATE {text:$intention, label:'intention'})
    -[:FOLLOWS_INSIGHT*1..20]-(n1:H_UPDATE {label:'insight'})) 
RETURN *
\end{verbatim}

\begin{figure}
    \centering
    \includegraphics[width=\columnwidth]{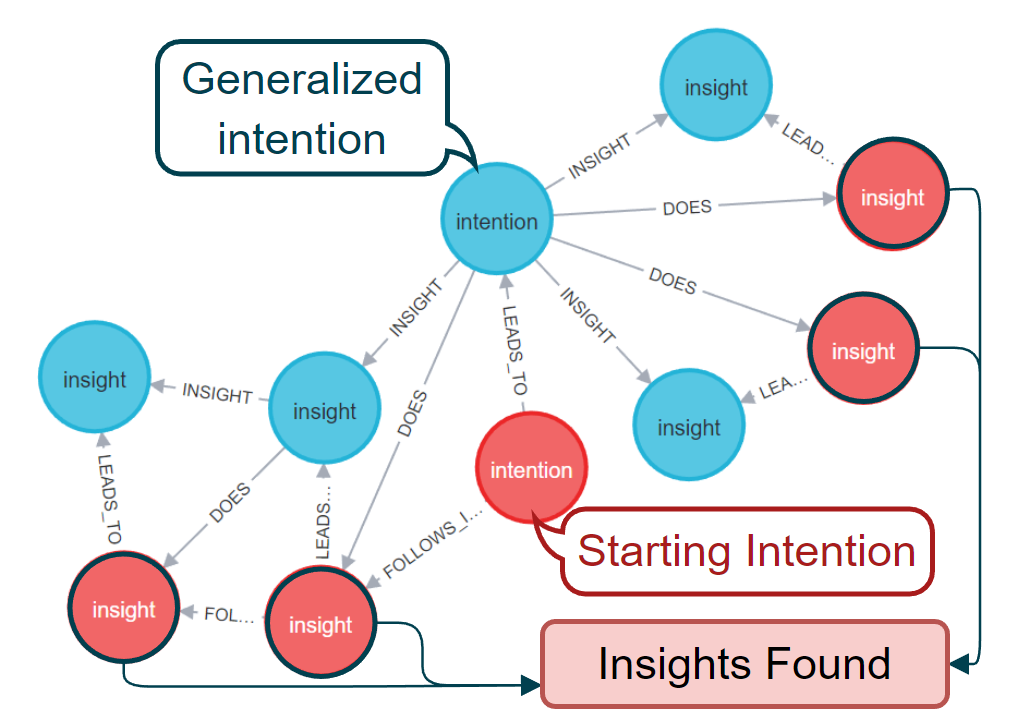}
    \includegraphics[width=0.3\columnwidth]{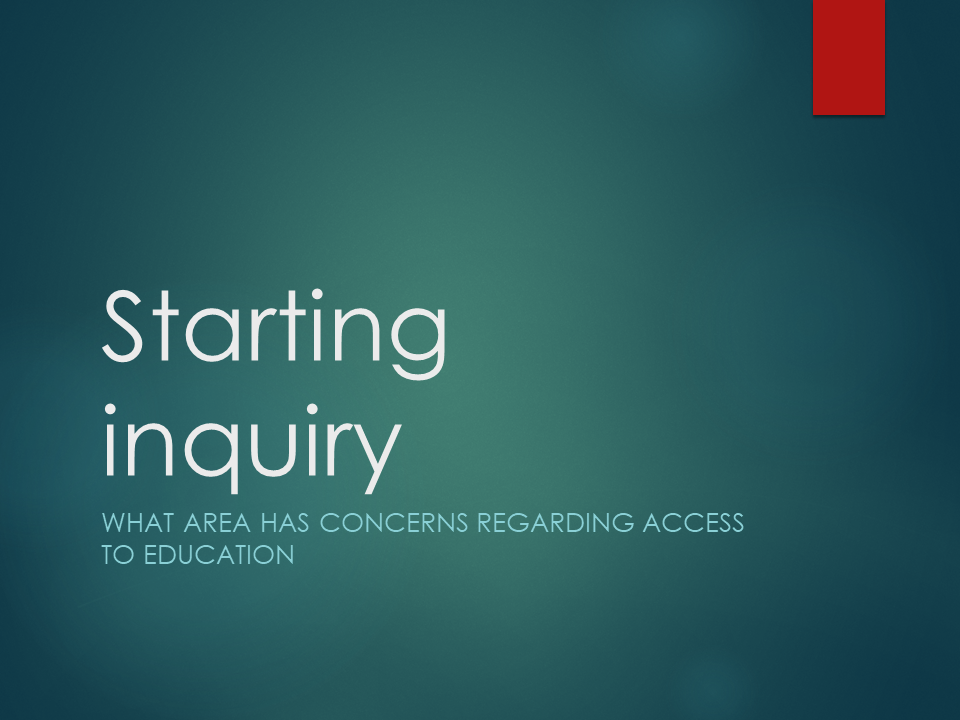}
    \includegraphics[width=0.3\columnwidth]{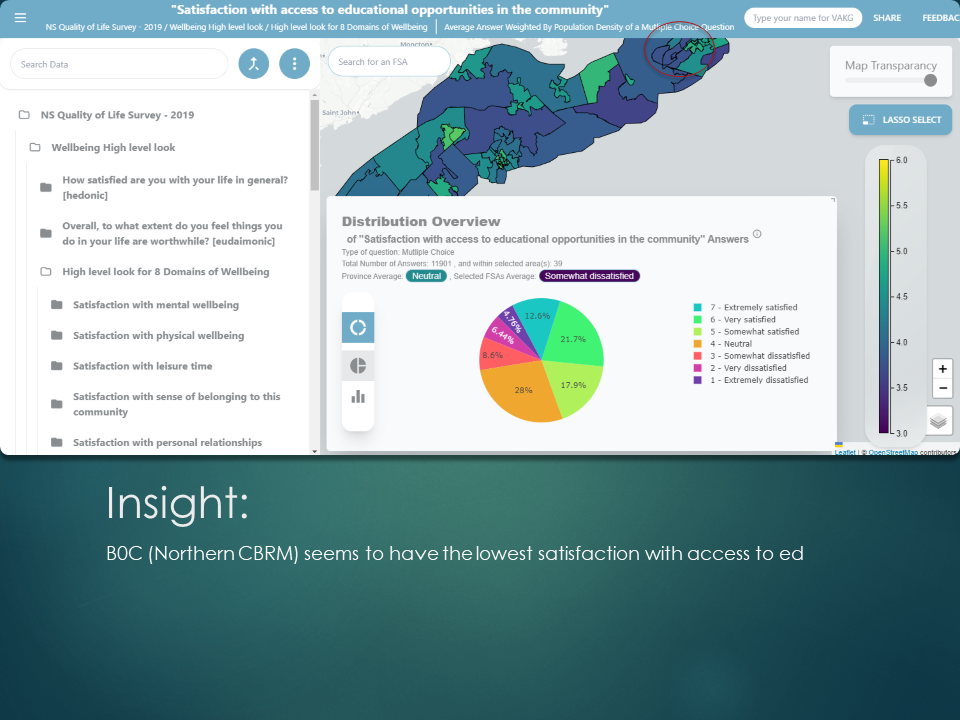}
    \includegraphics[width=0.3\columnwidth]{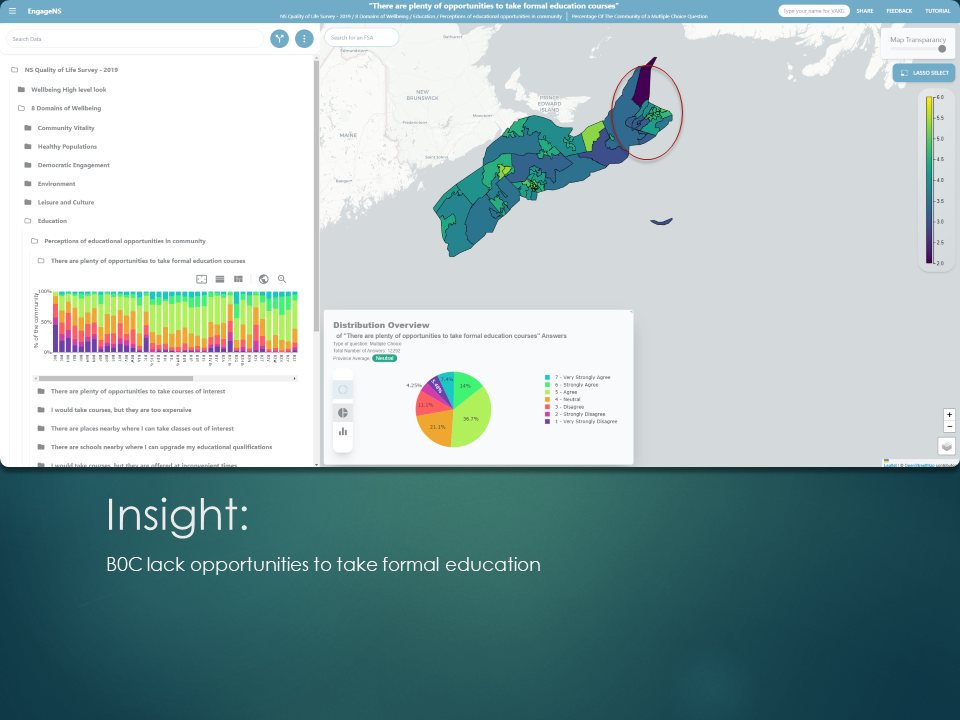}
    \includegraphics[width=0.4\columnwidth]{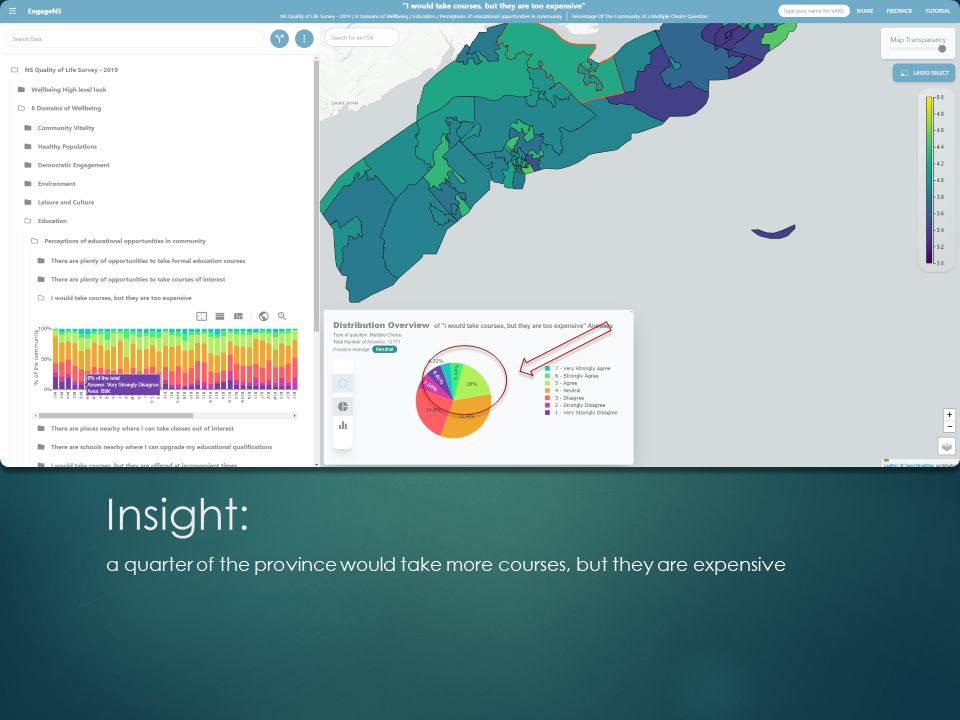}
    \includegraphics[width=0.4\columnwidth]{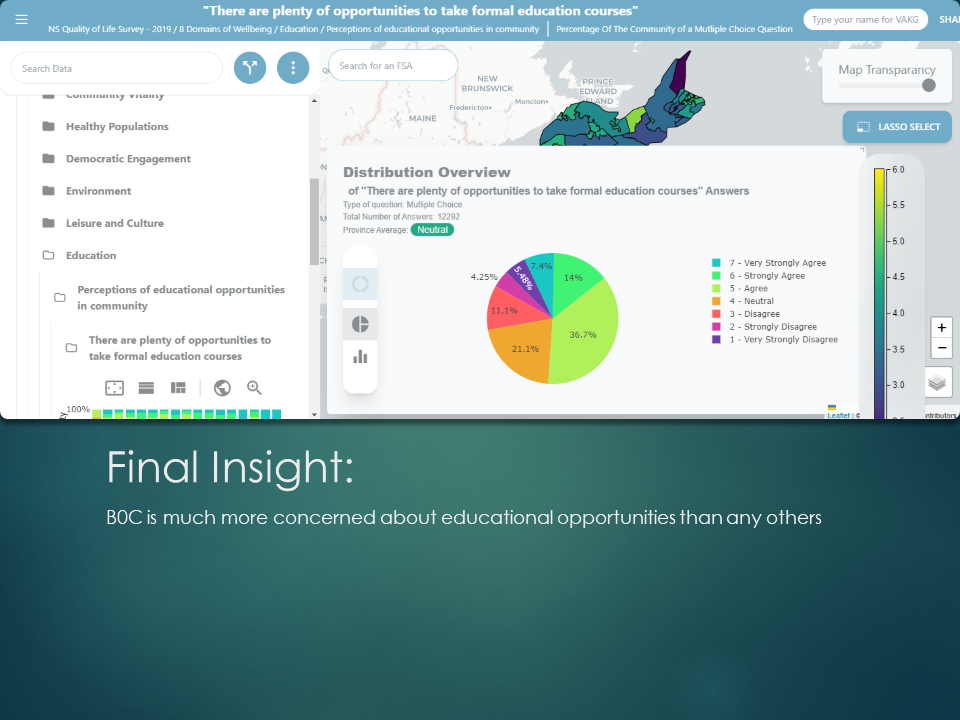}
    \caption{Graph visualization from Neo4j~\cite{noauthororeditorneo4j} when collecting all insights from all users given a starting intention. In this example, four insights were found from said intention, two of which were from the same user who wrote the intention (two to the left) and two from other users (two to the right). The slide deck generated on the bottom describes the four insights found given the original intention. }
    \label{fig:findinsights}
\end{figure}

Optional parameters could be included, such as $RETURN * LIMIT 1$ to only return the \textit{first} intention the user had or $RETURN * ORDER BY n1.created DESC LIMIT 1$ to only return the \textit{last} intention the user had. Now, in order to find \textit{all insights} from \textit{all users} which originated in a given intention, we use the query below. The result of running this query is exemplified in Fig.~\ref{fig:findinsights}.

\begin{verbatim}
MATCH intention_path=
    ((n:H_UPDATE {text:$intention, label:'intention'})
    -[:LEADS_TO]->(i)-[:INSIGHT*0..20]-(j)-[:LEADS_TO]
    -(n1:H_UPDATE {label:'insight'}))
return nodes(intention_path)
\end{verbatim}
\label{cy:intentionpath}

\begin{figure}
    \centering
    \includegraphics[width=\columnwidth]{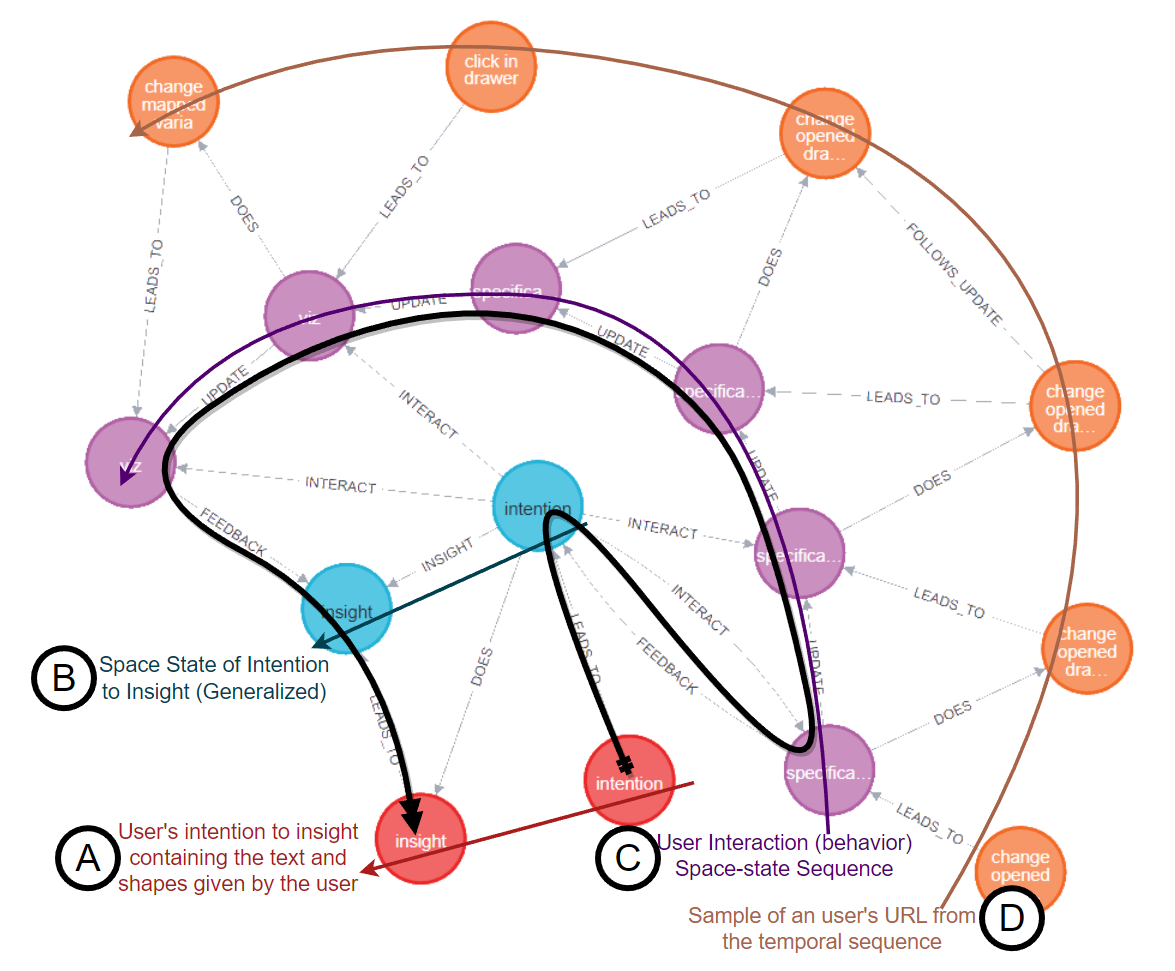}
    \includegraphics[width=0.22\columnwidth]{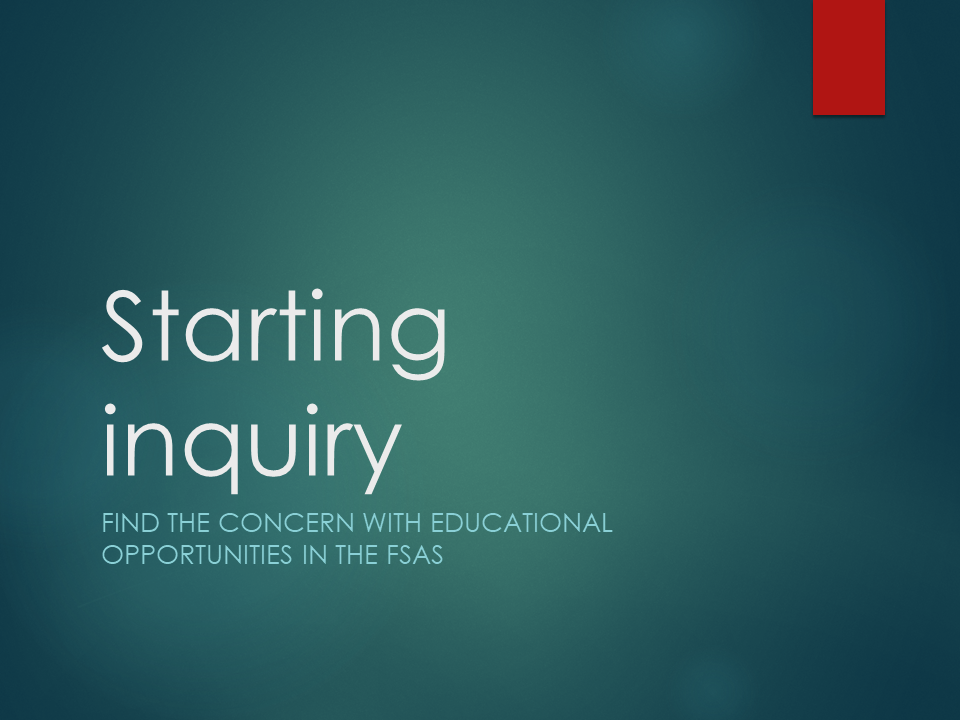}
    \includegraphics[width=0.22\columnwidth]{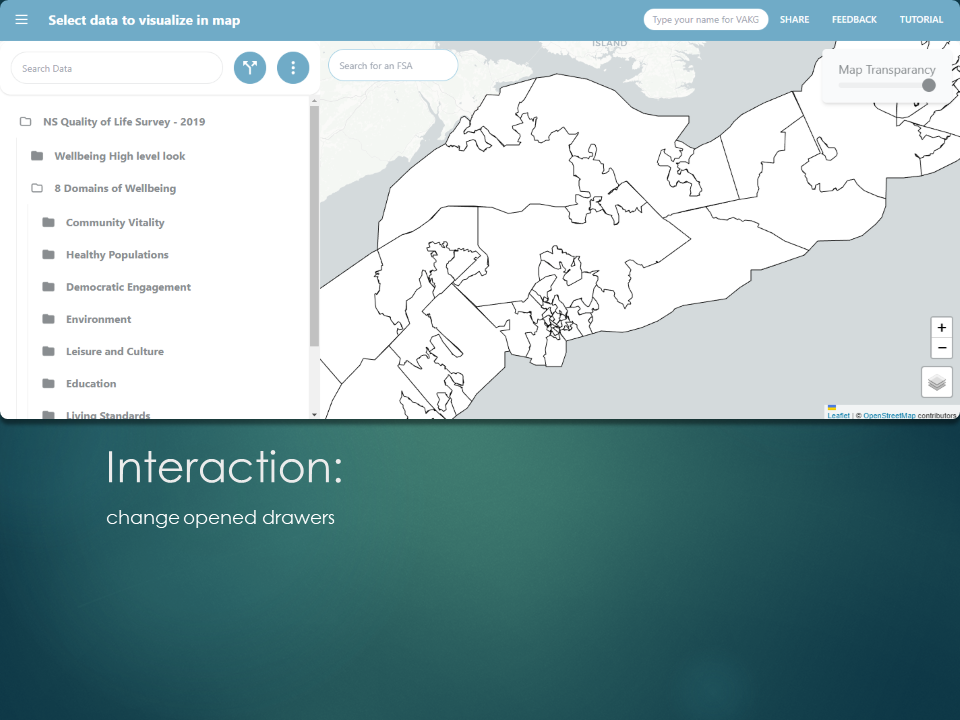}
    \includegraphics[width=0.22\columnwidth]{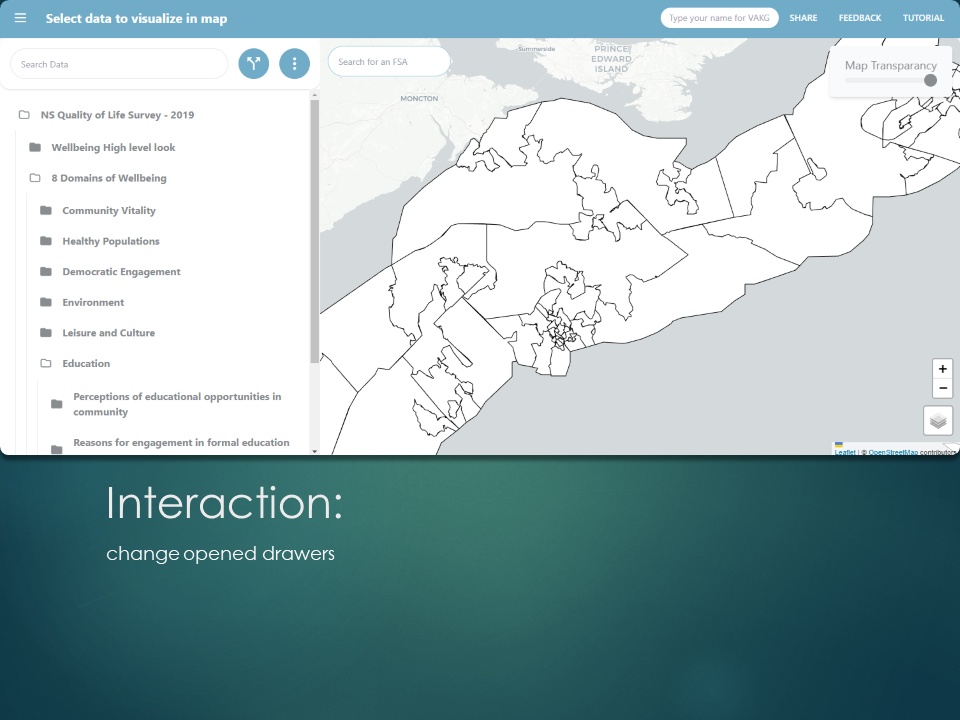}
    \includegraphics[width=0.22\columnwidth]{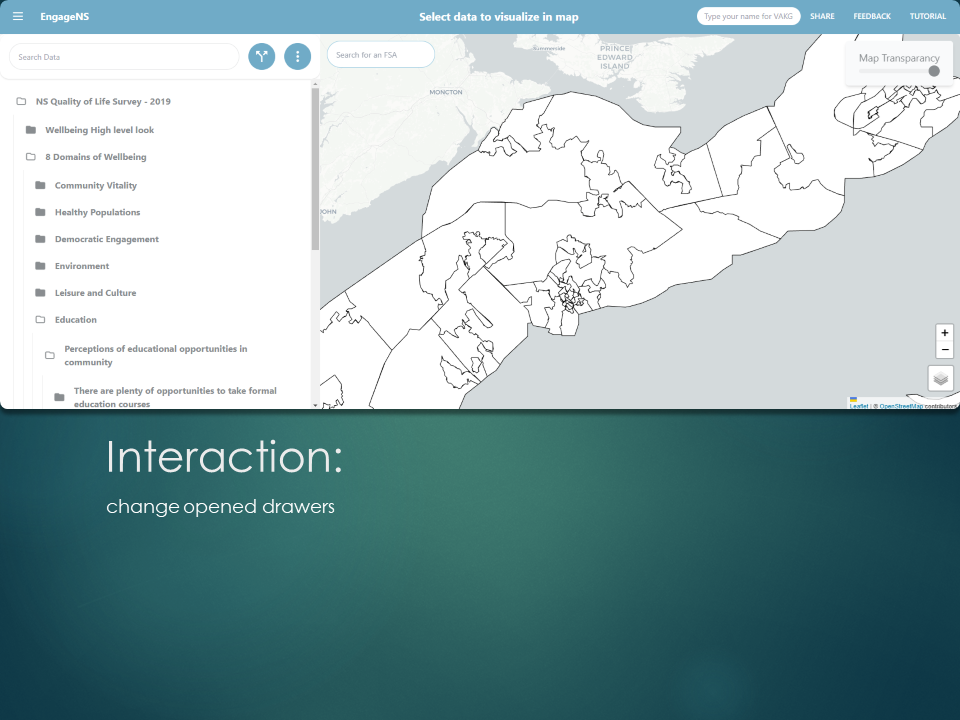}
    \includegraphics[width=0.22\columnwidth]{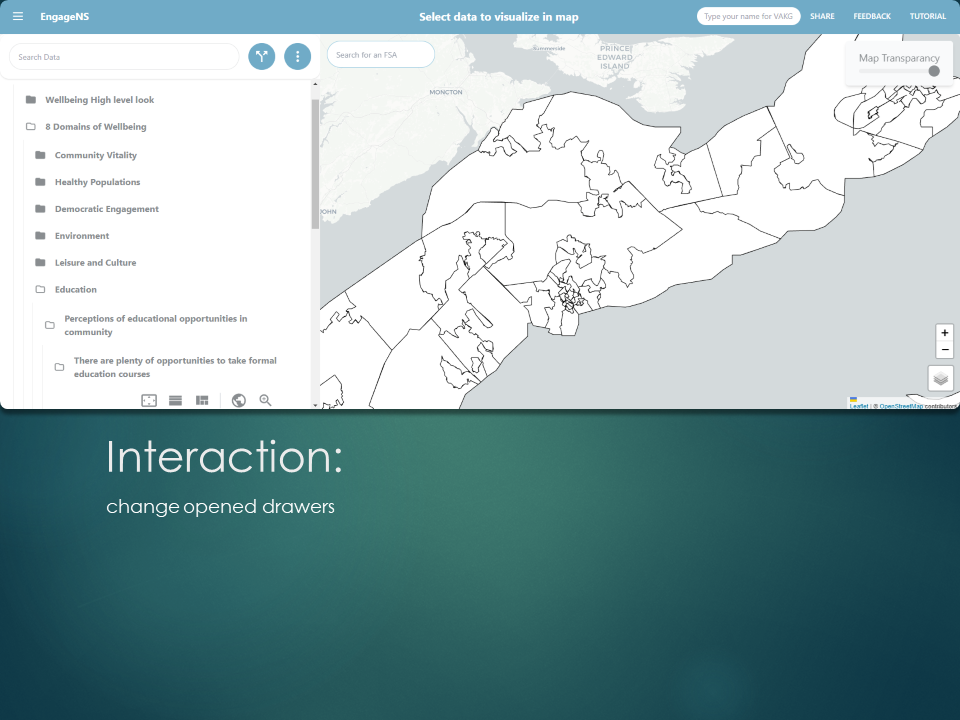}
    \includegraphics[width=0.22\columnwidth]{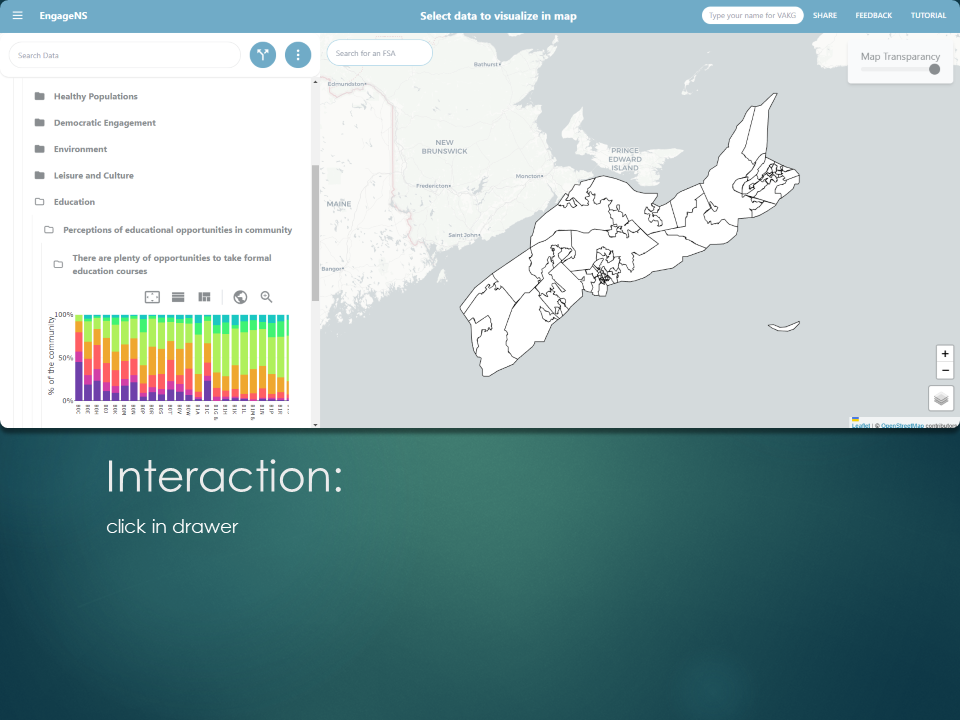}
    \includegraphics[width=0.22\columnwidth]{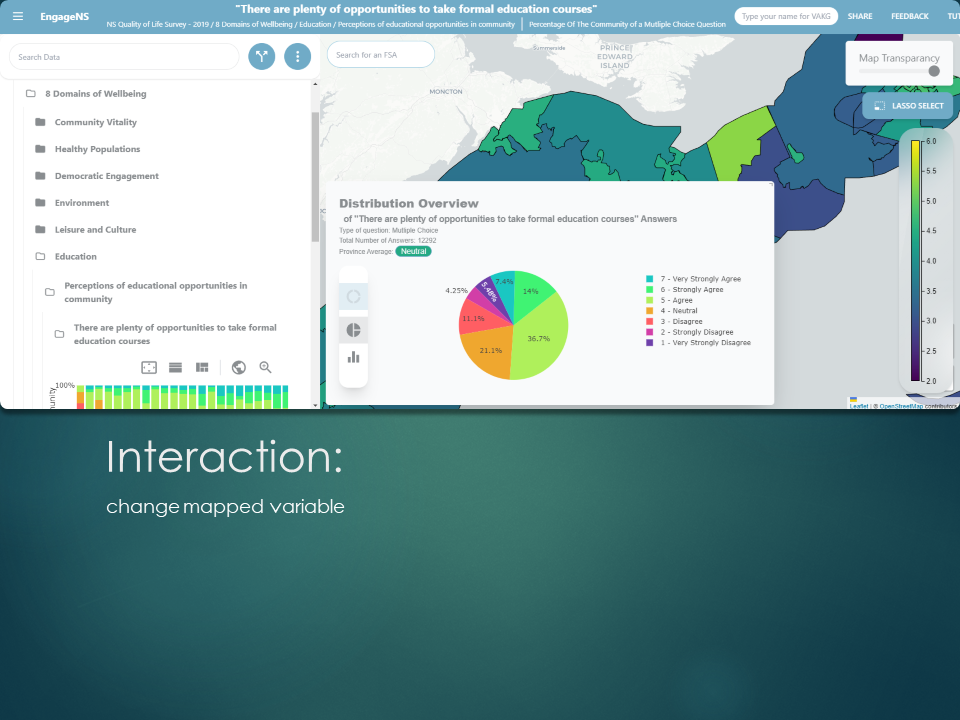}
    \includegraphics[width=0.22\columnwidth]{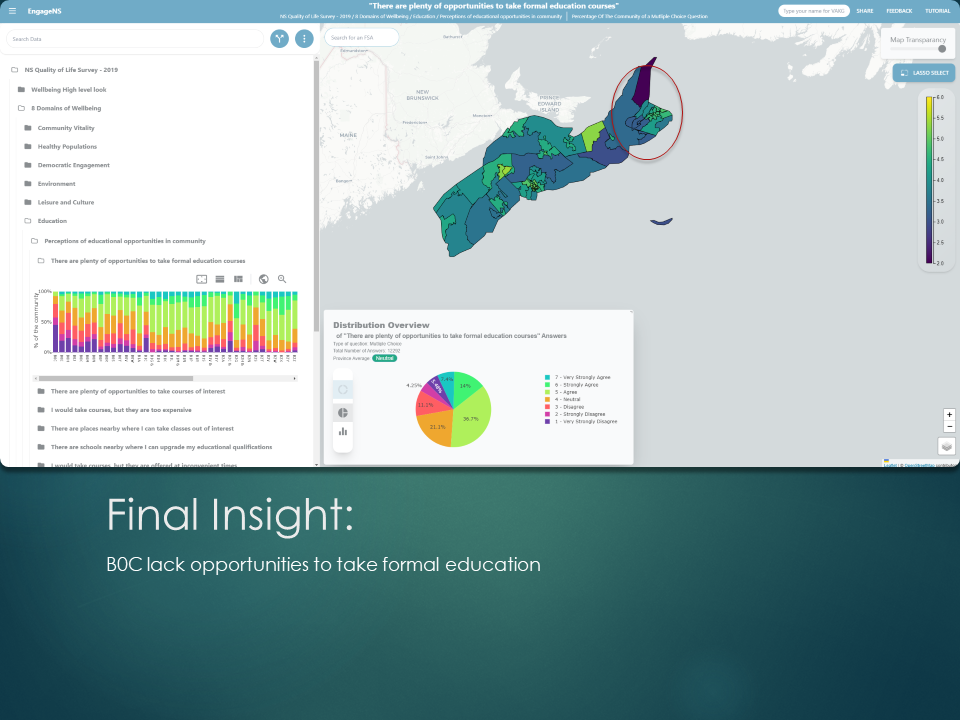}
    \caption{Above is a portion of KG extracted from \textit{Knowledge-Decks} when generating a slide deck, which contains all interactions done in the shortest between an intention and an insight. Visualization created with Neo4J browser~\cite{noauthororeditorneo4j}. The graph's path highlighted in black is used to generate the slide deck below, which is to be read from left to right and top to bottom. }
    \label{fig:slidedecksampleflow}
\end{figure}

So far, we have not collected nor identified the user's behavior between intention and insight. Following VAKG's structure, we can discover the user's behavior between any pair of intention ($\$intention$) and insight ($\$insight$) by finding the user's behavior path ($behavior\_path$). The $behavior\_path$ output of the query below describes the complete path starting from a user's insight (\textsc{Human} temporal sequence), into an insight space state (\textsc{Human} space state), and runs through all \textsc{Computer} space states until the path can connect to the intention which started it all, as shown in Fig.~\ref{fig:slidedecksampleflow}. Note that within this query, the $intention\_path$ retrieves the \textit{shortest path} of interaction events from intention to insight calculated from all users' paths combined. Therefore, this path may be and usually is, shorter than any of the individual users' paths.

\begin{verbatim}
MATCH intention_path=
    ((n:H_UPDATE {text:$insight, label:'insight'})
    -[:LEADS_TO]-(i)<-[:INSIGHT*0..20]-(j)-[:LEADS_TO]
    -(n1:H_UPDATE {text:$intention, label:'intention'}))
MATCH behavior_path=
    ((n1)-[:LEADS_TO]-()-[:FEEDBACK]-(c)-[:UPDATE*0..20]
    -(m:C_STATE)-[:FEEDBACK]-()-[:LEADS_TO]-(n))
WITH nodes(behavior_path) AS px 
    ORDER BY length(behavior_path) LIMIT 1
UNWIND px AS mainPath
OPTIONAL MATCH (mainPath)-[:LEADS_TO]-(interactions:C_UPDATE)
RETURN mainPath, interactions
\end{verbatim}
\label{cy:behaviorpath}


Other cypher queries were also developed, such as: which insight was the one \textit{most found} among different users, what is the \textit{longest} acyclic path recorded between an intention and a given insight, and what are \textit{all intentions} which unimpededly culminated in a given insight (with paths). These and other queries are available in a public repository~\cite{leonardo_christino_2021_5750019}. \textit{Knowledge-decks} also allows custom cypher queries to be run, allowing for extensive flexibility in its usage in other use cases.
%

With the collected paths of each of these queries, we can retrace the user's behavior using the information from the \textsc{Computer} temporal sequence nodes. For this, we sample one \textsc{Computer} temporal sequence node (orange nodes in Fig.~\ref{fig:slidedecksampleflow}) to obtain the actual user's interactions, including URLs, to recall each step of the process. The line with ``OPTIONAL MATCH'' in the query above collects the $interactions$ nodes for this purpose. Finally, we obtain the shapes drawn by the user from the final insight node (red ``insight'' nodes in Fig.~\ref{fig:slidedecksampleflow}).

To build a slide deck with the extracted data, \textit{Knowledge-Decks} navigates through the path generating one slide per node visited in the path. For instance, the slide deck from the graph in Fig.~\ref{fig:slidedecksampleflow} would start with a slide stating the user's intention, several slides with screenshots of the website, and a final slide with the insight reached. On the other hand, the slide deck of the graph~\ref{fig:findinsights} would, instead, have one slide per insight. Each slide can optionally include any collected information, such as the insight/intention text, text keywords, event name, URL, etc. For exemplification purposes, we have only included the insight or intention text and/or event name and a screenshot of the VA tool at that specific point in time.

To generate screenshots of the VA tools at a certain state, \textit{Knowledge-Decks} uses \textit{Selenium}~\cite{gojare2015analysis} to visit and take screenshots from the website and python-pptx~\cite{pptxpython} to generate the slides in PowerPoint format. Finally, any shapes drawn by users, such as circles or arrows, are added to their respective slides as PowerPoint shapes. The two slide decks generated from the graphs of Figs.~\ref{fig:findinsights} and ~\ref{fig:slidedecksampleflow} are below each respective figure.

The resulting slide deck can, of course, be edited by anyone and used in presentations or in any other way PowerPoints are used to retell the knowledge discovery story of the specific cypher query. Indeed, as part of our evaluation, we found that once the slide decks are generated and read by an expert, they were able to aggregate all insights throughout the slide deck and manually create a new \textit{conclusion slide} which concludes the knowledge discovery story with a description of what new knowledge was found.

\section{Application and Evaluation}

To evaluate our system, we have attached it to two VA tools that fit our pre-determined requirements (see Sec~\ref{sec:goals}), one of which is discussed here. \textit{Engage Nova Scotia}~\cite{graham} is an independent not-for-profit organization that started in June 2012. Its Quality of Life initiative, born in 2017, led to a survey in 2019 with 230 questions about the quality of life and well-being of residents across the province. Almost 13,000 people responded. From this data, the well-being mapping tool (WMT)~\cite{christino} was built to allow the population to access and explore the survey results. In their own words: the tool allows for visualization of the entire survey through maps and charts, which has already led to many actionable insights by the population, companies, and government entities. We can see that WMT matches our requirements due to its use of visualizations to better understand and explore the available data.

\subsection{Configuring VA tools for data collection}

In order to attach \textit{Knowledge-Decks} to WMT, we first need to define what data we are required to harness from the tool and its users to populate our KG. From Sec.~\ref{sec:modeling}, we see that in Fig.~\ref{fig:flow} there are five orange shapes: insight $P$, intention $E$, specification $S$, visualization $V$, and analysis $A$. We also see that $A$, $S$, and $V$ are \textsc{Machine}-related while $P$ and $E$ are \textsc{Human}-related (Fig.~\ref{fig:flow}). By following the definitions of these four taxonomies~\cite{christino2022data, federico2017role}, we defined what data was to be collected from each of the tools in Tab.~\ref{tab:propertymapdata}.

\begin{table}
\label{tab:propertymapdata}
\caption{Data stored in the Well-being Mapping Tool (WMT) KG node's contents (property maps). }
\begin{tabular}{p{16mm} | p{60mm}} 
 \textbf{Node Type} & \textbf{Data collected from the WMT}\\ [0.5ex] 
 \hline\hline
 \textsc{Human} temporal sequence & label (insight $P$ or intention $E$), created time, URL, screen size, text, keywords, shapes drawn, user id, and analysis id \\ 
 \hline
 \textsc{Human} state-space & label (insight $P$ or intention $E$), created time, last updated time, and keywords time \\
 \hline
 \textsc{Computer} state-space & label (specification $S$, visualization $V$ or analysis $A$), created time, last updated time, the status of the hierarchical structure to the left, bar chart's visualization schema (stacked, grouped, maximized, etc.), map position, map zoom, map areas selected, math operation used, and question visualized in map \\
 \hline
 \textsc{Computer} temporal sequence & event name, created time, URL, user id, analysis id, and all the same data from the related Computer state-space \\
\end{tabular}
\end{table}

\subsection{Evaluation through User Interaction}



To collect the required information, we asked $9$ users to perform pre-defined tasks with the tool to collect data and build the KGs. In short, after the participants were shown an introductory video, the survey asked participants to fill out a demographics questionnaire with $7$ questions, perform $5$ tasks, and fill out a set of Likert-style questions related to the participants' experience. The complete questionnaire can be found as supplemental material. The tasks were used to provide a common starting point and a common goal to all participants. During the tasks, participants were asked to write any intentions and insights they had, being part of or not related to the task at hand. Indeed, all tasks were exploratory in nature, and participants were asked and encouraged to freely roam the tool if they so wished. Among the participants, there were three experts on WMT, two non-experts, and two of unknown expertise level.

The data was fed to our system, and the resulting knowledge graph was vastly complex, with more than 900 nodes. Nevertheless, a preliminary analysis of the results showed that users had a total of 21 unique intentions and 26 unique insights. On the computer side, there were a total of 252 events or 143 unique events.
%
%
We also looked through all insights related to Q1 (first task) and found that the $9$ participants reached insights with an average of 20.9 interactions (machine-side events), and on average, 1.44 insights were reached by each participant within Q1.  Similarly, across all tasks, users had an average of 1.13 intentions and 1.29 insights per task. Overall, users had more insights than intentions. Also, on average, each intention took 6.11 interactions (machine-side events) to reach some insight.

\subsection{Slide Deck Generation}

We generated several slide decks from the collected data. Each slide deck type aims to tell a story from the collective experience of the users. As discussed in Sec.~\ref{sec:system}, our approach has different pre-defined queries, each of which tells a specific type of story. After generating the slide decks below, we discussed with an expert, who is proficient with the tool and its domain, whether and how she would use the generated slide decks and questioned her whether the slide decks provided new insights they were unaware of before the experiment.
%

\noindent\textbf{Collection of Insights given an Intent}: By using the $intention\_path$ query of Sec.~\ref{cy:intentionpath} to collect all insights users had given a known intention, we generated a total of 15 unique slide decks. One of such slides was done from the intention ``what area has concerns regarding access to education'' and had four related insights as seen in Fig.~\ref{fig:findinsights}. The resulting slide deck, shown in Fig.~\ref{fig:findinsights} bottom, discloses a list of four independent insights from three different users who had the same intention. From the slide decks above, we noticed that some of the slide decks displayed insights that are not directly to the intention of origin. 
%
%
In part, this is expected since, naturally, users can not be expected to only have insights that specifically answer their intention. Since analyzing the text content of insights and intentions and their correlation was not a goal in \textit{Knowledge-Decks}, we opted to focus on telling the users' stories until their last insight. For this, another pre-defined query generates the slide deck of the path taken from the intention in question to the \textbf{last} or furthermost insight within the KG. This new slide deck may include other insights the users had during their analysis, but if the user found an answer to their intention, this slide deck will tell the story of how the insight was reached, including all other insights along the way. As one can expect, these slide decks are larger, but they recall much more of the users' experiences.

\noindent\textbf{Behavior between intention and insight}: To generate a slide deck of how the VA tool was used to reach a given insight, we utilize the $behavior\_path$ query of Sec.~\ref{cy:behaviorpath}. Here, an important distinction has to be made: the story generated by these slide decks does not necessarily follow any user's behavior but instead uses the collection of all behaviors and extracts the shortest path from them. That is, in these slide decks, each slide may have originated from different users. However, it ultimately follows the optimal path of behavior (interaction events) between the intention and the insight. That said, a total of $61$ possible paths from intention to insight were found, each of which has \textit{one shortest behavior path} and many larger ones. For instance, the slide deck of Fig.~\ref{fig:slidedecksampleflow} was generated from the $behavior\_path$ query considering ``B0C lack opportunities to take formal education'' as the insight of interest and the ``Find the concern with educational opportunities in the FSAs'' as the intention of interest. Note that the insight originally came from a different user from the intention.

\begin{figure}
    \centering
    \includegraphics[width=0.28\columnwidth]{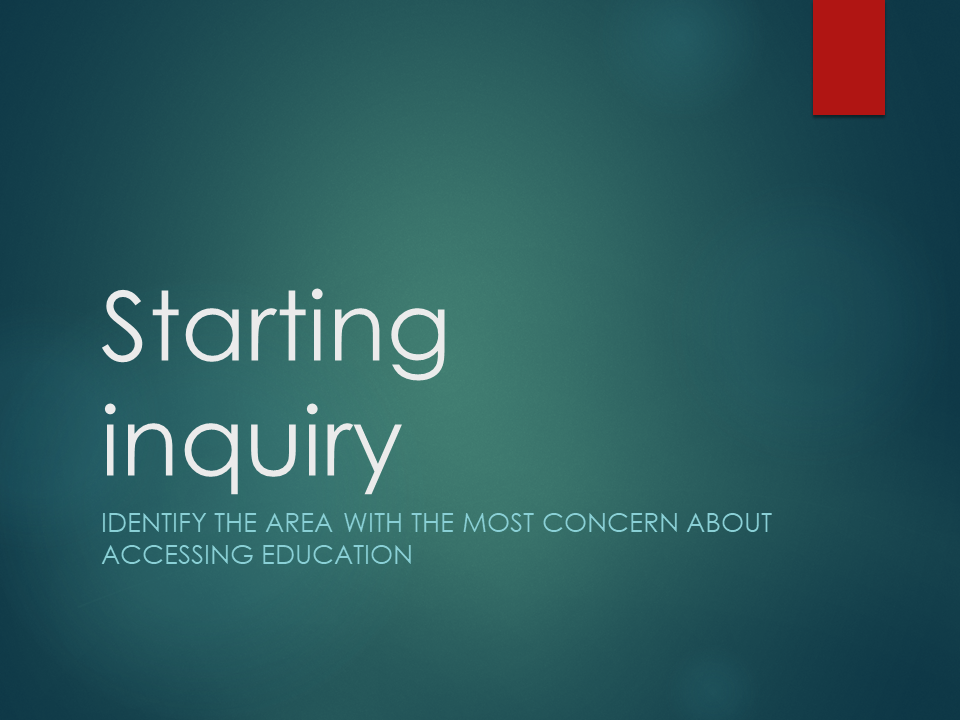}
    \includegraphics[width=0.28\columnwidth]{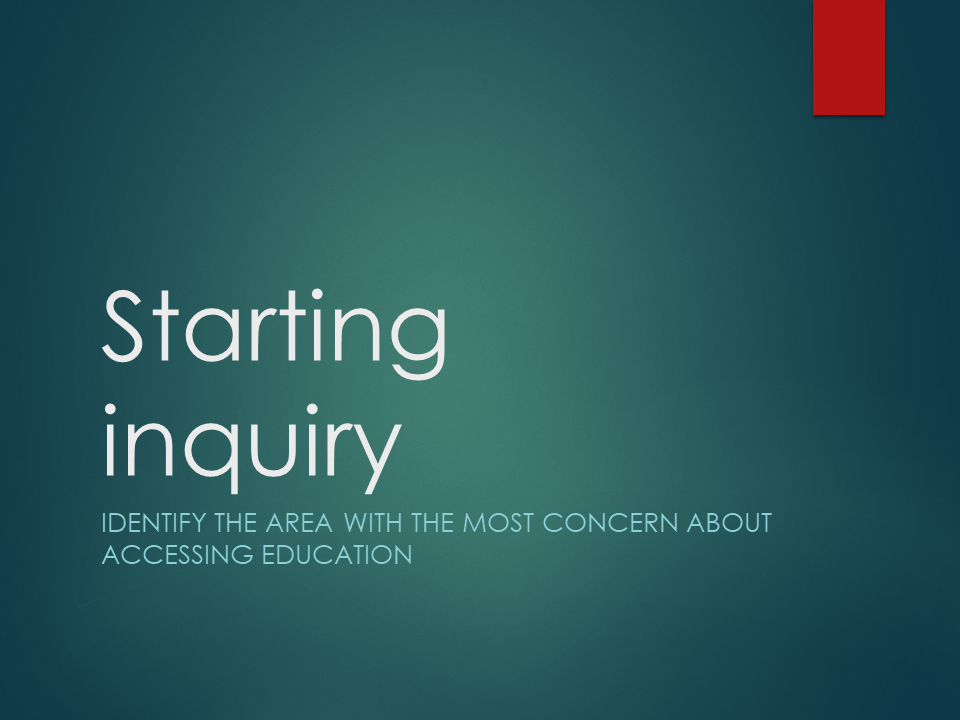}
    \includegraphics[width=0.28\columnwidth]{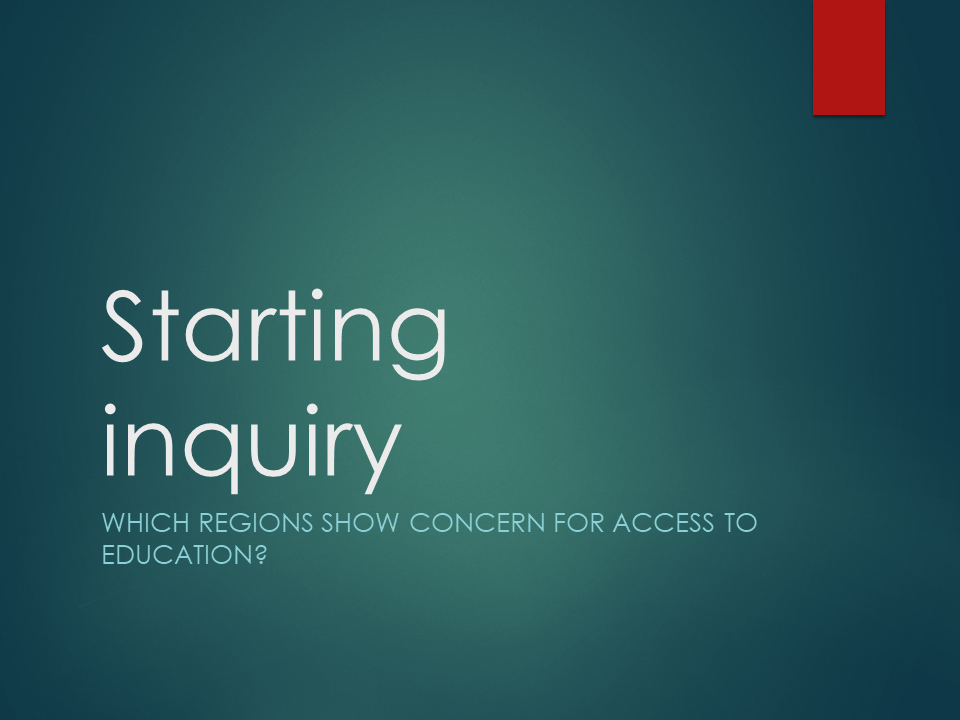}
    \includegraphics[width=0.28\columnwidth]{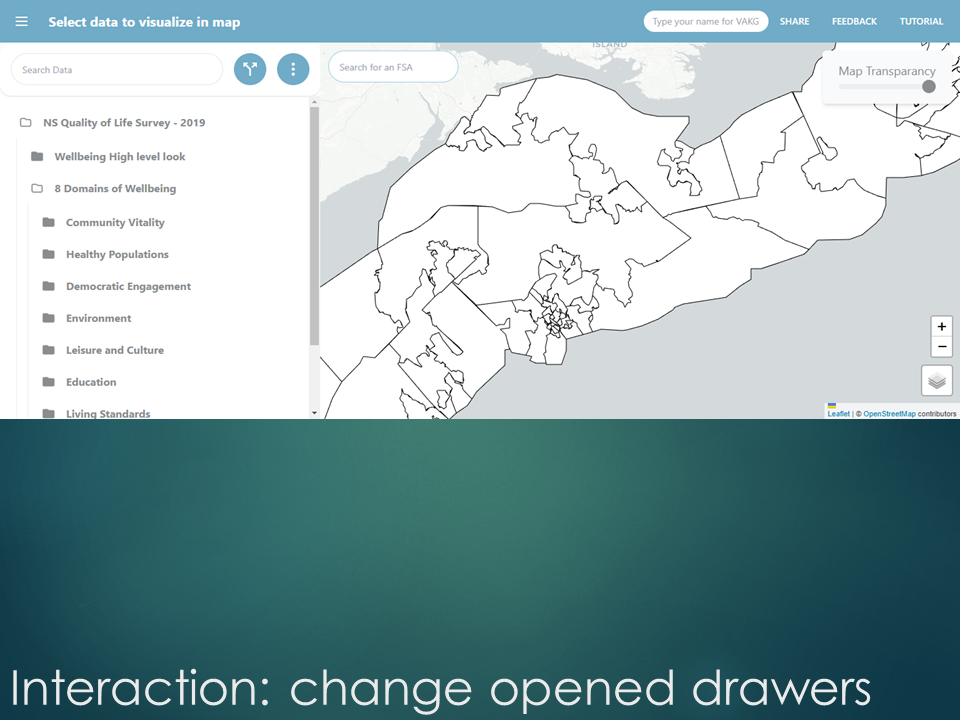}
    \includegraphics[width=0.28\columnwidth]{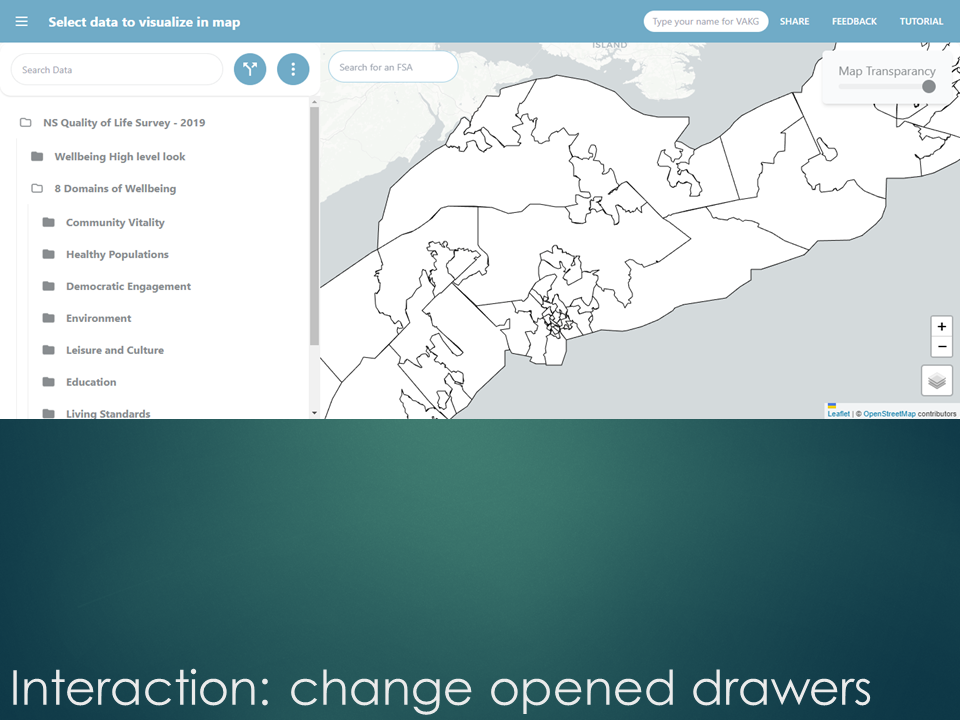}
    \includegraphics[width=0.28\columnwidth]{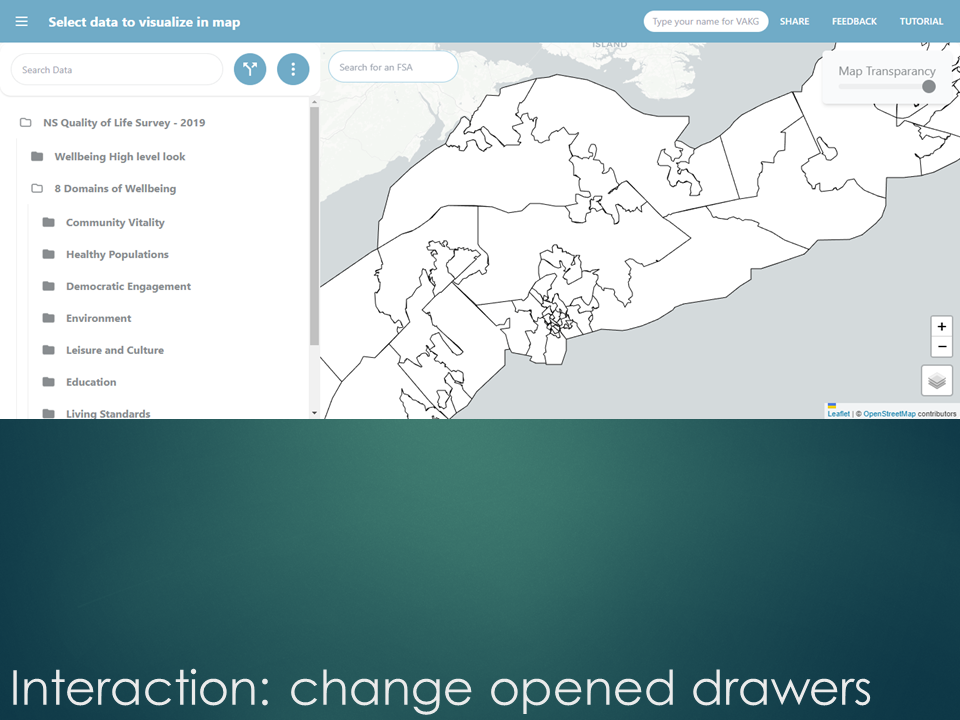}
    \includegraphics[width=0.28\columnwidth]{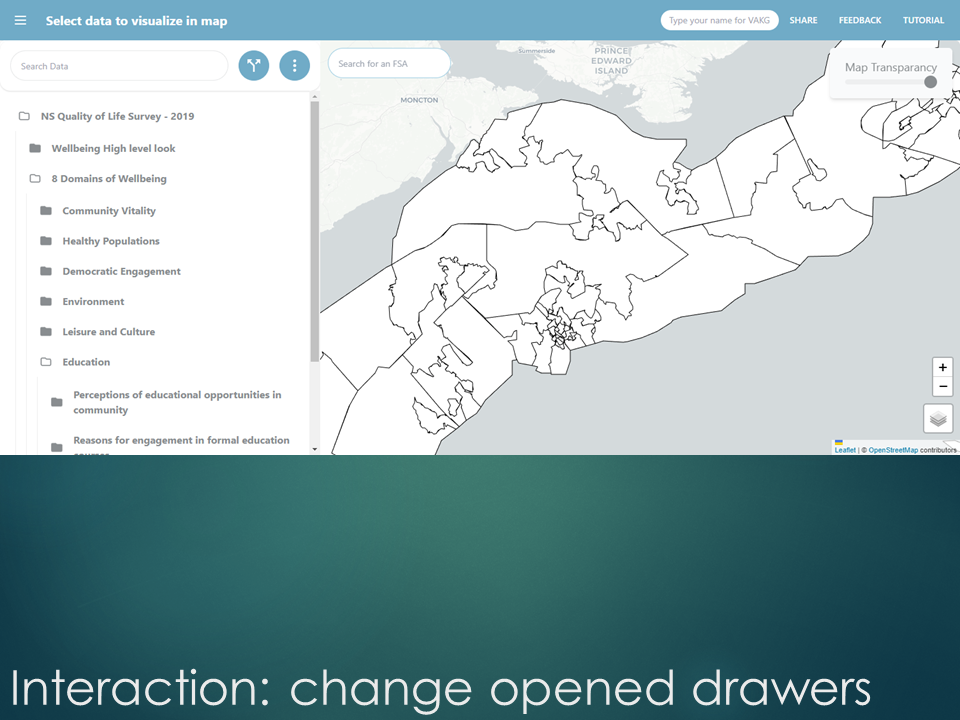}
    \includegraphics[width=0.28\columnwidth]{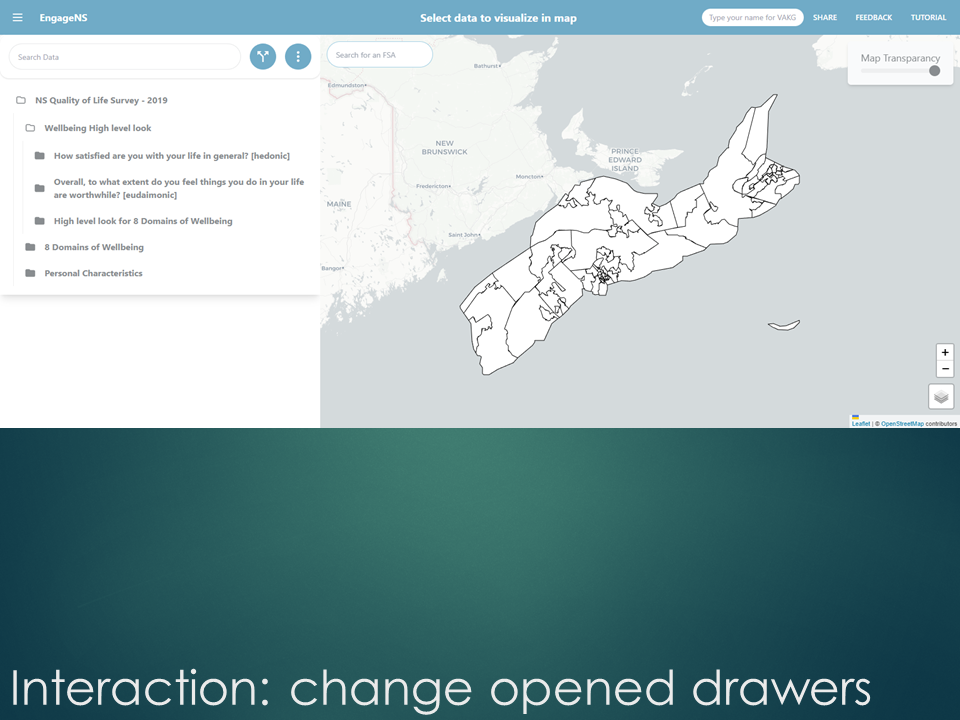}
    \includegraphics[width=0.28\columnwidth]{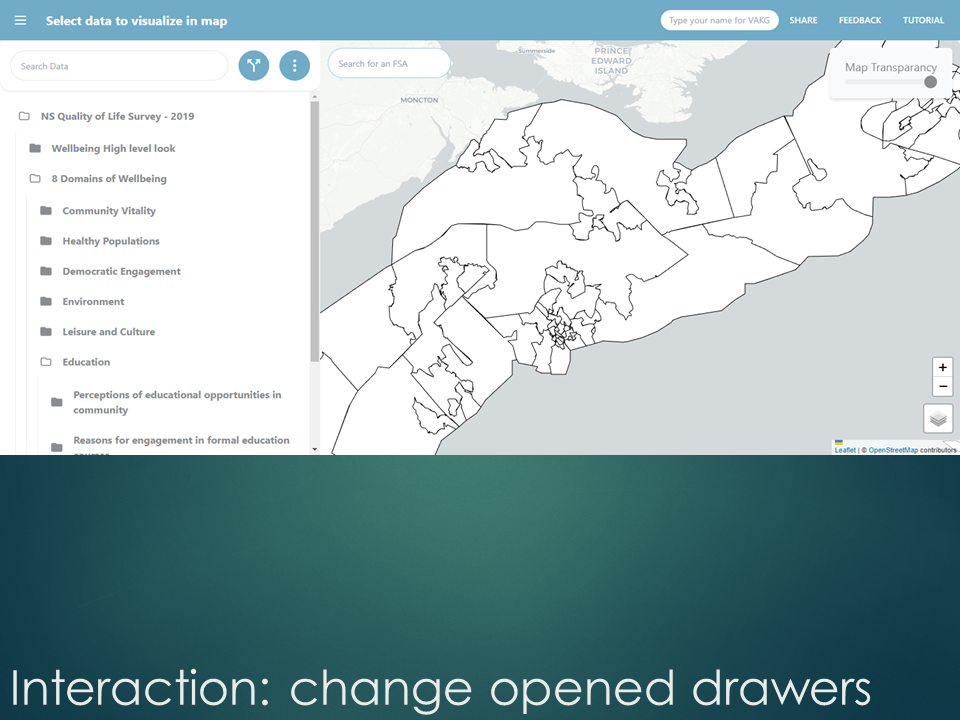}
    \includegraphics[width=0.28\columnwidth]{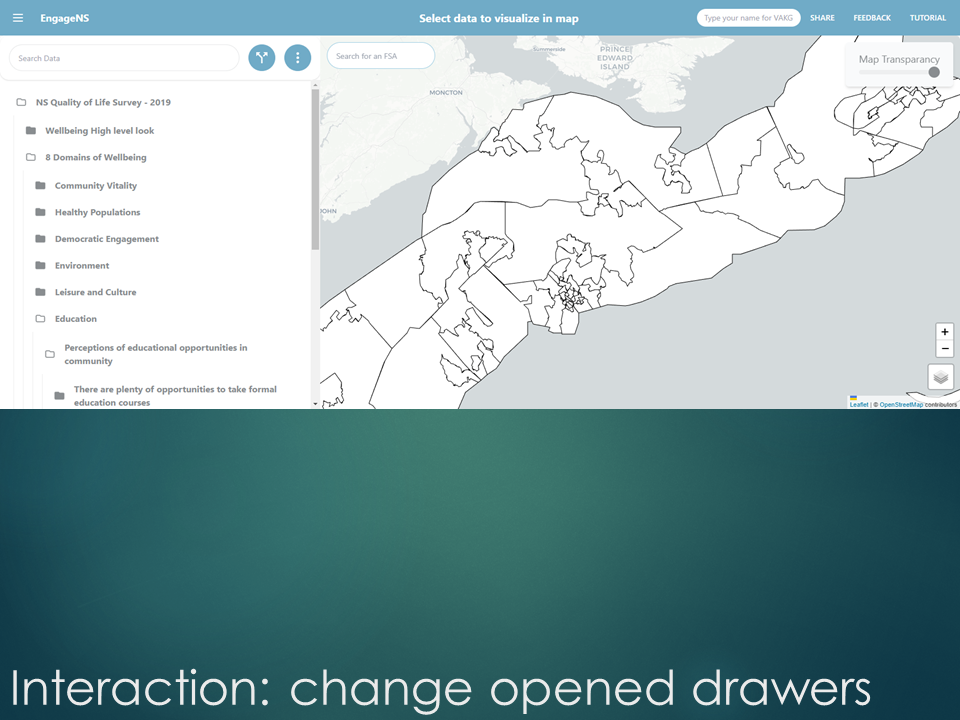}
    \includegraphics[width=0.28\columnwidth]{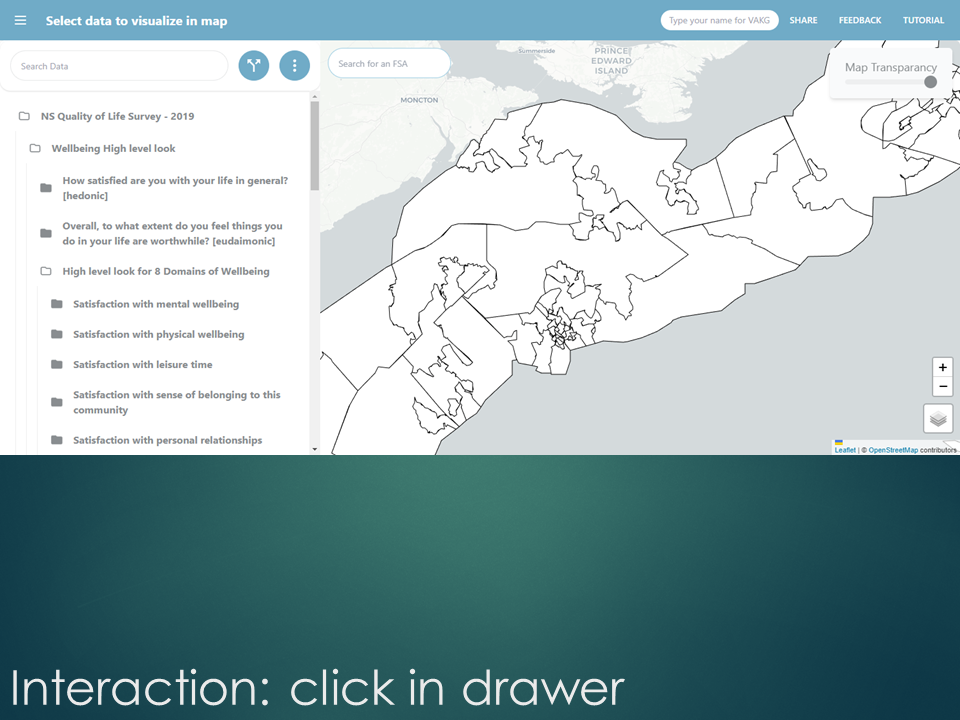}
    \includegraphics[width=0.28\columnwidth]{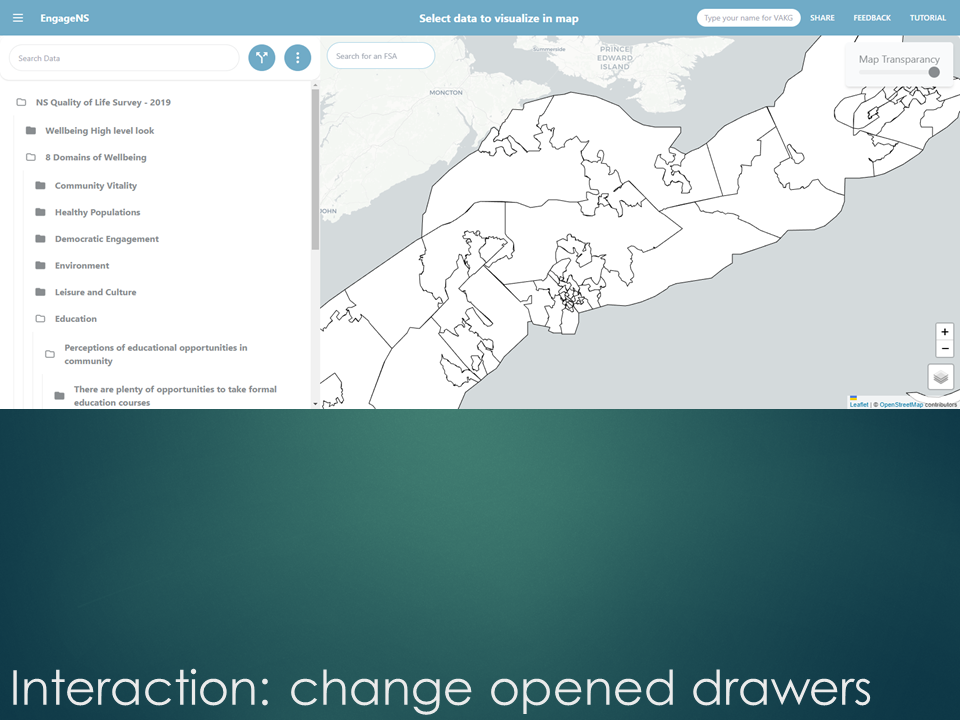}
    \includegraphics[width=0.28\columnwidth]{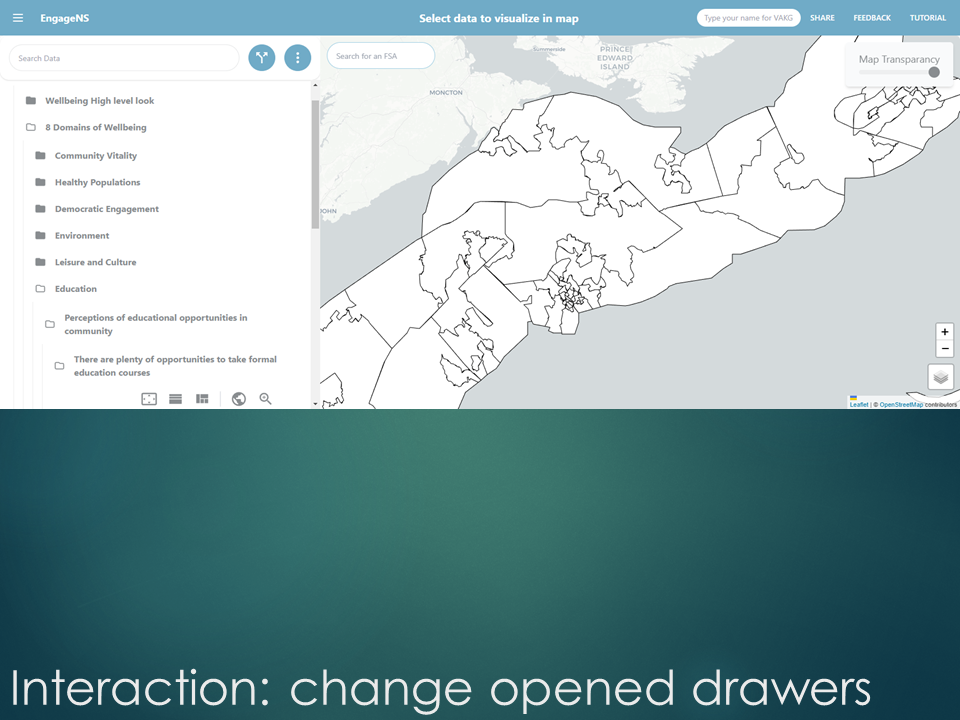}
    \includegraphics[width=0.28\columnwidth]{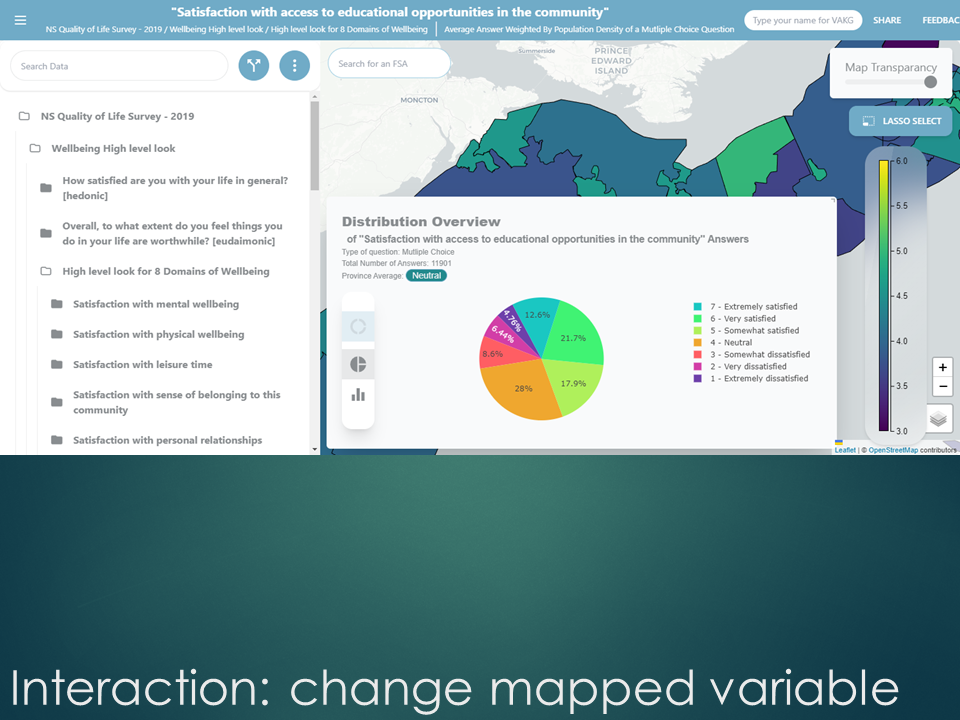}
    \includegraphics[width=0.28\columnwidth]{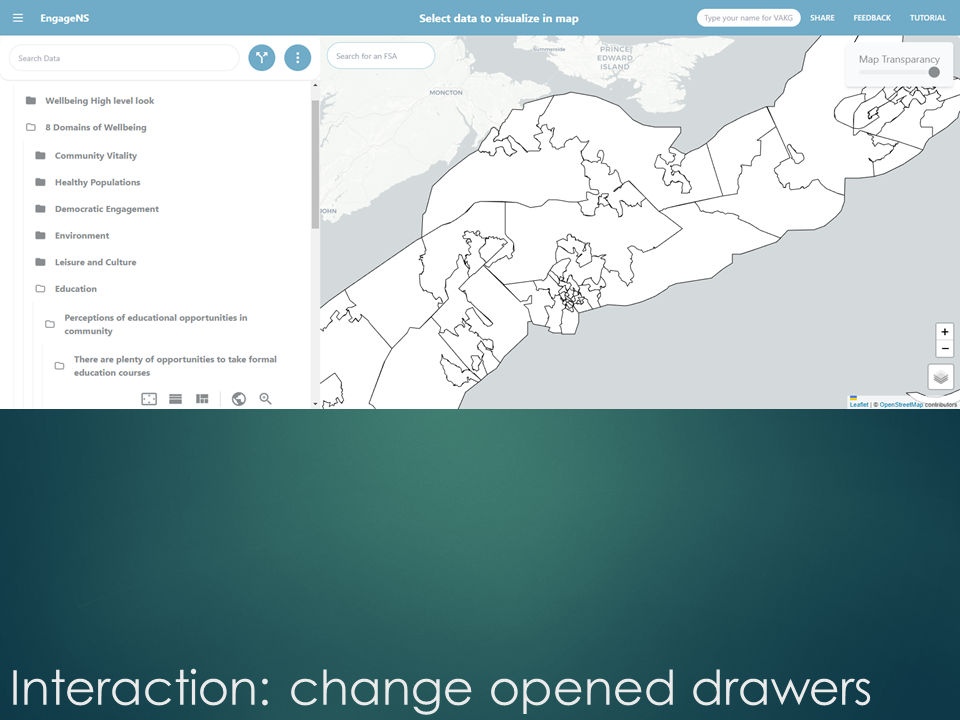}
    \includegraphics[width=0.28\columnwidth]{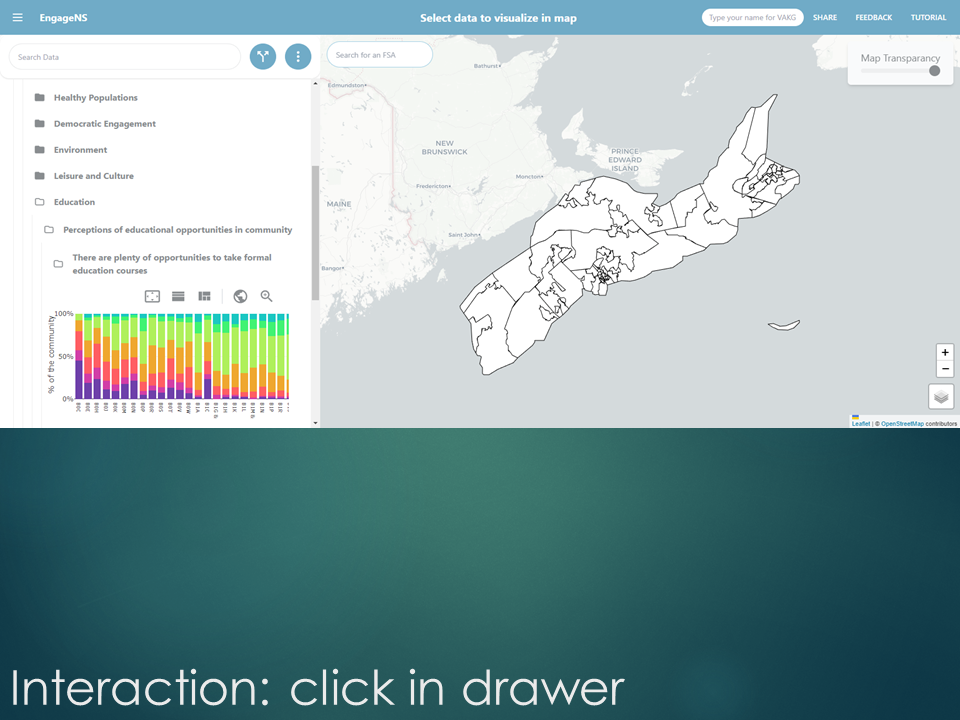}
    \includegraphics[width=0.28\columnwidth]{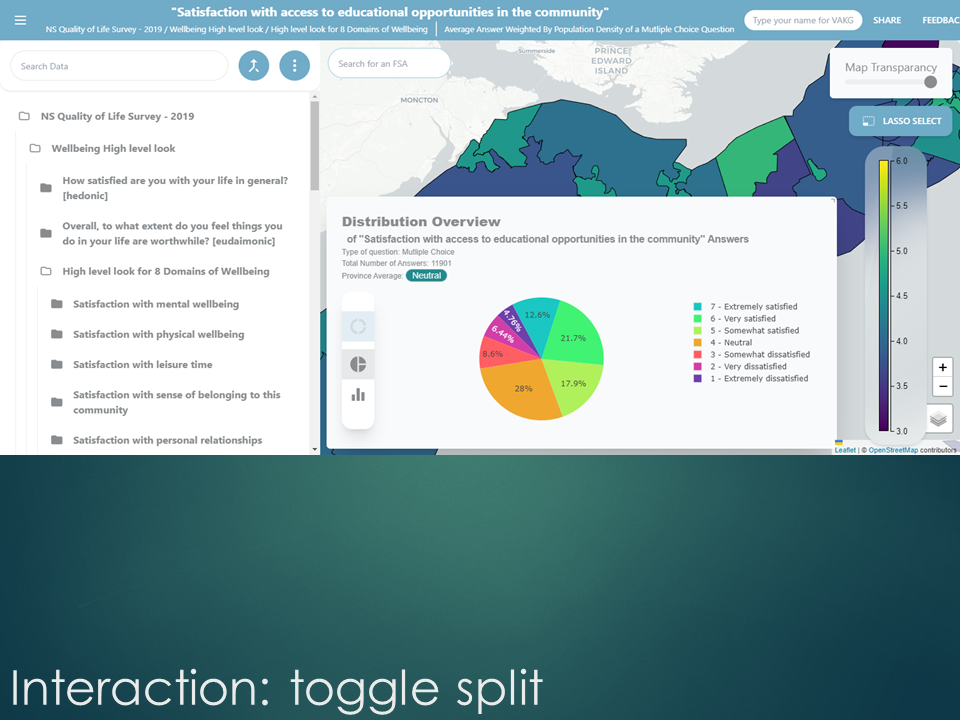}
    \includegraphics[width=0.28\columnwidth]{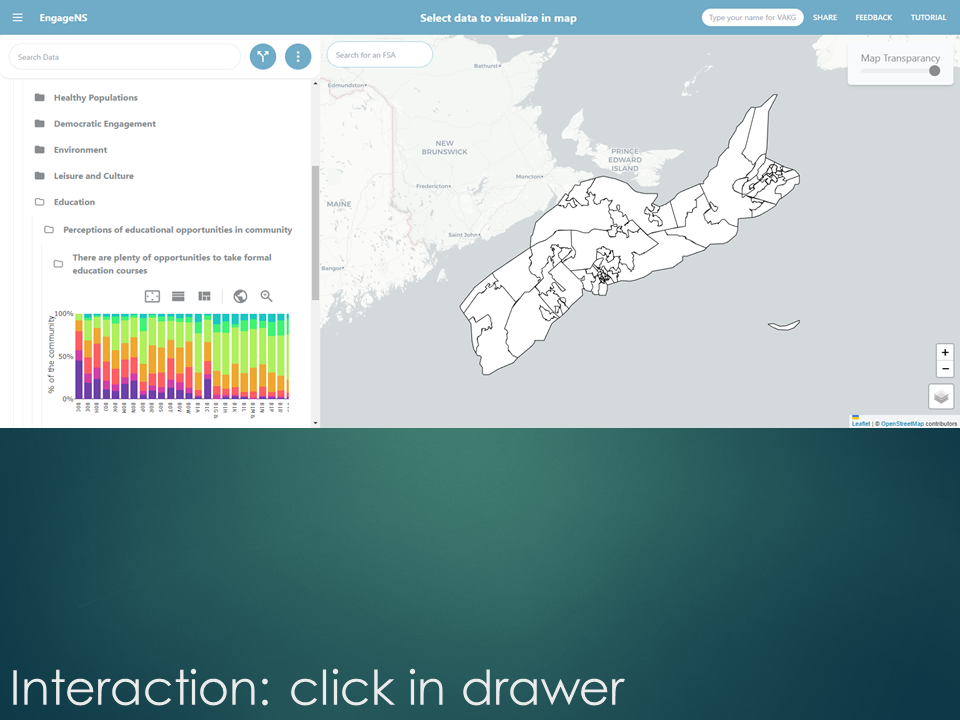}
    \includegraphics[width=0.28\columnwidth]{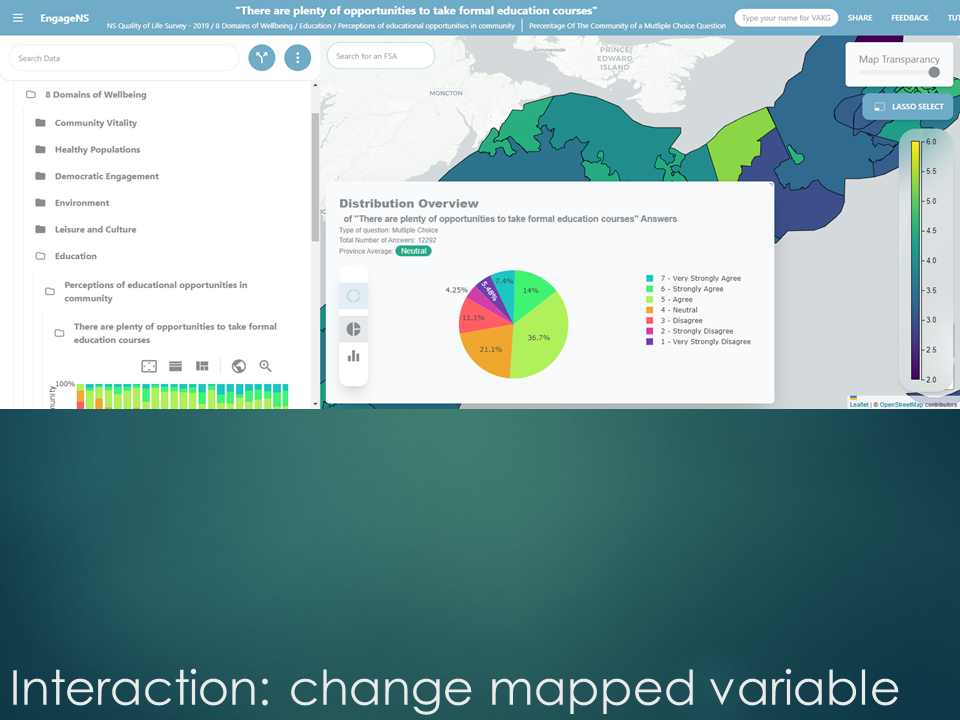}
    \includegraphics[width=0.28\columnwidth]{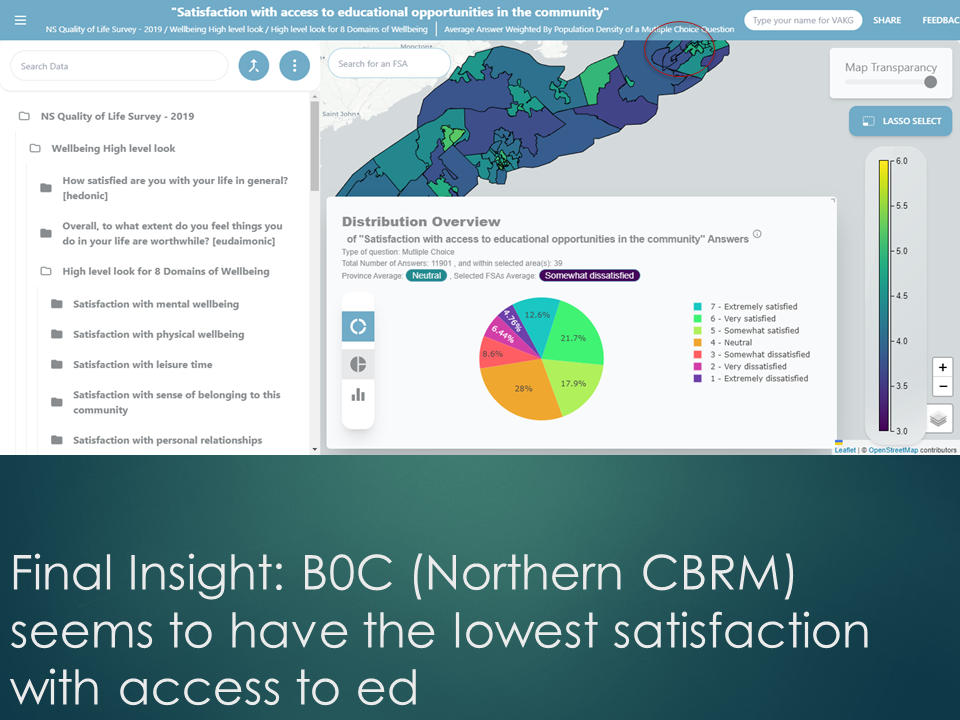}
    \includegraphics[width=0.28\columnwidth]{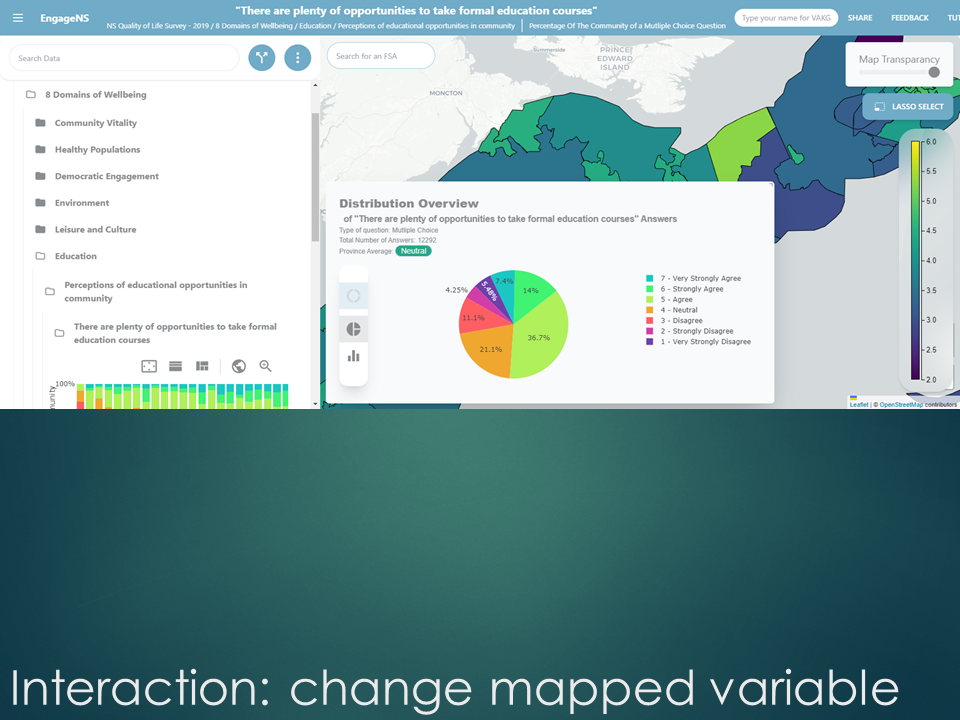}
    \includegraphics[width=0.28\columnwidth]{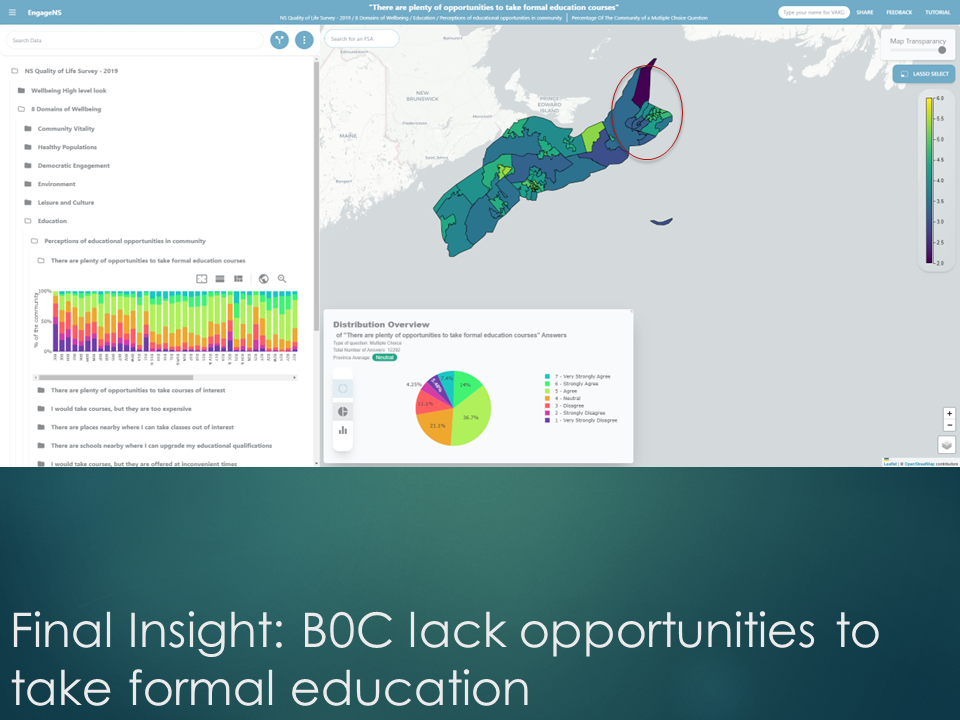}
    \includegraphics[width=0.28\columnwidth]{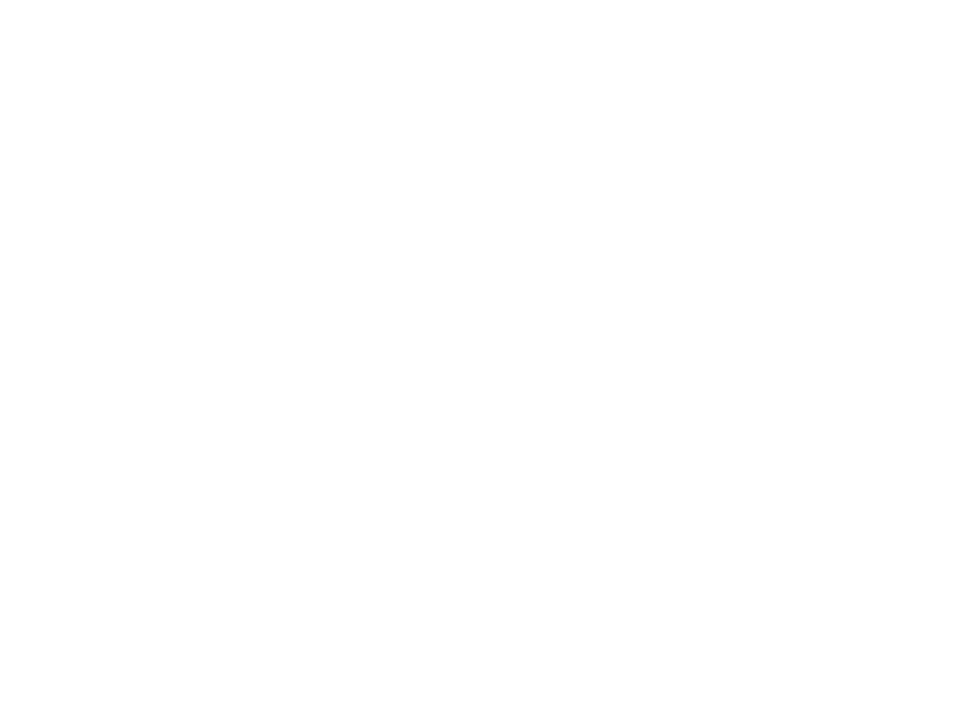}
    \includegraphics[width=0.28\columnwidth]{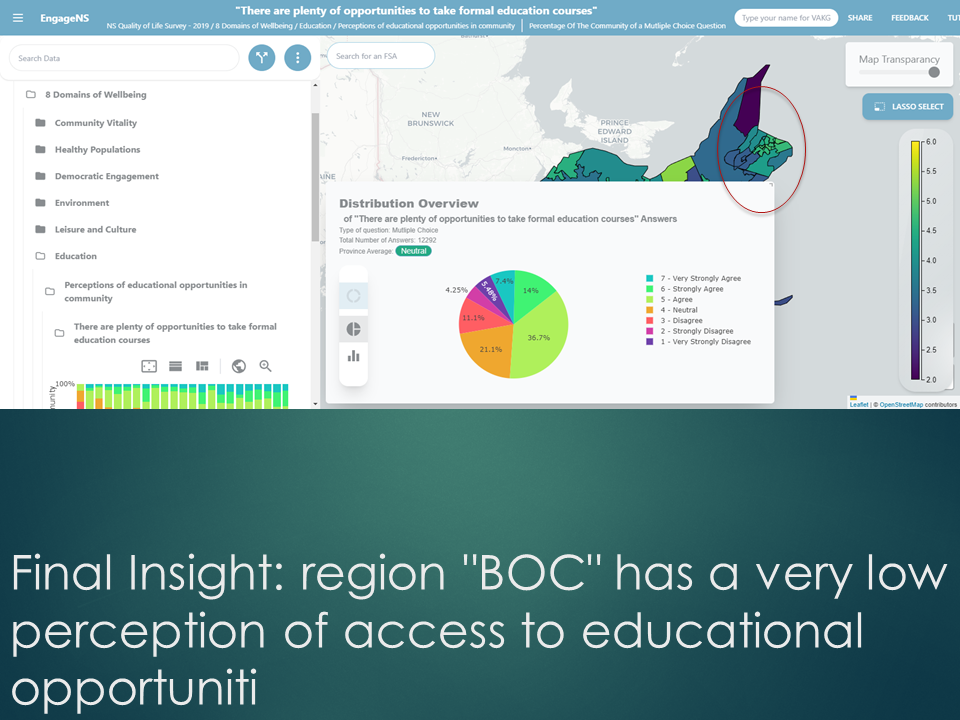}
    \caption{Three slide decks extracted from the insight WMT task 1.  }
    \vspace{-4mm}
    \label{fig:slidedeckeduc}
\end{figure}

\noindent\textbf{Behavior to reach an insight}: We may also ask \textit{Knowledge-Decks} to detect which intention led to some insight of interest automatically. We extracted three slide decks from three different user insights from Q1, as shown in Fig.~\ref{fig:slidedeckeduc}. From top to bottom, the slides follow the user's intention (first slide), behavior, and insight (last slide). We see that the left and center slide decks were detected to have the same \textit{closest} intention within the graph. Also, notice that although left and right paths had $6$ interaction events each and the Center $5$, all slide decks show that, when comparing the screenshots of each slide, \textit{the first interaction event was the same}, diverging from the second interaction onward. On that note, all interaction events of the left and right slide decks were the same, only diverging in the last slide, which displays the insight itself. Of course, a similar cypher query can also be used to generate the slide deck of a \textit{single} user's insights and their own intentions or to automatically find the closest \textit{insight} given an intention of interest.


Several other types of slide decks were extracted, but due to space constraints, they were redacted. Also, since further analysis or slide decks could be done to compare the tools, \textit{Knowledge-Decks} provide a direct Cypher API to perform custom analysis and the generation of custom slide decks.

%
%

%



\noindent\textbf{Expert opinion and evaluation}: We showed the slide decks described so far to an expert with the tool who works at Engage Nova Scotia and has extensive experience in the objective of WMT, including having to create many presentations about the tool, its usage, and of insights collected from the tool. She was first shown the slide deck of all insights from an intention (Fig.~\ref{fig:findinsights}) and was explained how the slide deck was generated. We asked for her opinion regarding how well-suited the slide deck was to be used as part of a presentation she may need to present. 

She showed significant interest in the slide deck. She said that a large amount of time would be saved by using \textit{Knowledge-Decks} to ask company employees to search for insights and, with the collected data, create a draft presentation containing all insights from the employees. She also noted that the overall slide deck layout is already very well suited to be used to present a list of insights and praised the advantage of using PowerPoint as the export method since she can simply change the design of the whole slide deck with one button to fit the company's presentation template. She also provided several feedback points, such as saying that the image could be even larger, the terminology of ``insight'' and ``intention'' may not be easily understood by some employees, and discussed how the insight texts were not suited as-is to be used as a presentation and would need spellchecking to refactor it into a more formal and complete speech. Additionally, she said that since the insights may not be directly related to the intention, editing the slide deck after the fact is required. She also noted that it would be ideal to have a prelude slide that explains how to read the slide decks and a conclusion slide summarizing the collection of insights to conclude the presentation. However, she agreed that these two slides are better suited to be written manually by herself or whoever will be doing the presentation.

She was also shown the interactivity slide deck (Fig.~\ref{fig:slidedecksampleflow}) and was once again impressed with its potential usage. She noted that these slides would be great to onboard new employees to the company or to be used in workshops or tutorials in technical-focused meetings. We then explained that the path shown by the slides is the optimal path between insight and intention, to which she noted that it is great that \textit{Knowledge-Decks} replace manual labor involved in a pre-meeting of stakeholders to discuss and coalesce experiences to optimize the slides into its most efficient story. She said this pre-meeting would focus only on defining the slide deck style, such as font and background type/color. However, she also noted that the current slide descriptions are only well understood by experienced users, such as herself. Ideally, \textit{Knowledge-Decks} should consider the tool's context and usage to generate a better-tailored description of each slide with less technical language. Nevertheless, she also noted that this slide could be used to describe how any of the insights of the previous slide deck was reached, or in other words, this slide deck would work well as a deep dive into a slide deck that lists all insights which originated from a given intention, such as the one described earlier. Overall, she said that \textit{Knowledge-Decks} would indeed be of significant help in creating drafts of presentations which would then be modified and edited into their final form within PowerPoint, including the removal of any potentially redundant or unwanted slides.


The evaluation of \textit{Knowledge-Decks} with a second VA tool called \textit{Bay-Of-Fundy Marine Life Tool}, which allows exploration of Marine Life, is included as supplemental material.

\section{Discussions, Limitations and Future Work}
\label{sec:future}

Our system has shown how to generate diverse knowledge discovery narratives from a KG. However, the modeling of our system is largely based on VAKG's methodology, which is unique in its applicability potential. This means we could not find another equivalent methodology to compare against. KGs have been very sparsely used for this kind of application, and no similar process would allow us to build a system based on VA's theoretical knowledge models. We propose \textit{Knowledge-Decks}' modeling strategy with the hope that other modeling strategies will emerge to be compared against ours.

On the other hand, many works use graph networks and natural language processing. Although \textit{Knowledge-Decks} has reached its goal, we recognize that more advanced processes could be applied. For instance, our keyword generation is done by lemma extraction but could also be performed by topic modeling~\cite{dahir2021query}, KeyBERT~\cite{keybertgithub}, or other machine learning models. Similarly, our KG could be used in many more potential graph network analyses, including advanced procedures of graph completion and recommendations. We judged it better to use simpler techniques in our work to allow for easier reproducibility and broader applicability of our approach. However, we intend to investigate the usage of more advanced algorithms like the ones listed in Sec.~\ref{sec:related} when we model new systems for knowledge discovery storytelling in the future.
%

One of our approach's core complexity is the need to retool the existing VA tools with the connection to implementation. Although this decision was by design, we recognize that not all tools are so easily modifiable to be attached to external libraries like ours. Some related works have shown to be better than ours in that regard, but they usually require the complete development of a new tool or the need for external user monitoring. Future works may investigate other ways to collect user intention, insight, and behavior without any modification on the VA tool's part, such as through browser extensions or by recording and processing the video/audio feed of users while they use tools and automatically processing them into a usable KG. On another note, our methodology can also be applicable to different use cases and goals, such as providing recommendations retelling previous users' experiences within the VA tool itself. Indeed, some participants provided us with valuable feedback on other potential uses of our methodology other than slide decks. Of course, these have their own goals and limitations different from ours, so we digress.

Another key issue when handling user data is the privacy concern, especially on the users' part, regarding how user data is being utilized. In our system and throughout our study, we maintained complete anonymity, but privacy and accountability is a concern usually raised by users and researchers alike~\cite{xu2015analytic}. Although no participant raised this issue during our tests, we see the attempts of companies and research institutions to restrict or limit the collection of user data of any kind. Further evaluation is required to say that \textit{Knowledge-Decks} can be used in commercial scenarios. Nevertheless, \textit{Knowledge-Decks} does not explicitly attempt to solve this issue other than that only collecting anonymous information.

Finally, certain design choices may be of concern. For instance, using slide decks to narrate the user's knowledge discovery process has its issues~\cite{park2022storyfacets}, such as being seen as useful only for presentations to novice users. As noted by the expert, the slide decks may also contain slides that may be unrelated to the rest when, for instance, a user had an insight unrelated to their intention. However, the ubiquitousness of PowerPoint slides for presentation purposes, including infographics and storytelling, must be considered. PowerPoint files are also easily editable, allowing for the inclusion or exclusion of information, such as a summary slide or the removal of an unrelated insight slide, to be handled on a case-by-case basis. As StoryFacet~\cite{park2022storyfacets} says, adding animations, filtering, and other features of PowerPoint or similar software would also potentially enhance the slide decks generated, but these modifications are already beyond our original goal. Also, the design used to implement our system was based on a specific type of VA tool, a limitation we recognize. Yet, we believe that many tools fit our requirements, providing therefore a good starting point for further research. \textit{Knowledge-Decks} also allows the use of custom queries to generate other types of slide decks. Therefore, although our approach can be considered limited due to the type of tools it can be attached to and how the slide decks are generated, it provides a reasonable basis for other works to expand upon.

\section{Conclusion}\label{sec:conclusion}

We have presented \textit{Knowledge-Decks}, a novel approach that generates slide decks of user knowledge discovery processes using VA applications and tools. \textit{Knowledge-Decks} collect user intentions, behavior, and insights during users' knowledge discovery sessions and automatically structure the data into a KG modeled based on the VA tools. By executing specific pre-defined Cypher queries, \textit{Knowledge-Decks} can extract paths from a KG and format them as PowerPoint slide decks that tell a knowledge discovery story. \textit{Knowledge-Decks} was evaluated by being attached to two existing VA tools where users were asked to perform $5$ pre-defined exploratory-based tasks. By collecting user intentions, behavior (interactions), and insights, three main types of stories were told: what insights were reached given a particular intention of interest; which intentions and behavior led to a particular insight; and how insights were reached given a specific user intention. 
%
%
The slide decks of these stories were shown to experts, which validated their usefulness as a draft presentation resource that can be used to investigate the user stories or to be used as a first or draft version of an actual presentation to be used.

\section*{Acknowledgment}
The authors acknowledge the support of the Natural Sciences and Engineering Research Council of Canada (NSERC). This work was supported by Engage Nova Scotia and Mitacs through the Mitacs Accelerate program (funding reference number IT28167).

\bibliographystyle{eurovisDefinitions/eg-alpha-doi}  
\bibliography{main}

\end{document}



\maketitle

\section{Other Tool}

\noindent\textbf{Bay-Of-Fundy Marine Life Tool (MLT)}: The \textit{Ocean Tracking Network} (OTN)~\cite{iverson2019ocean} is a global aquatic animal tracking, technology, data management, and partnership platform headquartered at Dalhousie University in Canada. OTN and its partners are using electronic tags to track more than 200 keystones, commercially important and endangered species worldwide. In order to better understand the existing data on Stripped Bass and Shark detection throughout the Bay of Fundy in Nova Scotia, researchers came together and developed a visual analytic tool where data can be displayed and explored. In their own words: by displaying raw and aggregated data in different interlinked visualizations, domain experts could understand both the marine life and the limitations of their own sensors when collecting data for decision-making.

From the two tool's descriptions (above and in the actual paper), we can see that although they are part of completely different domains (social sciences and marine biology) and focus on different types of data (multi-choice survey of QoL and marine life tag detection over time), both are similar in nature due to their aim to provide visualizations to better understand and explore the available data. Both are within the requirements to be used with our system. Here is the full property map table, including the two tools:

\begin{table}
\label{tab:propertymapdata}
\caption{Data stored in each of the tool's respective knowledge graph node's contents (property maps). }
\begin{tabular}{||p{12mm} | p{32mm} p{25mm} ||} 
 \hline
 Node Type & Well-being Tool & Marine Life Tool \\ [0.5ex] 
 \hline\hline
 Human temporal sequence & label (insight $P$ or intention $E$), created time, URL, screen size, text, keywords, shapes drawn, user id, and analysis id & same as Well-being tool \\ 
 \hline
 Human state-space & label (insight $P$ or intention $E$), created time, last updated time, and keywords time & same as Well-being tool \\
 \hline
 Computer state-space & label (specification $S$, visualization $V$ or analysis $A$), created time, last updated time, the status of the hierarchical structure to the left, bar chart's visualization schema (stacked, grouped, maximized, etc.), map position, map zoom, map areas selected, math operation used, and question visualized in map & label (specification $S$, visualization $V$ or analysis $A$), created time, last updated time, map position, map zoom, selected time-frame if any, and selected filters if any \\
 \hline
 Computer temporal sequence & event name, created time, URL, user id, analysis id, and all the same data from the related Computer state-space & event name, created time, URL, user id, analysis id, and all the same data from the related Computer state-space \\
 \hline
\end{tabular}
\end{table}

The data was fed to our system, and the resulting knowledge graph of both tools combined was vastly complex. Nevertheless, a preliminary analysis of the results showed that users had a total of 70 unique intentions (36 for WMT and 34 for MLT) and 54 unique insights (27 for each). From user input, there were 40 unique intentions (21 for WMT and 19 for MLT) and 53 unique insights (26 for WMT and 27 for MLT). On the machine side, there were a total of 458 events (252 WMT to 206 MLT) or 230 unique events (143 to 87). We can see from these numbers that although there were fewer interactions with MLT, it offered the same amount of insights. This could either mean that the tasks were simpler on MLT or that the users had to interact less with this tool to reach insights. 



\section{Screenshot of the Tools}
\begin{figure*}
    \centering
    \includegraphics[width=2\columnwidth]{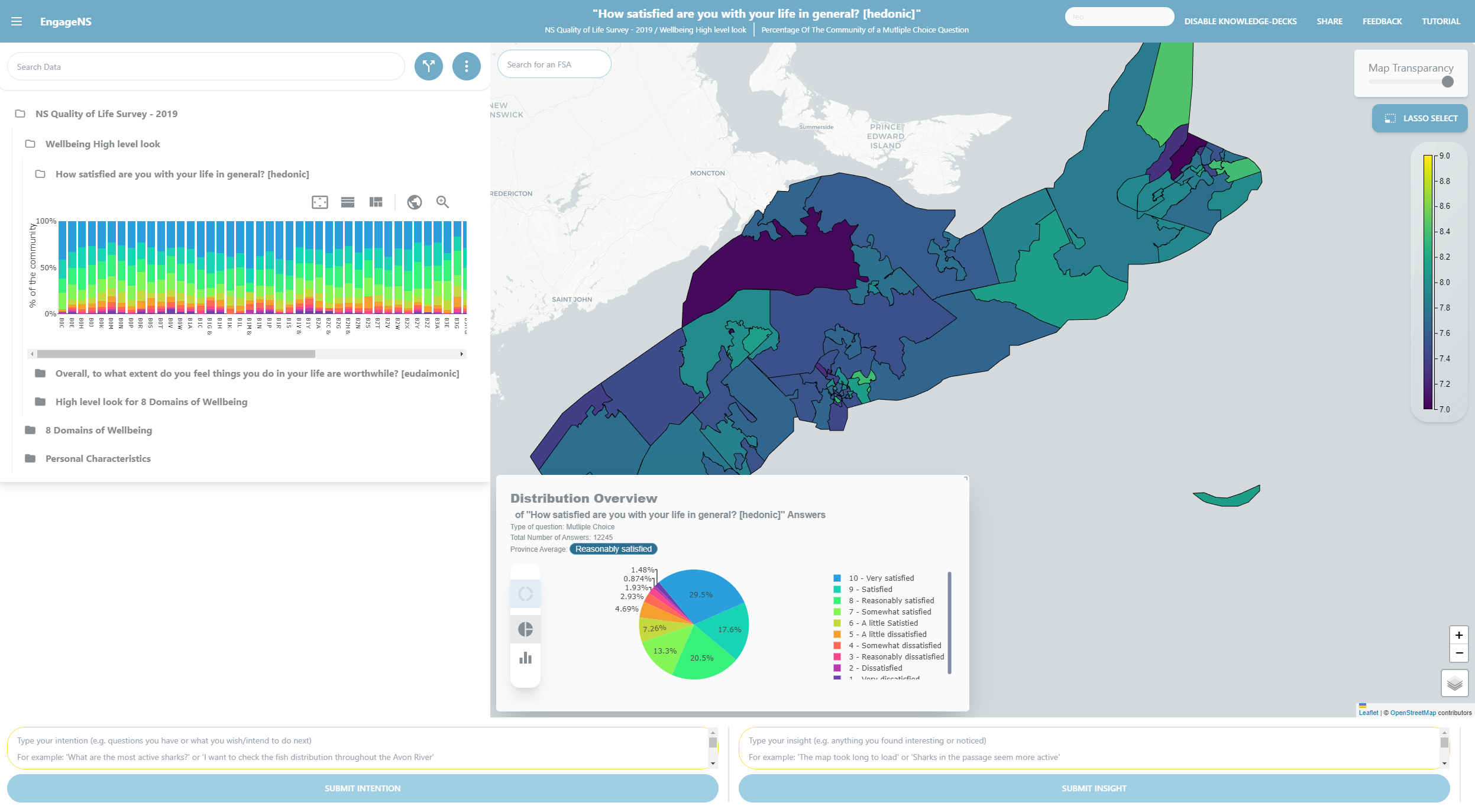}
    \caption{Screenshot of the Well-being Mapping Tool. Knowledge-Decks options, including the text input areas, are on the top right and the bottom part of the website.  }
    \label{fig:wmt}
\end{figure*}

\begin{figure*}
    \centering
    \includegraphics[width=2\columnwidth]{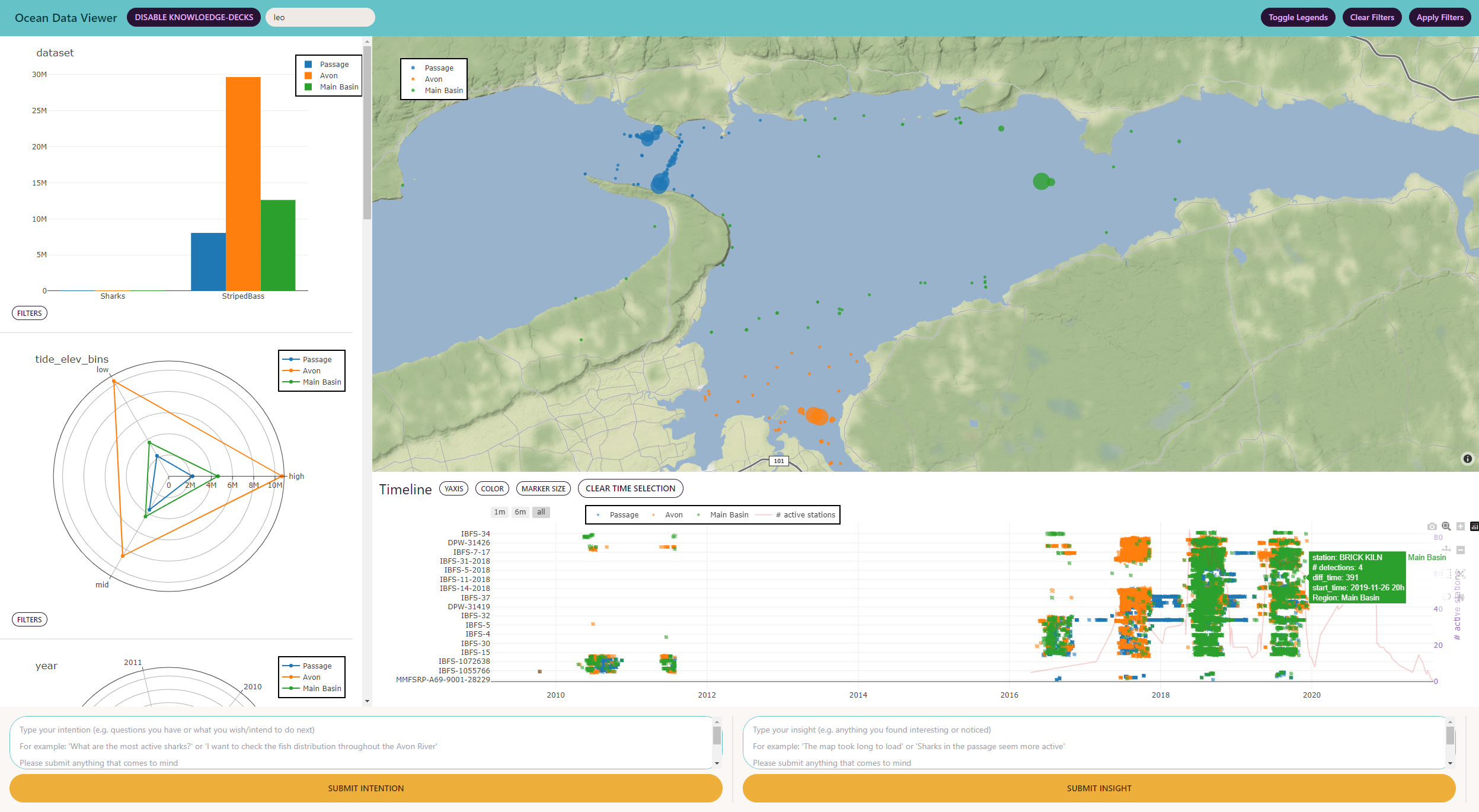}
    \caption{Screenshot of the Marine Life Tool. Knowledge-Decks options, including the text input areas, are on the top right and the bottom part of the website.  }
    \label{fig:mlt}
\end{figure*}

See Figs.~\ref{fig:wmt} and ~\ref{fig:mlt}.

\section{Questionnaire}

Appendix C–Demographic Questionnaire
\begin{itemize}
    \item How long have you used a computer?
    \begin{itemize}
    \item Less than one week
    \item 1 week to less than 1 month
    \item 1 month to less than 1 year
    \item 1 year to less than 2 years
    \item 2 years to less than 4 years
    \item 4 or more years
    \item I prefer not to answer
    \end{itemize}
    \item On average, how much time do you spend per week on a computer?
    \begin{itemize}
    \item Less than one hour
    \item One to less than 4 hours
    \item to less than 10 hours
    \item 10 to less than 20 hours
    \item 20 to less than 40 hours
    \item Over 40 hours
    \item I prefer not to answer
    \end{itemize}
    \item How comfortable are you at using interactive user interface?\begin{itemize}
    \item Extremely comfortable
    \item Very comfortable
    \item Comfortable
    \item Uncomfortable
    \item Very uncomfortable
    \item Extremely uncomfortable
    \item I prefer not to answer
    \end{itemize}
    \item How familiar are you with Data Analytics Tools, such as Microsoft Excel or Tableau?
    \begin{itemize}
    \item Very well
    \item Well
    \item Neutral
    \item Not well
    \item Not well at all
    \item I prefer not to answer
    \end{itemize}
    \item At what level do you think your understanding of written English is?
    \begin{itemize}
    \item Excellent
    \item Very good
    \item Good
    \item Acceptable
    \item Bad
    \item Very bad
    \item None
    \item I prefer not to answer
    \end{itemize}
    \item What is the highest level of education you have completed?
    \begin{itemize}
    \item Little or no formal education
    \item High school or equivalent
    \item College or university
    \item Master
    \item Doctoral
    \item Post-Doctoral
    \item I prefer not to answer
    \end{itemize}
    \item In case you have or are pursuing a degree, what is your primary area of study?
    \begin{itemize}
    \item Computer Science
    \item Information technology
    \item Internetworking
    \item Social Science
    \item Health Science
    \item Other
    \item I have no primary area of study
    \item I prefer not to answer
    \end{itemize}
\end{itemize}

Prelude - Video tutorial
Please view this video tutorial (link) only once and respond to the following statements about the visualization-based interface, using the given scale:

Questionnaire 1.1 - Pre-defined execution Questionnaire of the Quality of Life Tool
Now open the website at this link and attempt to answer the following question. As the video has shown, write the individual intentions and/or insights you discover in the process. If you find the answer, type as a final insight and then respond to the following statements about your experience using the given scale. Note that the process of finding the answer is what is important to us, so don’t worry about how well you answer the question, just attempt to within the allotted time of 5min.

Question: Which FSA shows the most concern with access to educational opportunities?
	
Question	Answers
I was able to follow the steps without any problems	Strongly Disagree	Somewhat Disagree	Neutral	Somewhat Agree	Strongly Agree
I was able to quickly understand what I needed to do to perform the required steps in the webapp	Strongly Disagree	Somewhat Disagree	Neutral	Somewhat Agree	Strongly Agree
While performing the steps, I totally ignored other information not relevant to the question	Strongly Disagree	Somewhat Disagree	Neutral	Somewhat Agree	Strongly Agree
I believe I was able to answer the requested question 	Strongly Disagree	Somewhat Disagree	Neutral	Somewhat Agree	Strongly Agree

Questionnaire 1.2 - Pre-defined execution Questionnaire of the Quality of Life Tool
Now open the website at this link and attempt to answer the following question. As the video has shown, write the individual intentions and/or insights you discover in the process. If you find the answer, type as a final insight and then respond to the following statements about your experience using the given scale. Note that the process of finding the answer is what is important to us, so don’t worry about how well you answer the question, just attempt to within the allotted time of 5min.

Question: Is there a “happiest” community that shows the highest life satisfaction score?
	
Question	Answers
I was able to follow the steps without any problems	Strongly Disagree	Somewhat Disagree	Neutral	Somewhat Agree	Strongly Agree
I was able to quickly understand what I needed to do to perform the required steps in the webapp	Strongly Disagree	Somewhat Disagree	Neutral	Somewhat Agree	Strongly Agree
While performing the steps, I totally ignored other information not relevant to the question	Strongly Disagree	Somewhat Disagree	Neutral	Somewhat Agree	Strongly Agree
I believe I was able to answer the requested question 	Strongly Disagree	Somewhat Disagree	Neutral	Somewhat Agree	Strongly Agree

Questionnaire 1.3 - Pre-defined execution Questionnaire of the Quality of Life Tool
Now open the website at this link and attempt to answer the following question. As the video has shown, write the individual intentions and/or insights you discover in the process. If you find the answer, type as a final insight and then respond to the following statements about your experience using the given scale. Note that the process of finding the answer is what is important to us, so don’t worry about how well you answer the question, just attempt to within the allotted time of 5min.

Question: Where do parents experience barriers to access recreation in terms of facilities not offering childcare?

Question	Answers
I was able to follow the steps without any problems	Strongly Disagree	Somewhat Disagree	Neutral	Somewhat Agree	Strongly Agree
I was able to quickly understand what I needed to do to perform the required steps in the webapp	Strongly Disagree	Somewhat Disagree	Neutral	Somewhat Agree	Strongly Agree
While performing the steps, I totally ignored other information not relevant to the question	Strongly Disagree	Somewhat Disagree	Neutral	Somewhat Agree	Strongly Agree
I believe I was able to answer the requested question 	Strongly Disagree	Somewhat Disagree	Neutral	Somewhat Agree	Strongly Agree

Questionnaire 1.4 - Pre-defined execution Questionnaire of the Quality of Life Tool
Now open the website at this link and attempt to answer the following question. As the video has shown, write the individual intentions and/or insights you discover in the process. If you find the answer, type as a final insight and then respond to the following statements about your experience using the given scale. Note that the process of finding the answer is what is important to us, so don’t worry about how well you answer the question, just attempt to within the allotted time of 5min.

Question: Where do people tend to buy local food? (or, where is the lowest amount of those who “never” buy local food?)
	
Question	Answers
I was able to follow the steps without any problems	Strongly Disagree	Somewhat Disagree	Neutral	Somewhat Agree	Strongly Agree
I was able to quickly understand what I needed to do to perform the required steps in the webapp	Strongly Disagree	Somewhat Disagree	Neutral	Somewhat Agree	Strongly Agree
While performing the steps, I totally ignored other information not relevant to the question	Strongly Disagree	Somewhat Disagree	Neutral	Somewhat Agree	Strongly Agree
I believe I was able to answer the requested question 	Strongly Disagree	Somewhat Disagree	Neutral	Somewhat Agree	Strongly Agree

Questionnaire 1.5 - Pre-defined execution Questionnaire of the Quality of Life Tool
Now open the website at this link and attempt to answer the following question. As the video has shown, write the individual intentions and/or insights you discover in the process. If you find the answer, type as a final insight and then respond to the following statements about your experience using the given scale. Note that the process of finding the answer is what is important to us, so don’t worry about how well you answer the question, just attempt to within the allotted time of 5min.

Question: Where do people report the highest overall work life balance? 
	
Question	Answers
I was able to follow the steps without any problems	Strongly Disagree	Somewhat Disagree	Neutral	Somewhat Agree	Strongly Agree
I was able to quickly understand what I needed to do to perform the required steps in the webapp	Strongly Disagree	Somewhat Disagree	Neutral	Somewhat Agree	Strongly Agree
While performing the steps, I totally ignored other information not relevant to the question	Strongly Disagree	Somewhat Disagree	Neutral	Somewhat Agree	Strongly Agree
I believe I was able to answer the requested question 	Strongly Disagree	Somewhat Disagree	Neutral	Somewhat Agree	Strongly Agree

Questionnaire 2 – Quality of Life Tool’s Interface Features Questionnaire
Please respond to the following statements about the visualization-based interface, using the given scale:

Question	Answers
The tool was easy to use	Strongly Disagree	Somewhat Disagree	Neutral	Somewhat Agree	Strongly Agree
The questions were simple to answer	Strongly Disagree	Somewhat Disagree	Neutral	Somewhat Agree	Strongly Agree
I answered the questions confidently	Strongly Disagree	Somewhat Disagree	Neutral	Somewhat Agree	Strongly Agree
It was intuitive to write down my intentions during the process	Strongly Disagree	Somewhat Disagree	Neutral	Somewhat Agree	Strongly Agree
It was intuitive to write down my insights during the process	Strongly Disagree	Somewhat Disagree	Neutral	Somewhat Agree	Strongly Agree
I believe if I were to see other’s intentions, I would be able to better understand the question.	Strongly Disagree	Somewhat Disagree	Neutral	Somewhat Agree	Strongly Agree
I believe if I were to see others insights, I would be able to answer the question faster	Strongly Disagree	Somewhat Disagree	Neutral	Somewhat Agree	Strongly Agree
I believe if I were to see others insights, I would be able to answer the question with more precision	Strongly Disagree	Somewhat Disagree	Neutral	Somewhat Agree	Strongly Agree

Questionnaire 3.1 - Pre-defined execution Questionnaire of the Ocean Data Tool
Now open the website at this link and attempt to answer the following question. As the video has shown, write the individual intentions and/or insights you discover in the process. If you find the answer, type as a final insight and then respond to the following statements about your experience using the given scale. Note that the process of finding the answer is what is important to us, so don’t worry about how well you answer the question, just attempt to within the allotted time of 5min.

Question: Considering the cardinal directions (north, northeast, east, etc), where are most of the sensors (also called stations) located within the Bay of Fundy?
	
Question	Answers
I was able to follow the steps without any problems	Strongly Disagree	Somewhat Disagree	Neutral	Somewhat Agree	Strongly Agree
I was able to quickly understand what I needed to do to perform the required steps in the webapp	Strongly Disagree	Somewhat Disagree	Neutral	Somewhat Agree	Strongly Agree
While performing the steps, I totally ignored other information not relevant to the question	Strongly Disagree	Somewhat Disagree	Neutral	Somewhat Agree	Strongly Agree
I believe I was able to answer the requested question 	Strongly Disagree	Somewhat Disagree	Neutral	Somewhat Agree	Strongly Agree

Questionnaire 3.2 - Pre-defined execution Questionnaire of the Quality of Life Tool
Now open the website at this link and attempt to answer the following question. As the video has shown, write the individual intentions and/or insights you discover in the process. If you find the answer, type as a final insight and then respond to the following statements about your experience using the given scale. Note that the process of finding the answer is what is important to us, so don’t worry about how well you answer the question, just attempt to within the allotted time of 5min.

Question: In which year and month were fish most detected on the passage?
	
Question	Answers
I was able to follow the steps without any problems	Strongly Disagree	Somewhat Disagree	Neutral	Somewhat Agree	Strongly Agree
I was able to quickly understand what I needed to do to perform the required steps in the webapp	Strongly Disagree	Somewhat Disagree	Neutral	Somewhat Agree	Strongly Agree
While performing the steps, I totally ignored other information not relevant to the question	Strongly Disagree	Somewhat Disagree	Neutral	Somewhat Agree	Strongly Agree
I believe I was able to answer the requested question 	Strongly Disagree	Somewhat Disagree	Neutral	Somewhat Agree	Strongly Agree

Questionnaire 3.3 - Pre-defined execution Questionnaire of the Quality of Life Tool
Now open the website at this link and attempt to answer the following question. As the video has shown, write the individual intentions and/or insights you discover in the process. If you find the answer, type as a final insight and then respond to the following statements about your experience using the given scale. Note that the process of finding the answer is what is important to us, so don’t worry about how well you answer the question, just attempt to within the allotted time of 5min.

Question: How much did the tide affect the presence and number of sharks in the avon river in 2019?

Question	Answers
I was able to follow the steps without any problems	Strongly Disagree	Somewhat Disagree	Neutral	Somewhat Agree	Strongly Agree
I was able to quickly understand what I needed to do to perform the required steps in the webapp	Strongly Disagree	Somewhat Disagree	Neutral	Somewhat Agree	Strongly Agree
While performing the steps, I totally ignored other information not relevant to the question	Strongly Disagree	Somewhat Disagree	Neutral	Somewhat Agree	Strongly Agree
I believe I was able to answer the requested question 	Strongly Disagree	Somewhat Disagree	Neutral	Somewhat Agree	Strongly Agree

Questionnaire 3.4 - Pre-defined execution Questionnaire of the Quality of Life Tool
Now open the website at this link and attempt to answer the following question. As the video has shown, write the individual intentions and/or insights you discover in the process. If you find the answer, type as a final insight and then respond to the following statements about your experience using the given scale. Note that the process of finding the answer is what is important to us, so don’t worry about how well you answer the question, just attempt to within the allotted time of 5min.

Question: On what day of the week are sharks most and least active in 2018?
	
Question	Answers
I was able to follow the steps without any problems	Strongly Disagree	Somewhat Disagree	Neutral	Somewhat Agree	Strongly Agree
I was able to quickly understand what I needed to do to perform the required steps in the webapp	Strongly Disagree	Somewhat Disagree	Neutral	Somewhat Agree	Strongly Agree
While performing the steps, I totally ignored other information not relevant to the question	Strongly Disagree	Somewhat Disagree	Neutral	Somewhat Agree	Strongly Agree
I believe I was able to answer the requested question 	Strongly Disagree	Somewhat Disagree	Neutral	Somewhat Agree	Strongly Agree

Questionnaire 3.5 - Pre-defined execution Questionnaire of the Quality of Life Tool
Now open the website at this link and attempt to answer the following question. As the video has shown, write the individual intentions and/or insights you discover in the process. If you find the answer, type as a final insight and then respond to the following statements about your experience using the given scale. Note that the process of finding the answer is what is important to us, so don’t worry about how well you answer the question, just attempt to within the allotted time of 5min.

Question: How does the number of detected striped bass compare between the sensors running across the north shore of the main basin, the southern shore sensors and the avon river sensors?
	
Question	Answers
I was able to follow the steps without any problems	Strongly Disagree	Somewhat Disagree	Neutral	Somewhat Agree	Strongly Agree
I was able to quickly understand what I needed to do to perform the required steps in the webapp	Strongly Disagree	Somewhat Disagree	Neutral	Somewhat Agree	Strongly Agree
While performing the steps, I totally ignored other information not relevant to the question	Strongly Disagree	Somewhat Disagree	Neutral	Somewhat Agree	Strongly Agree
I believe I was able to answer the requested question 	Strongly Disagree	Somewhat Disagree	Neutral	Somewhat Agree	Strongly Agree

Questionnaire 4 – Ocean Tool’s Interface Features Questionnaire
Please respond to the following statements about the visualization-based interface, using the given scale:

Question	Answers
The tool was easy to use	Strongly Disagree	Somewhat Disagree	Neutral	Somewhat Agree	Strongly Agree
The questions were simple to answer	Strongly Disagree	Somewhat Disagree	Neutral	Somewhat Agree	Strongly Agree
I answered the questions confidently	Strongly Disagree	Somewhat Disagree	Neutral	Somewhat Agree	Strongly Agree
It was intuitive to write down my intentions during the process	Strongly Disagree	Somewhat Disagree	Neutral	Somewhat Agree	Strongly Agree
It was intuitive to write down my insights during the process	Strongly Disagree	Somewhat Disagree	Neutral	Somewhat Agree	Strongly Agree
I believe if I were to see other’s intentions, I would be able to better understand the question.	Strongly Disagree	Somewhat Disagree	Neutral	Somewhat Agree	Strongly Agree
I believe if I were to see others insights, I would be able to answer the question faster	Strongly Disagree	Somewhat Disagree	Neutral	Somewhat Agree	Strongly Agree
I believe if I were to see others insights, I would be able to answer the question with more precision	Strongly Disagree	Somewhat Disagree	Neutral	Somewhat Agree	Strongly Agree

Please give us more comments about your experience
Is there any way you expect that the intentions and insights could be used to empower you while you were answering the questions?

\section{Knowledge-Decks Ontology}
\begin{figure}
    \includegraphics[width=\columnwidth]{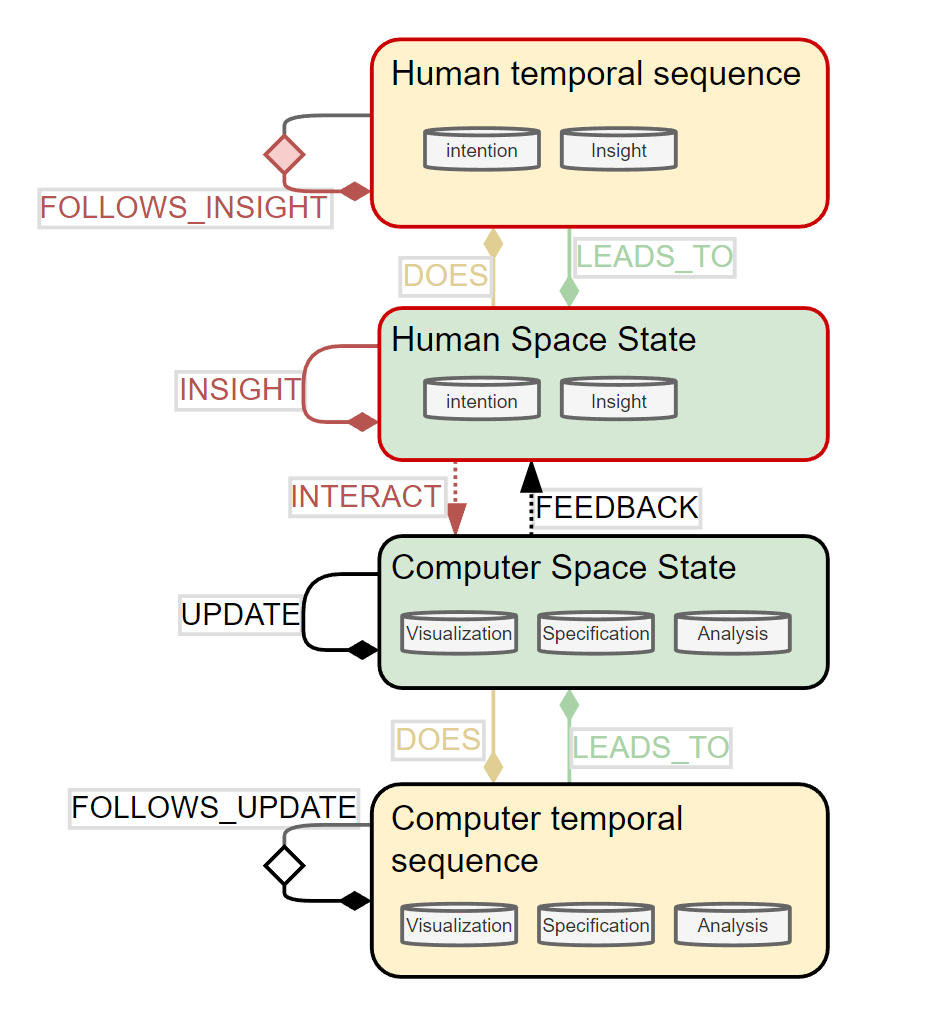}
    \caption{Ontology structure of the Knowledge Graph used in Knowledge-Decks. There are four classes of nodes, each containing property-maps as depicted above. The relationships between the classes are in capital letters and are used within the queries to specify which relationship is needed to be followed to extract the knowledge discovery paths.  }
    \label{fig:ontology}
\end{figure}

See Fig.~\ref{fig:ontology}.

\section{Slide Decks Generated}

\begin{figure*}
    \centering
    \includegraphics[width=0.3\textwidth]{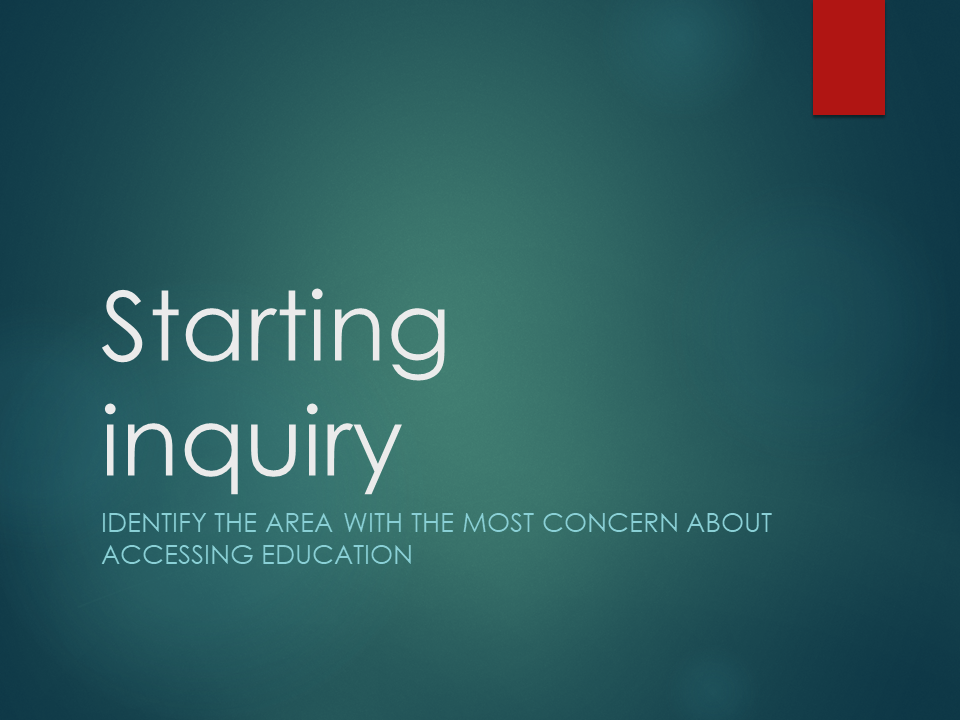}
    \includegraphics[width=0.3\textwidth]{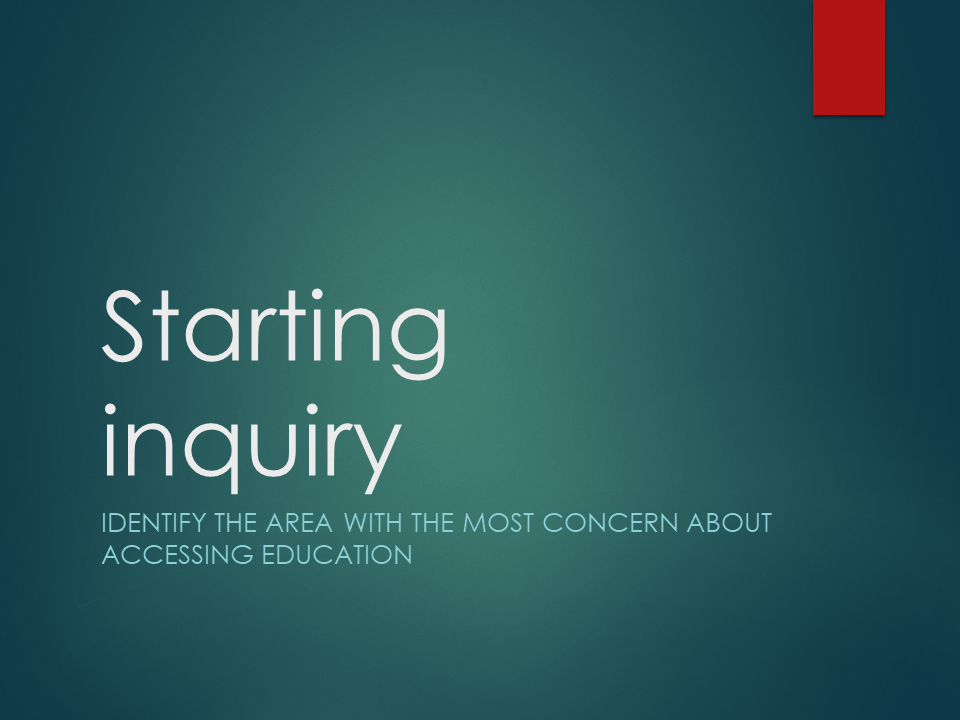}
    \includegraphics[width=0.3\textwidth]{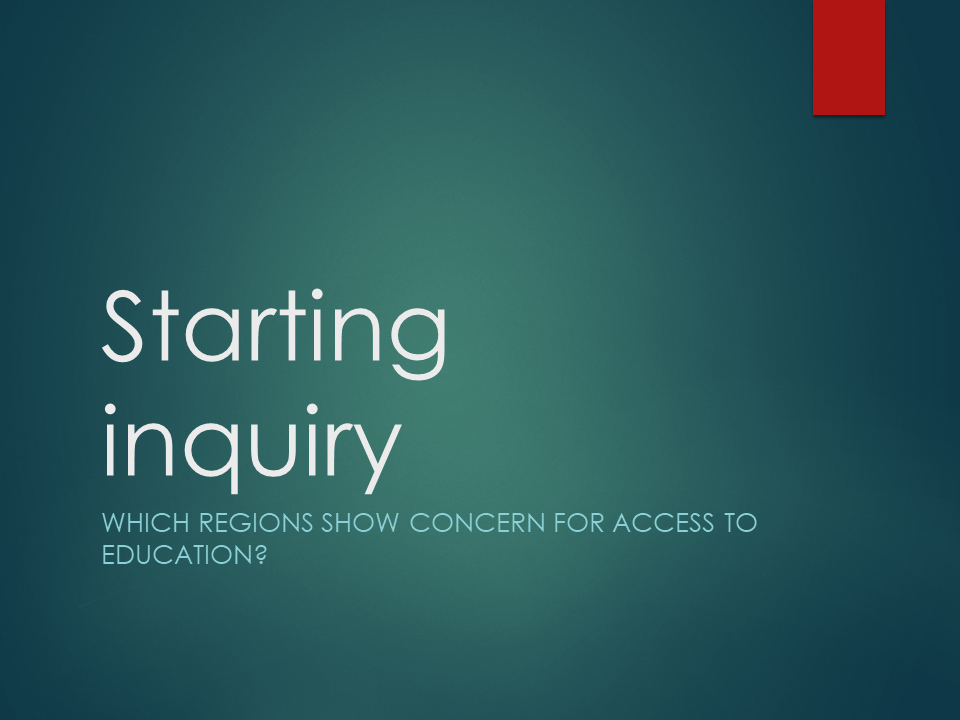}
    \includegraphics[width=0.3\textwidth]{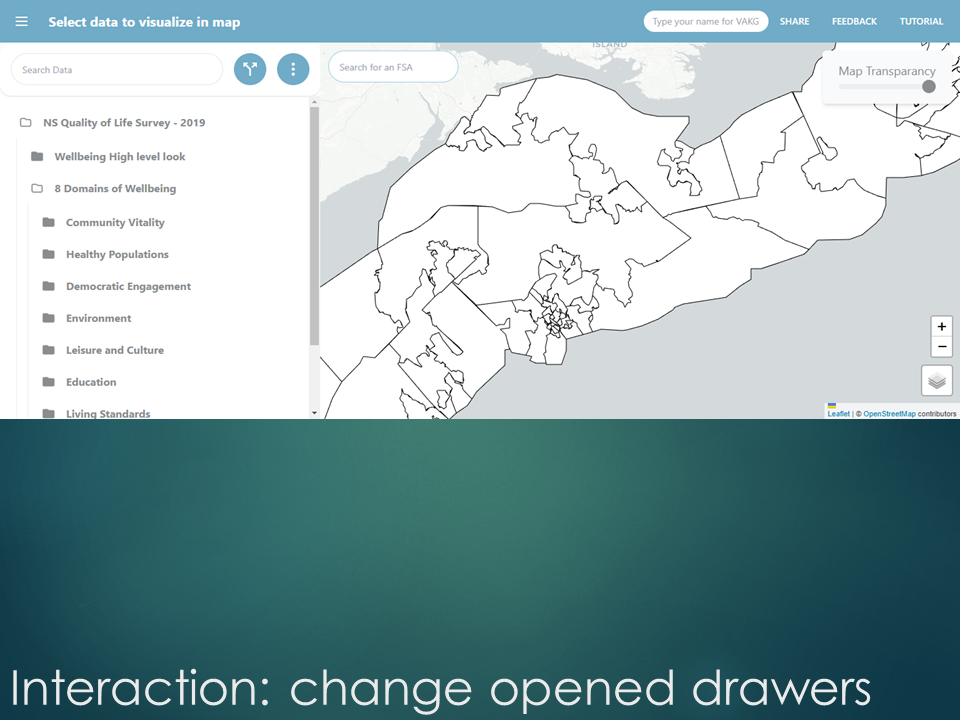}
    \includegraphics[width=0.3\textwidth]{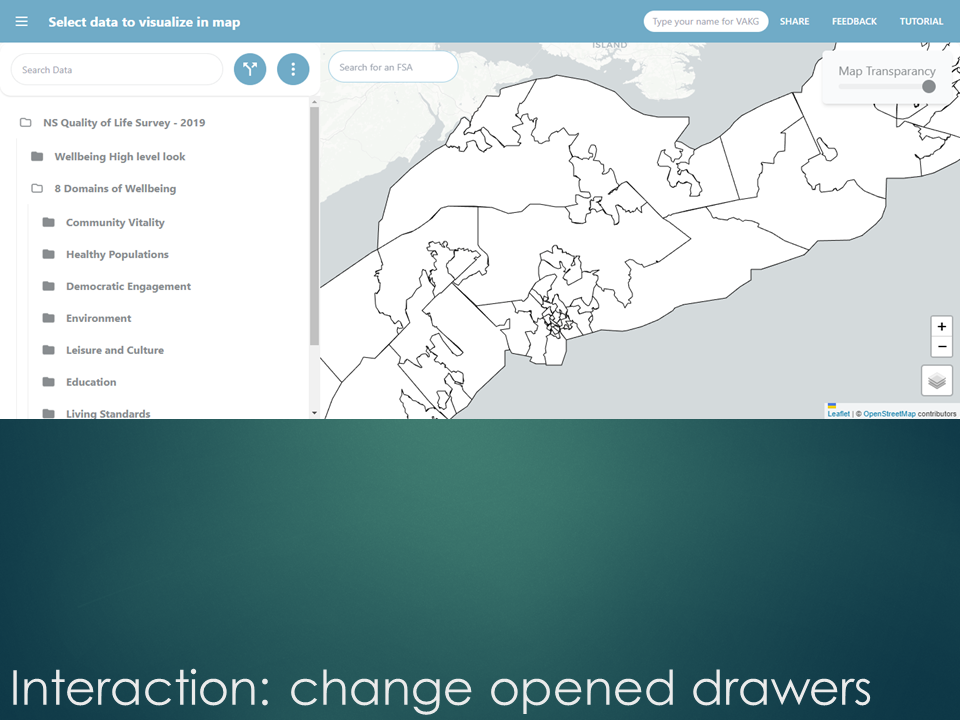}
    \includegraphics[width=0.3\textwidth]{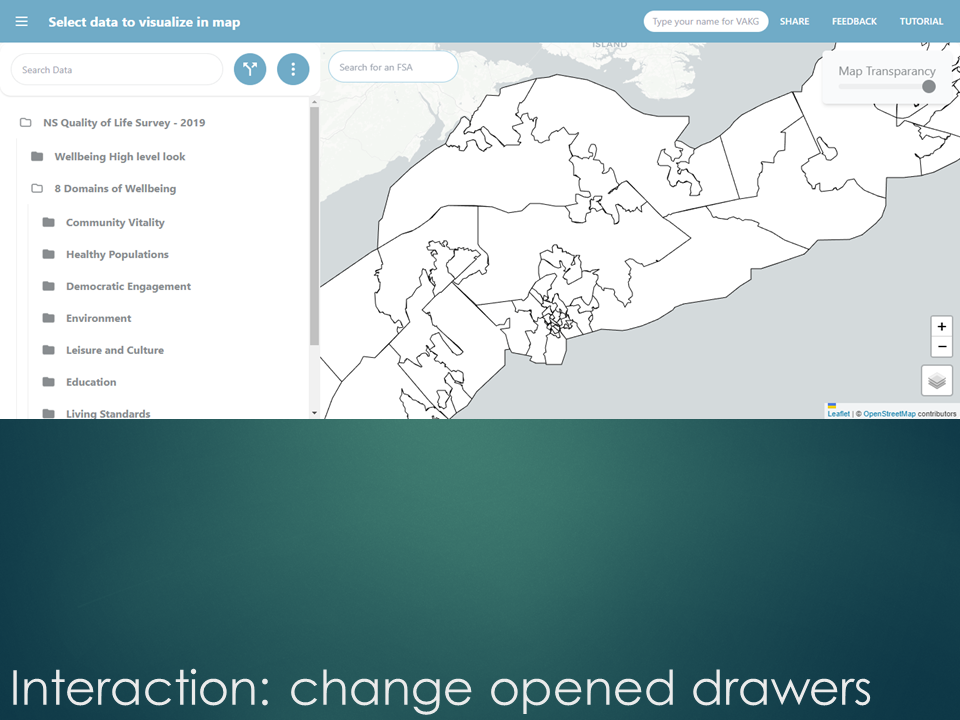}
    \includegraphics[width=0.3\textwidth]{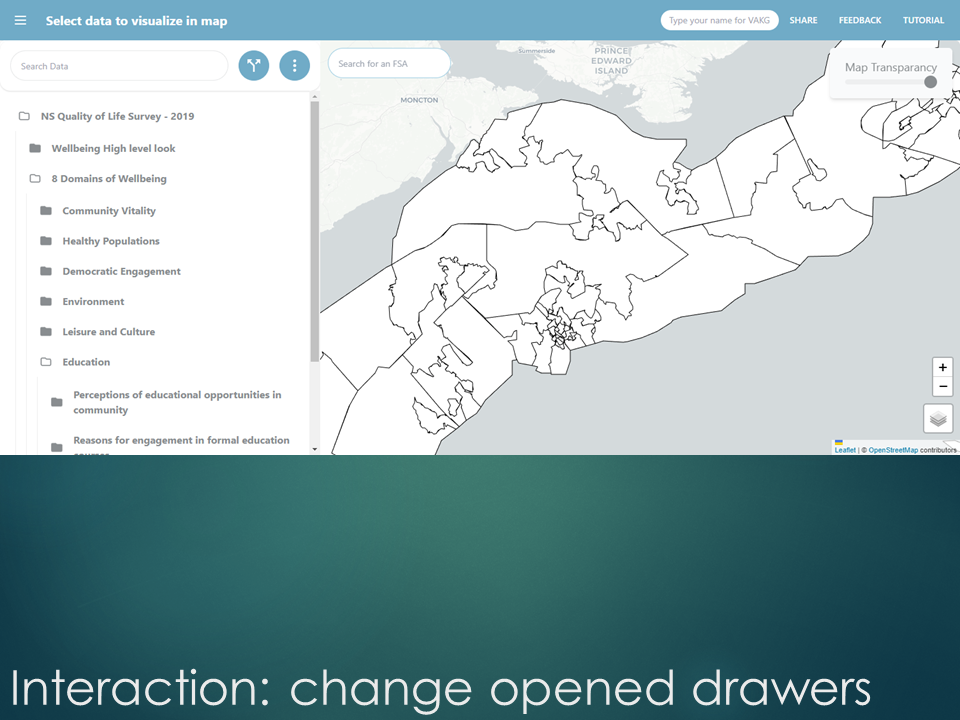}
    \includegraphics[width=0.3\textwidth]{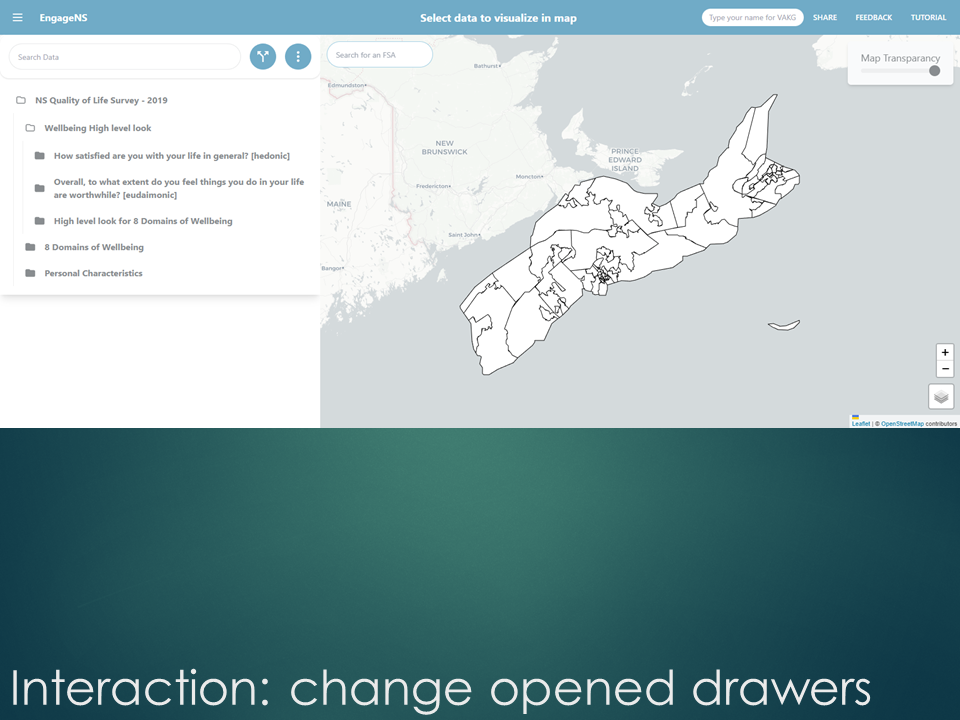}
    \includegraphics[width=0.3\textwidth]{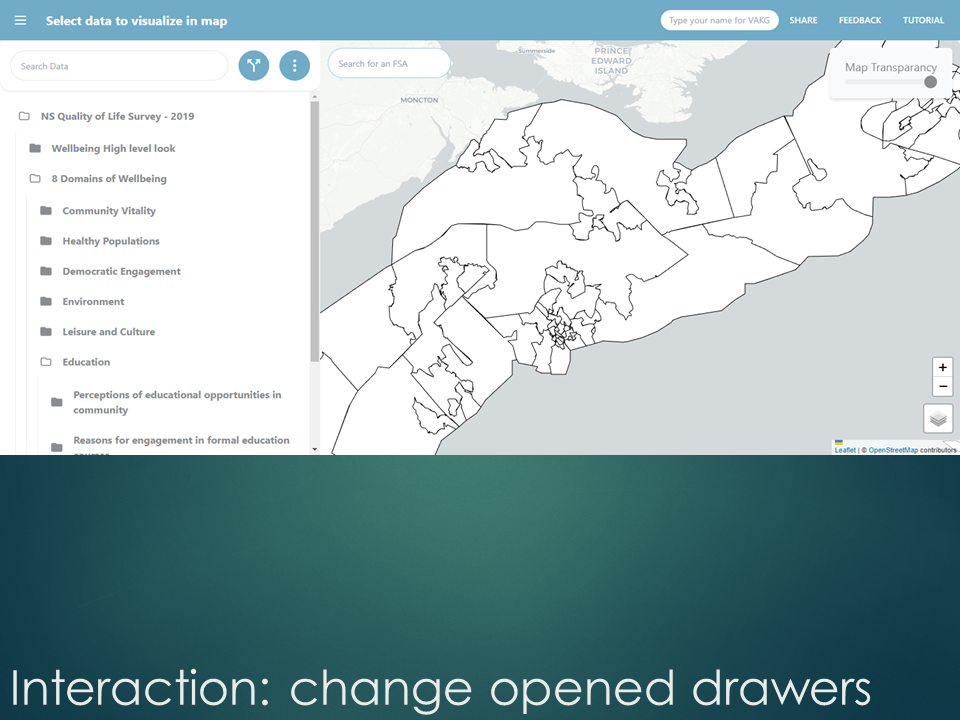}
    \includegraphics[width=0.3\textwidth]{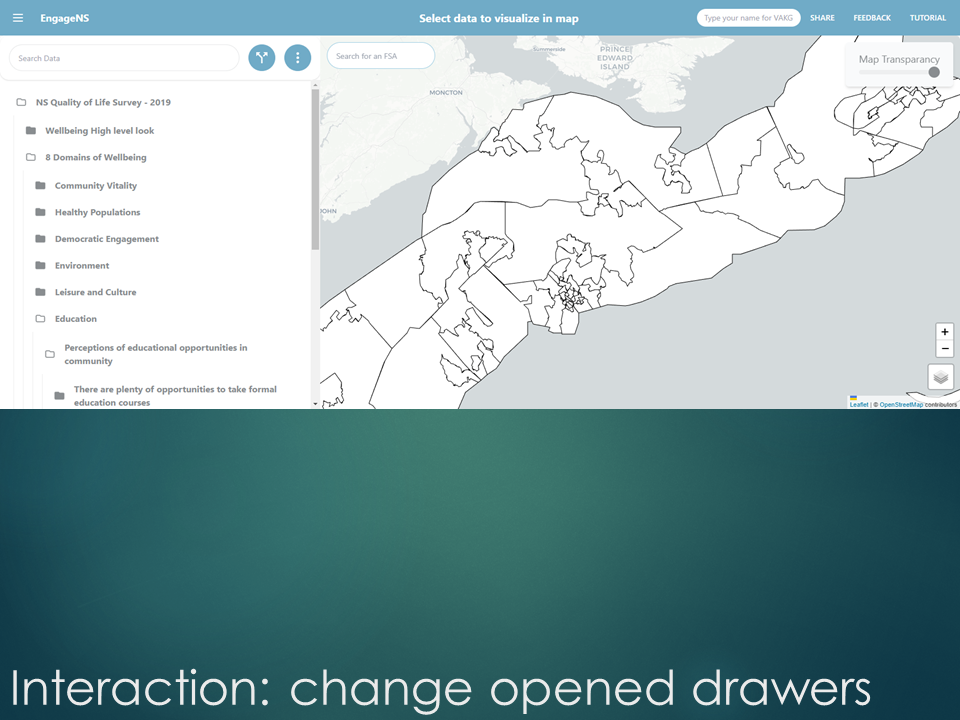}
    \includegraphics[width=0.3\textwidth]{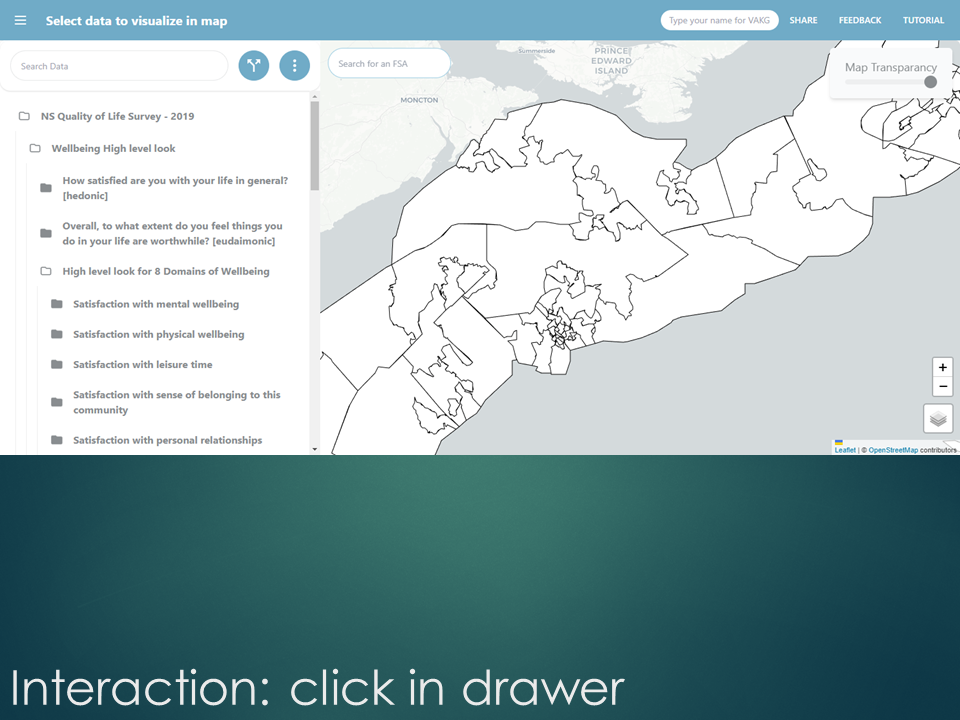}
    \includegraphics[width=0.3\textwidth]{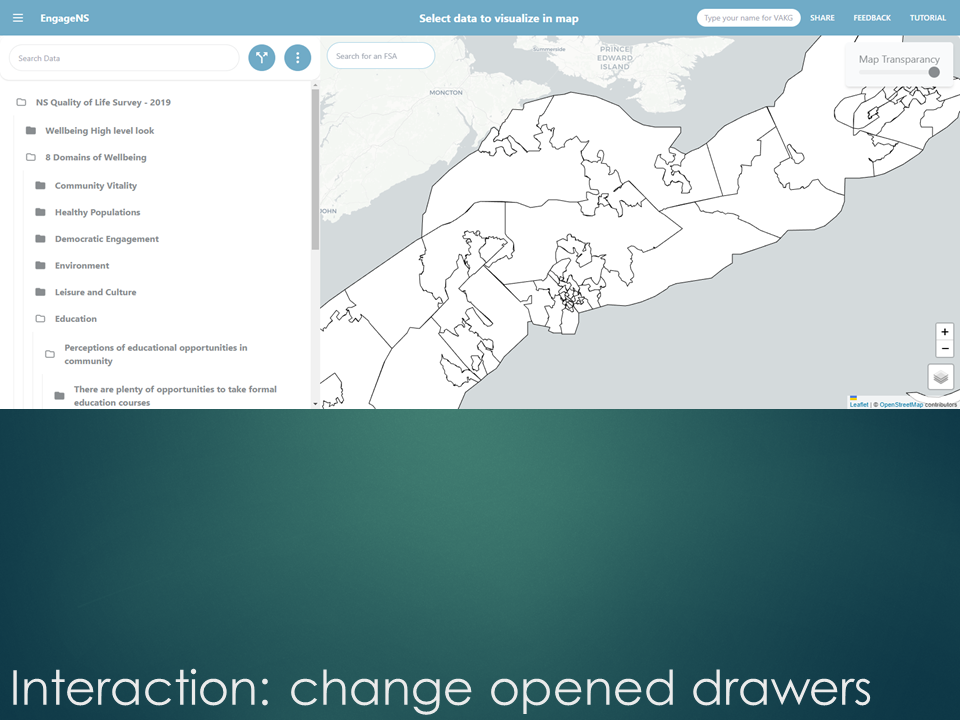}
    \includegraphics[width=0.3\textwidth]{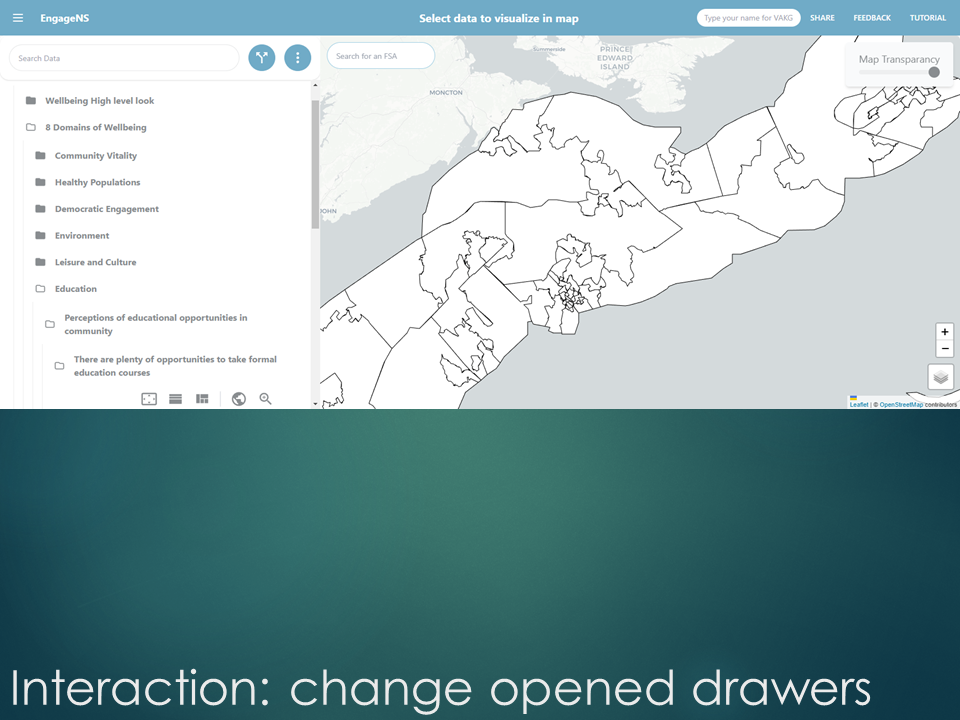}
    \includegraphics[width=0.3\textwidth]{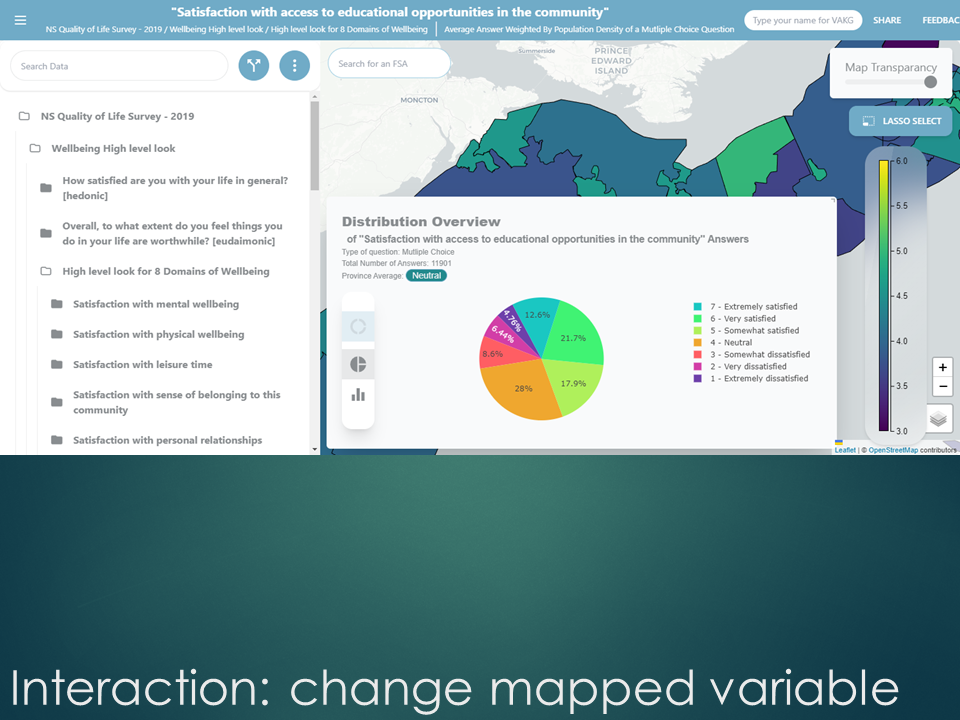}
    \includegraphics[width=0.3\textwidth]{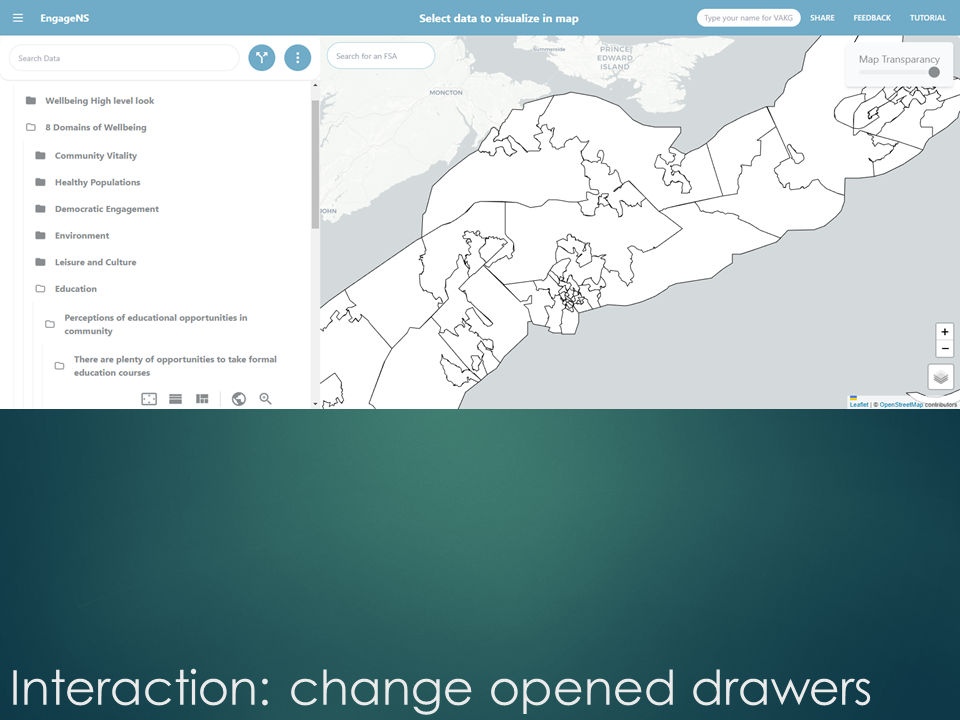}
    \includegraphics[width=0.3\textwidth]{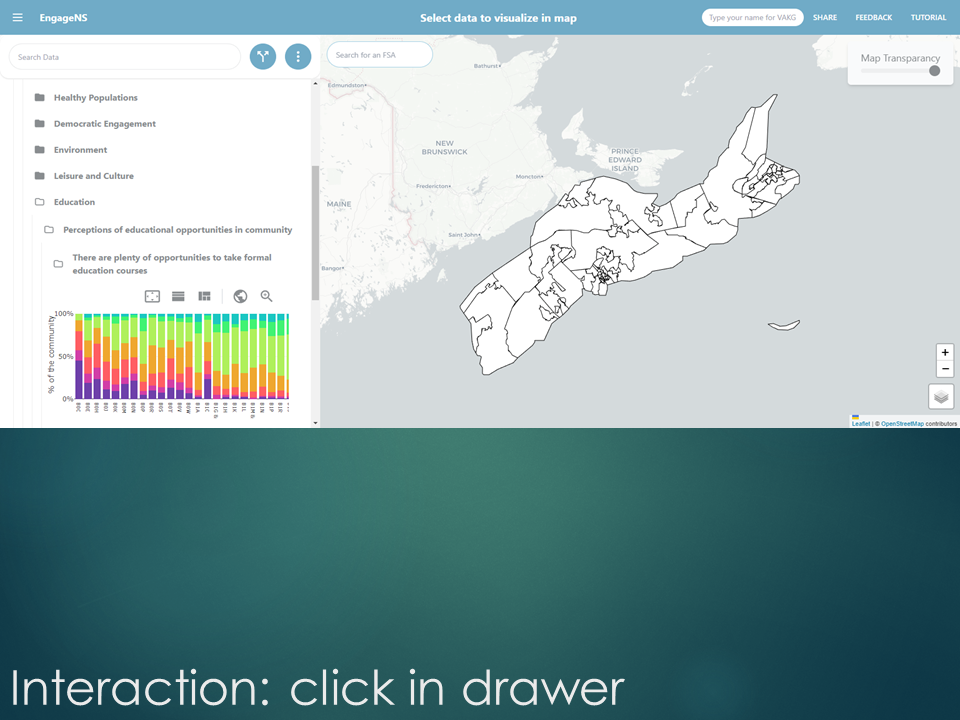}
    \includegraphics[width=0.3\textwidth]{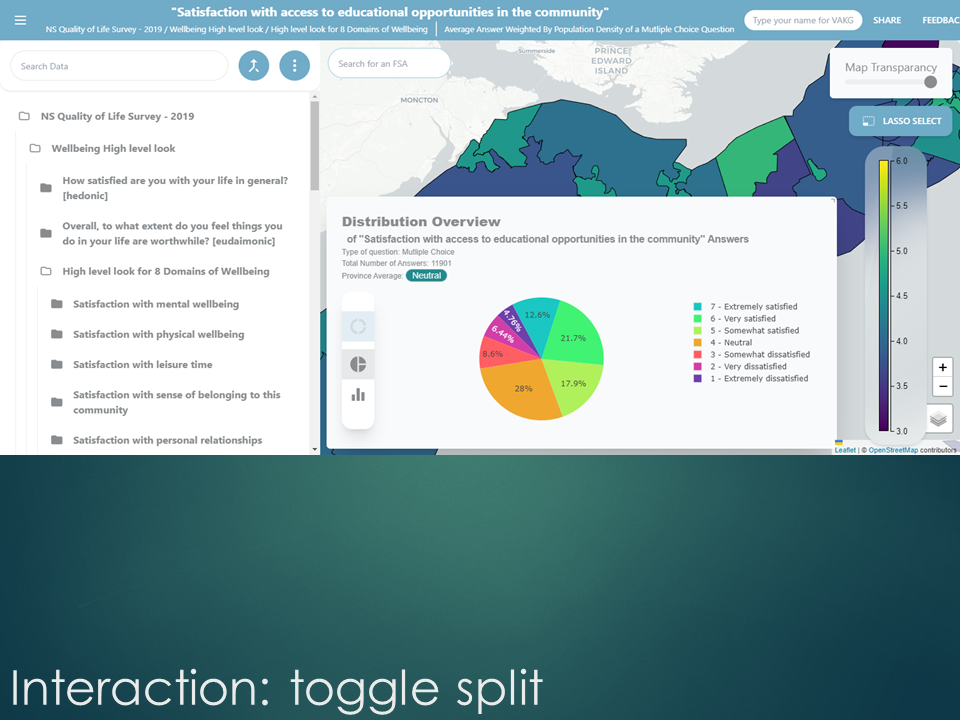}
    \includegraphics[width=0.3\textwidth]{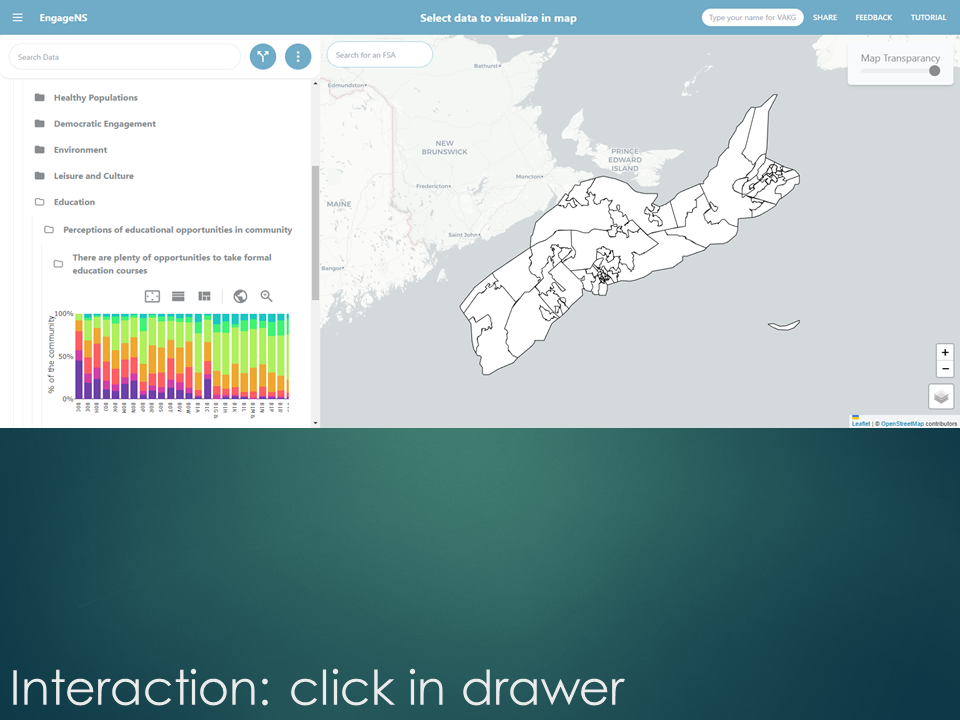}
    \caption{Three slide decks extracted from the insight WMT task 1 (part 1).  }
    \label{fig:slidedeckeduc}
\end{figure*} 

\begin{figure*}
    \centering
    \includegraphics[width=0.3\textwidth]{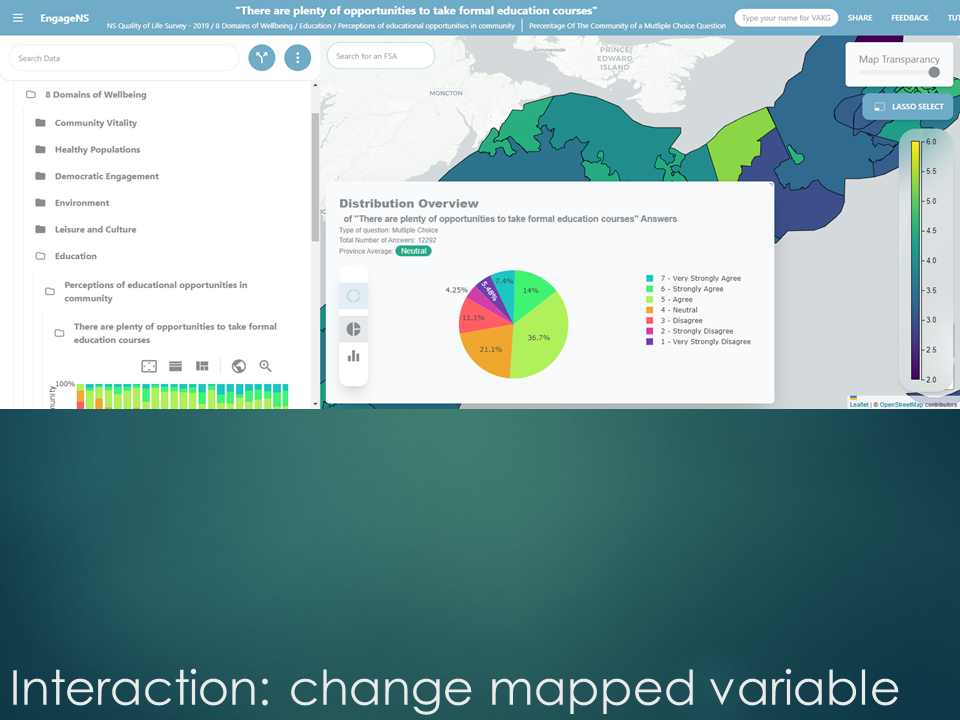}
    \includegraphics[width=0.3\textwidth]{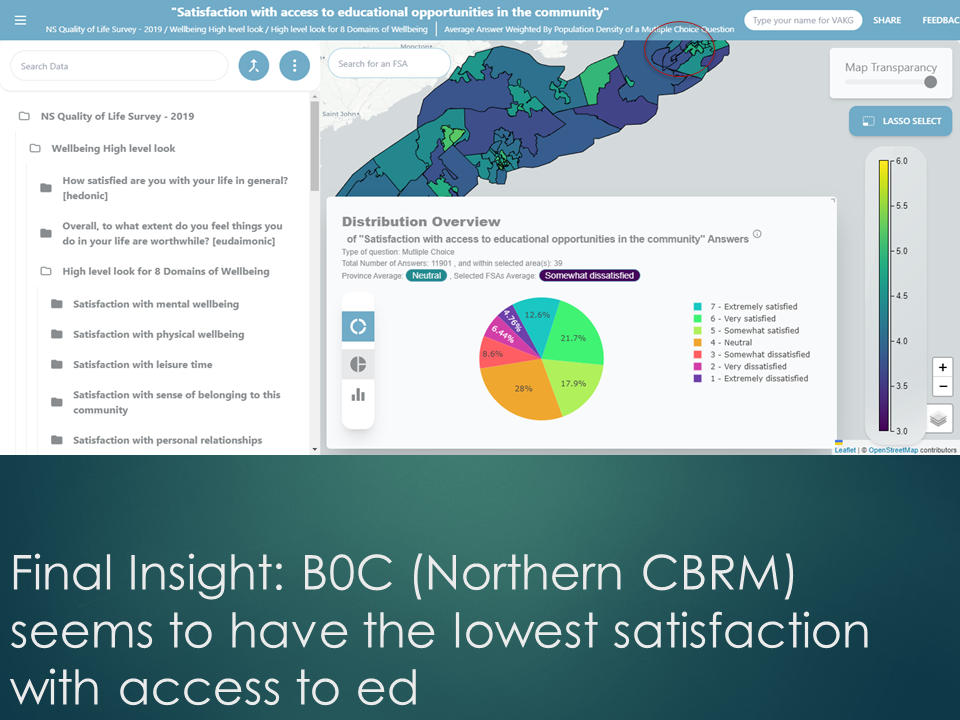}
    \includegraphics[width=0.3\textwidth]{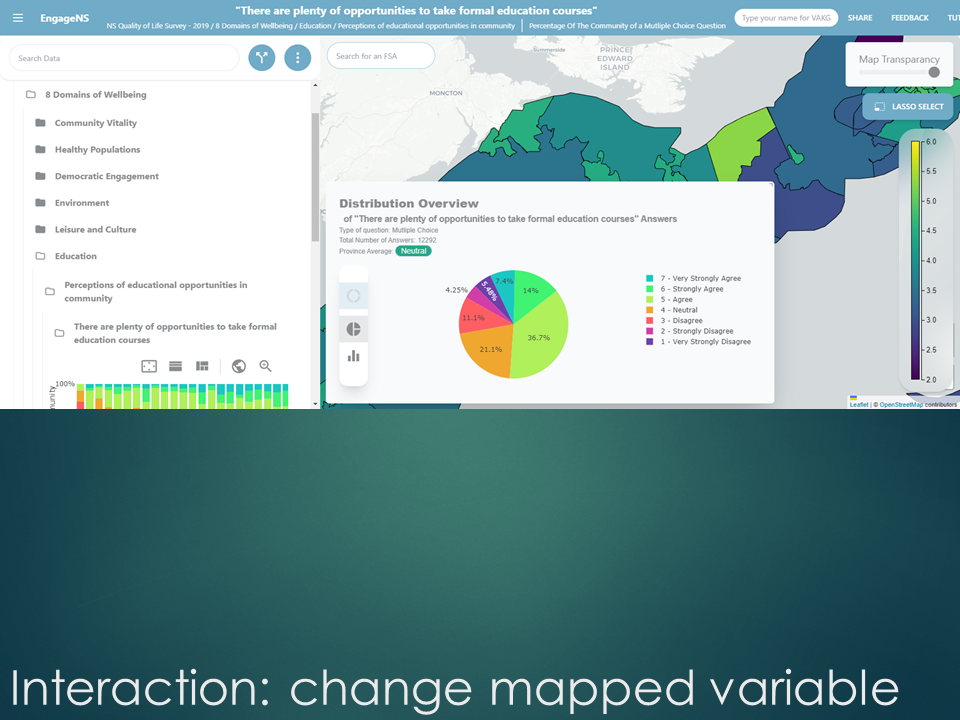}
    \includegraphics[width=0.3\textwidth]{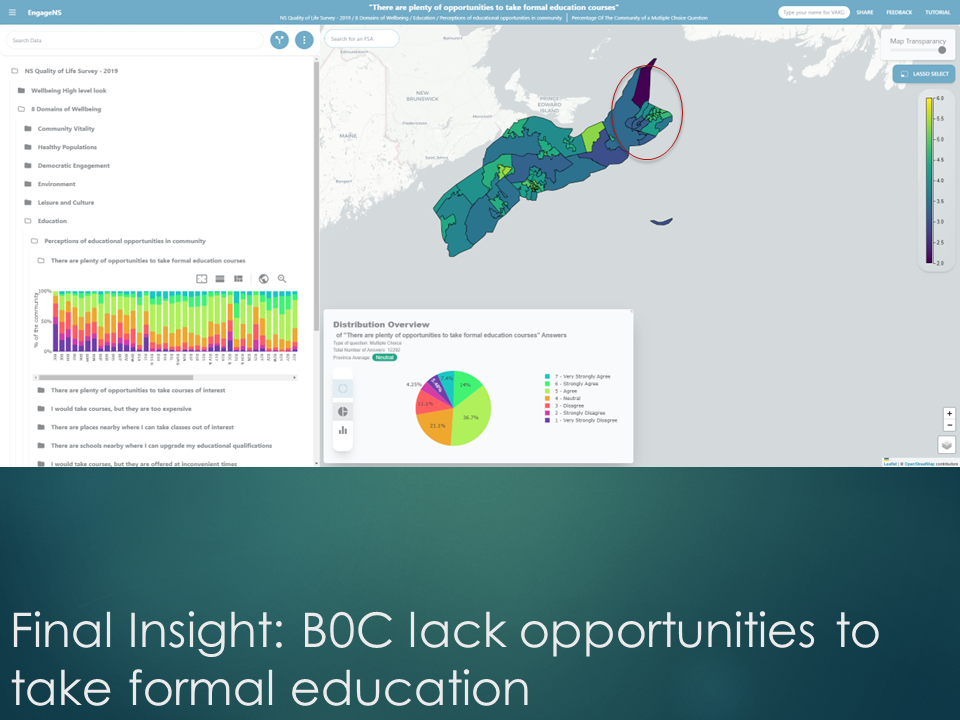}
    \includegraphics[width=0.3\textwidth]{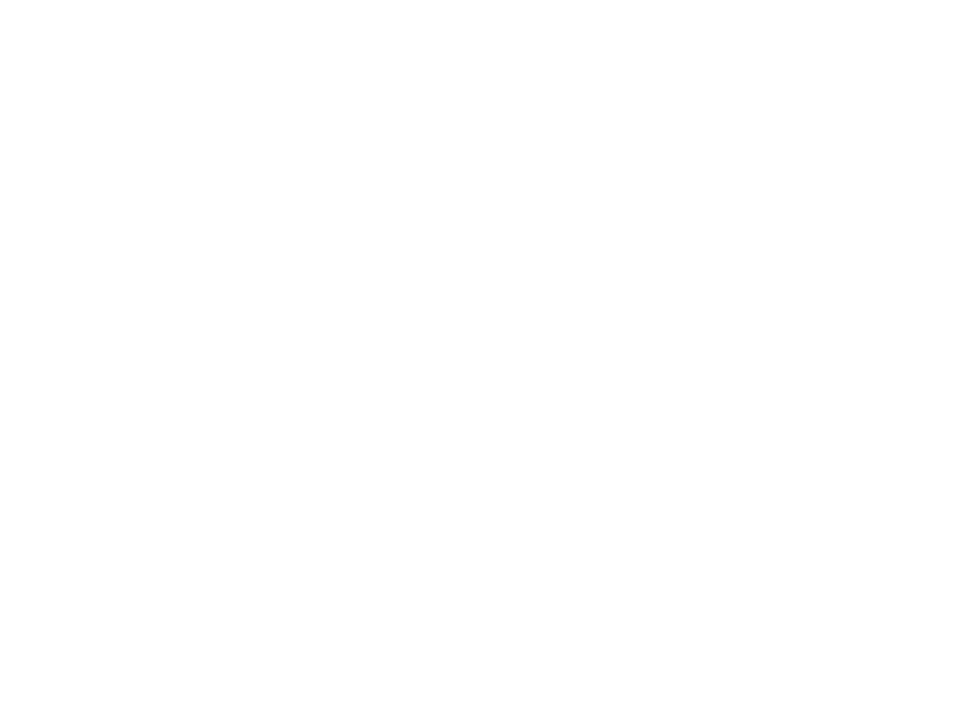}
    \includegraphics[width=0.3\textwidth]{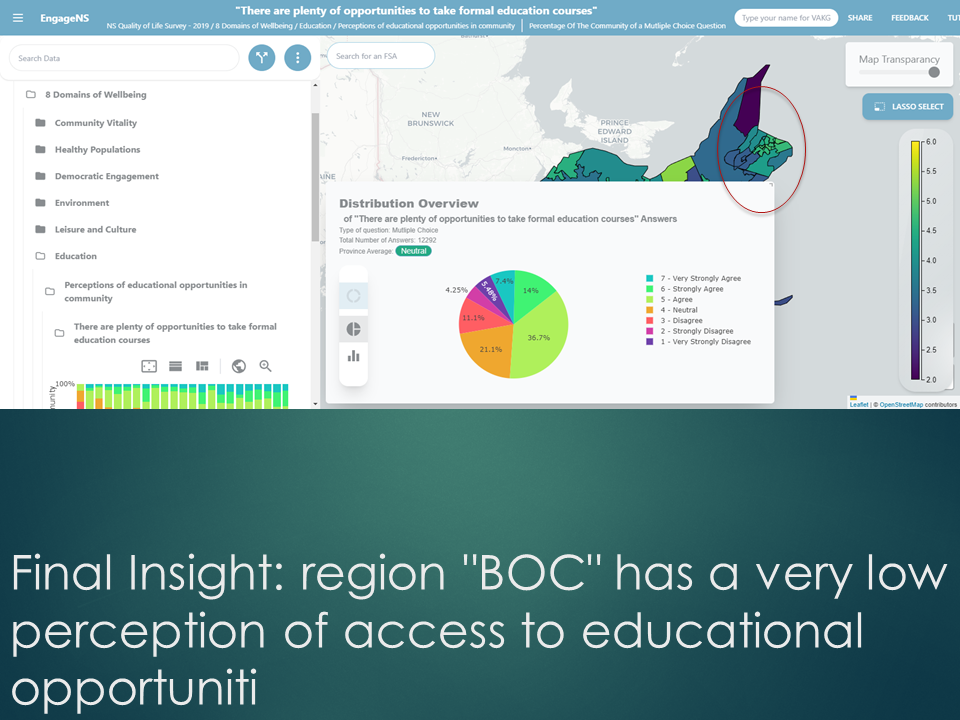}
    \caption{Three slide decks extracted from the insight WMT task 1 (part 2).  }
    \label{fig:slidedeckeduc}
\end{figure*}

\begin{figure*}
    \centering
    \includegraphics[width=\textwidth]{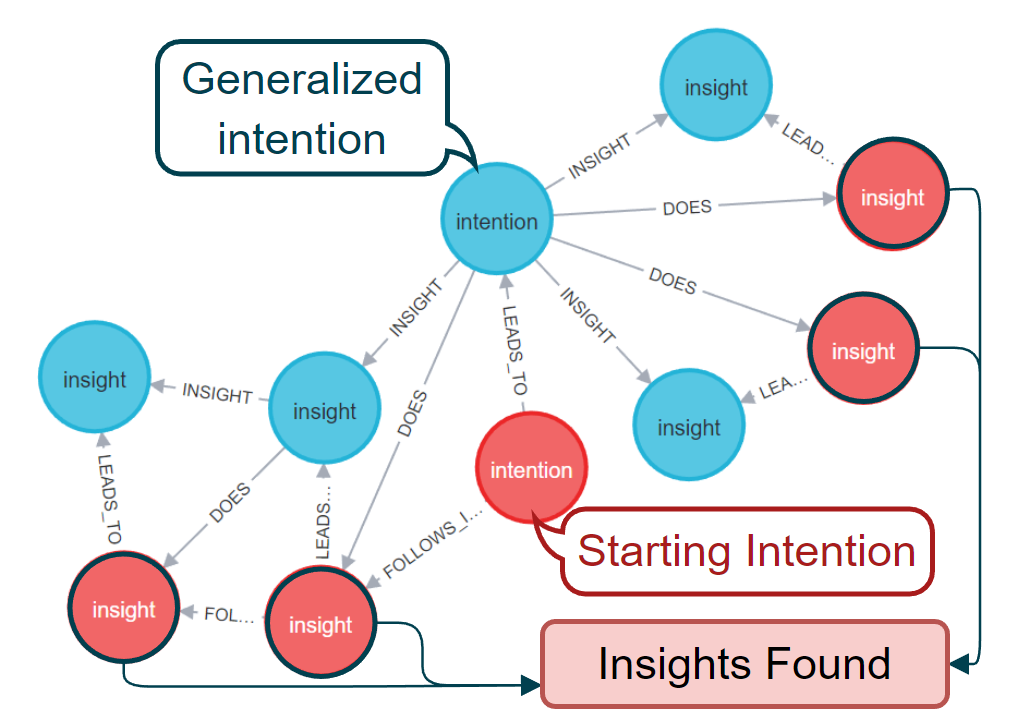}
    \includegraphics[width=0.33\textwidth]{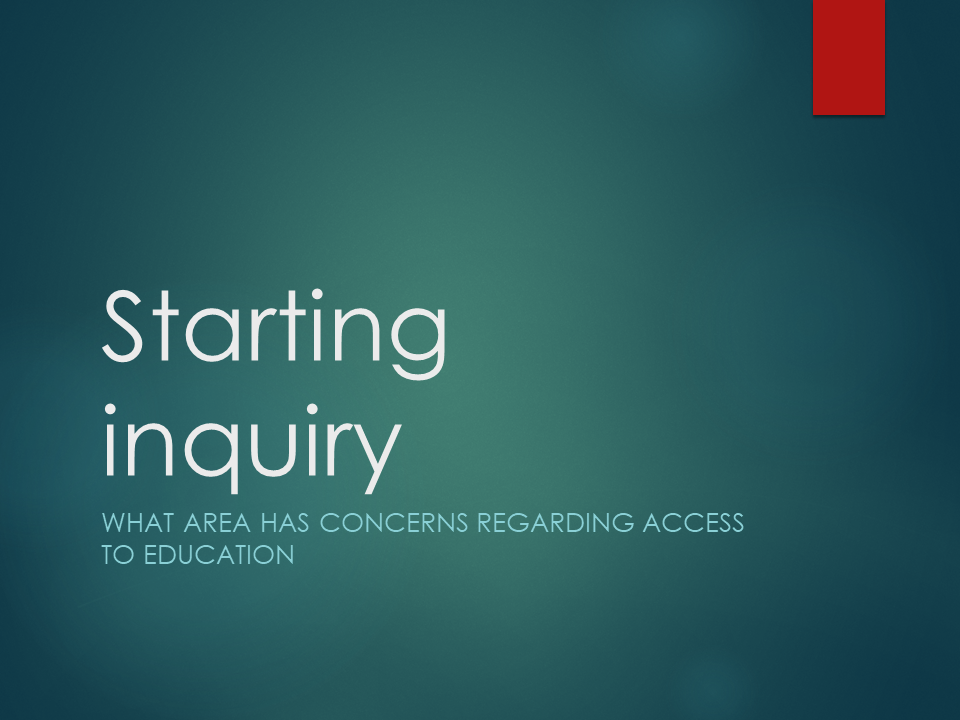}
    \includegraphics[width=0.33\textwidth]{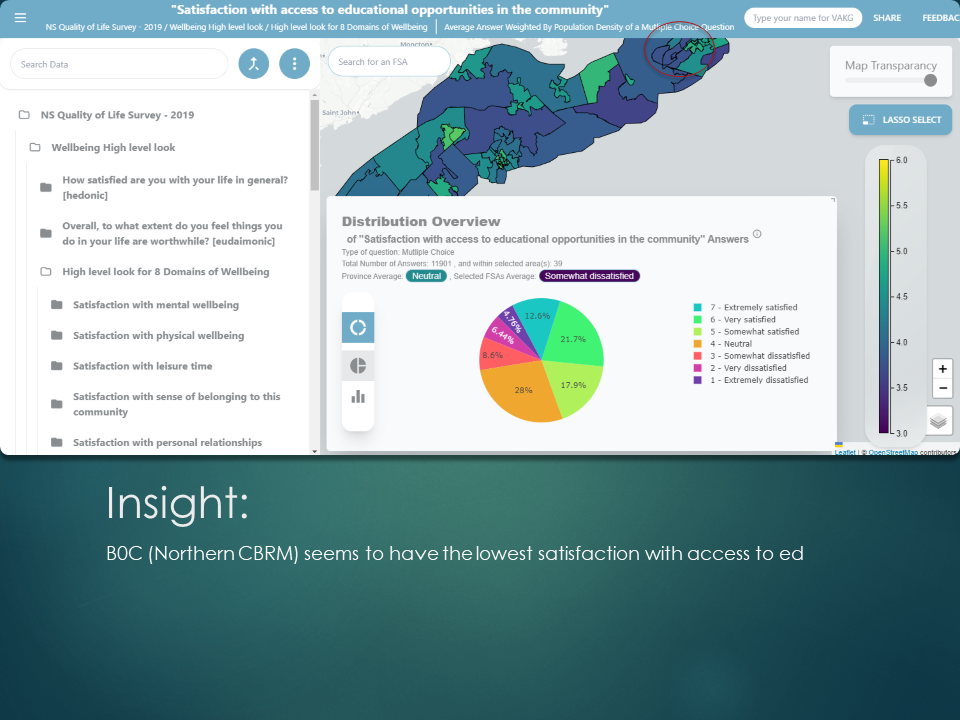}
    \includegraphics[width=0.33\textwidth]{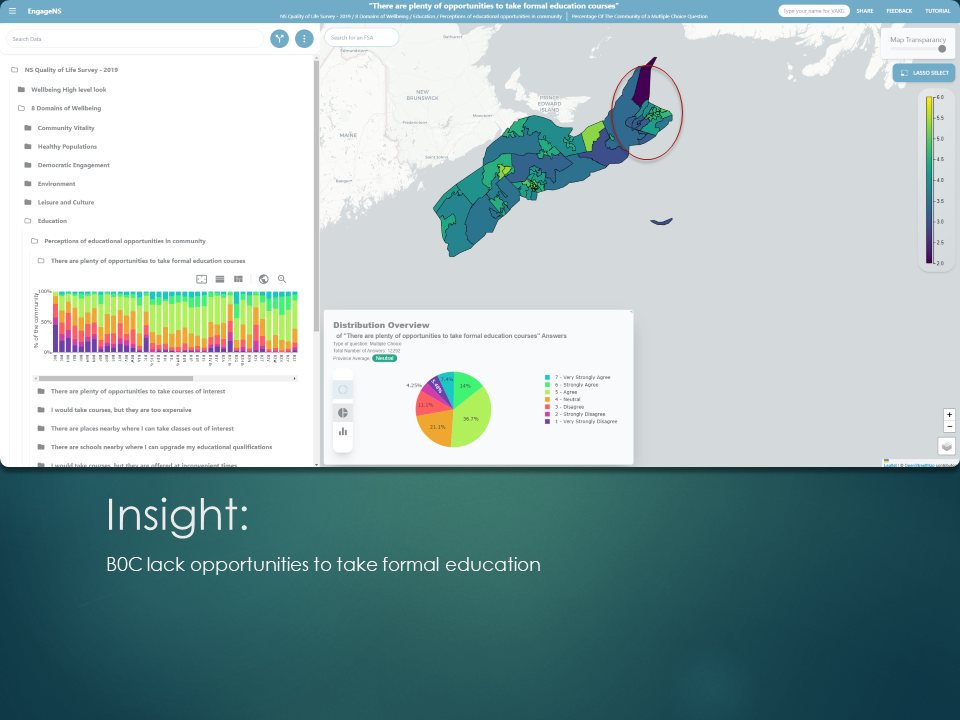}
    \includegraphics[width=0.4\textwidth]{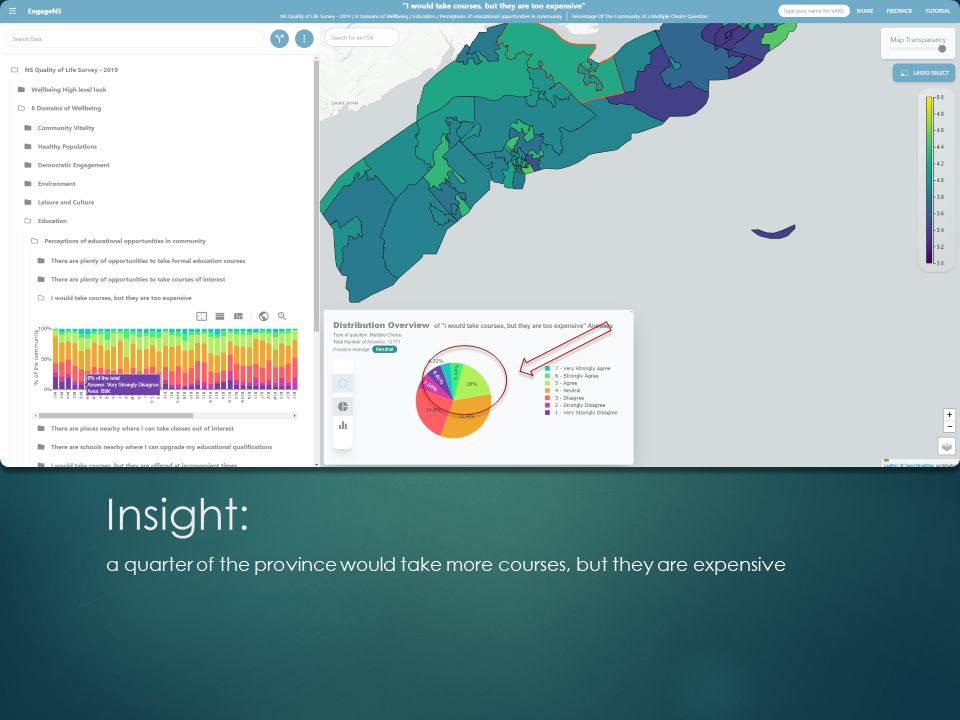}
    \includegraphics[width=0.4\textwidth]{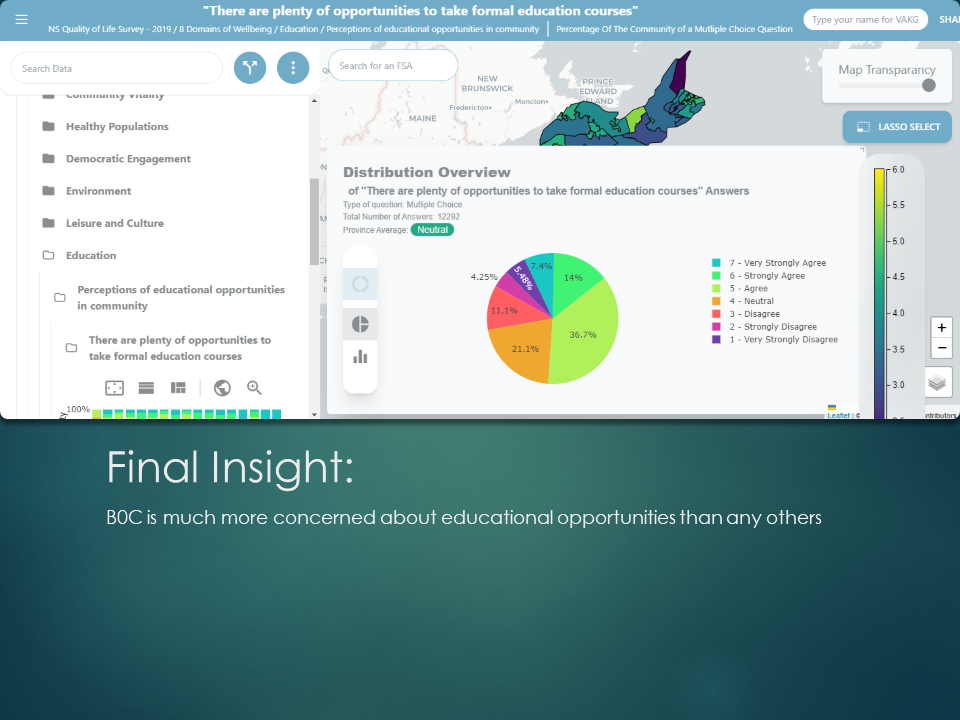}
    \caption{Graph visualization from Neo4j~\cite{noauthororeditorneo4j} when collecting all insights from all users given a starting intention. In this example, four insights were found from said intention, two of which were from the same user who wrote the intention (two to the left) and two from other users (two to the right). The slide deck generated on the bottom describes the four insights found given the original intention. }
    \label{fig:findinsights}
\end{figure*} 

\begin{figure*}
    \centering
    \includegraphics[width=\textwidth]{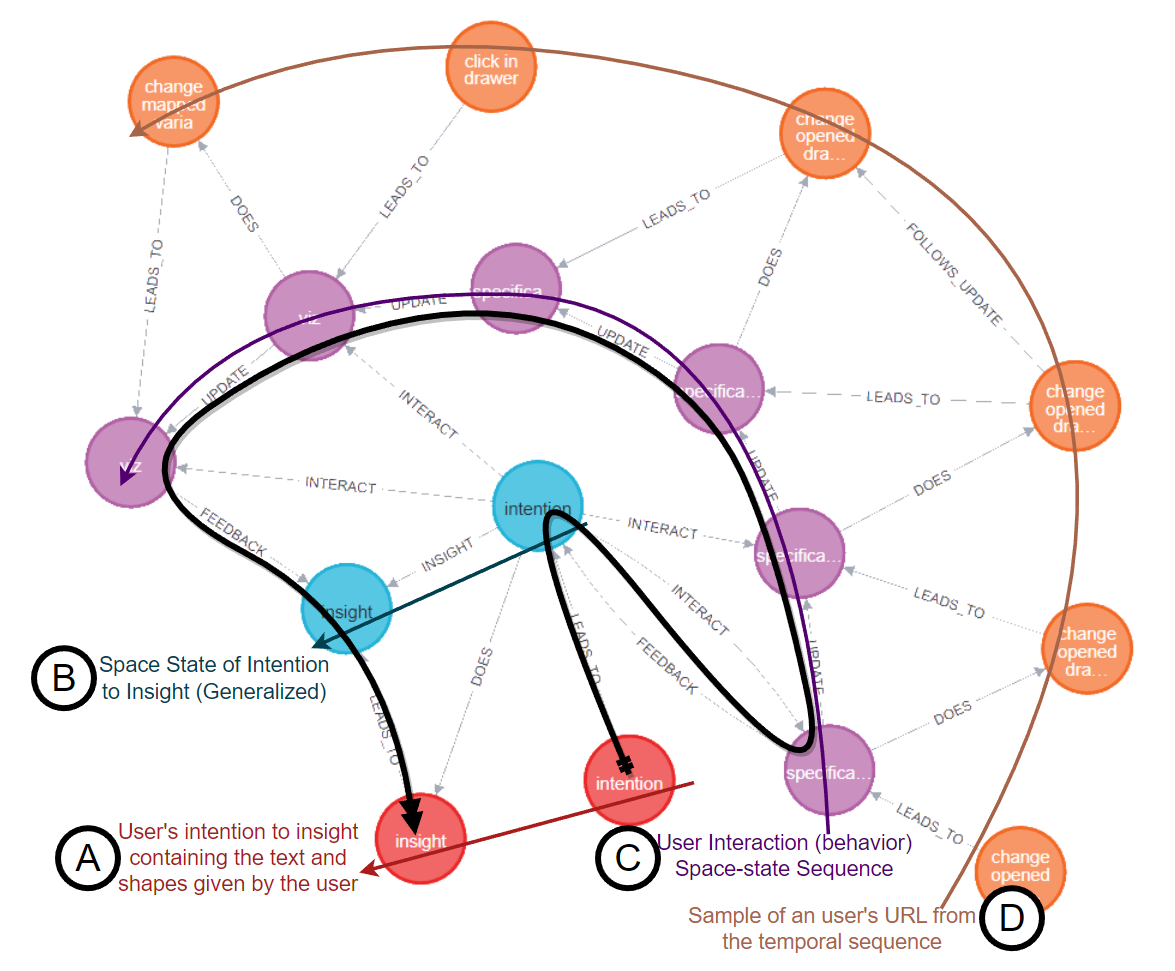}
    \caption{Above is a portion of KG extracted from \textit{Knowledge-Decks} when generating a slide deck, which contains all interactions done in the shortest between an intention and an insight. Visualization created with Neo4J browser~\cite{noauthororeditorneo4j}. The graph's path highlighted in black is used to generate the slide deck below, which is to be read from left to right and top to bottom (part 1). }
    \label{fig:slidedecksampleflow}
\end{figure*}

\begin{figure*}
    \centering
    \includegraphics[width=0.4\textwidth]{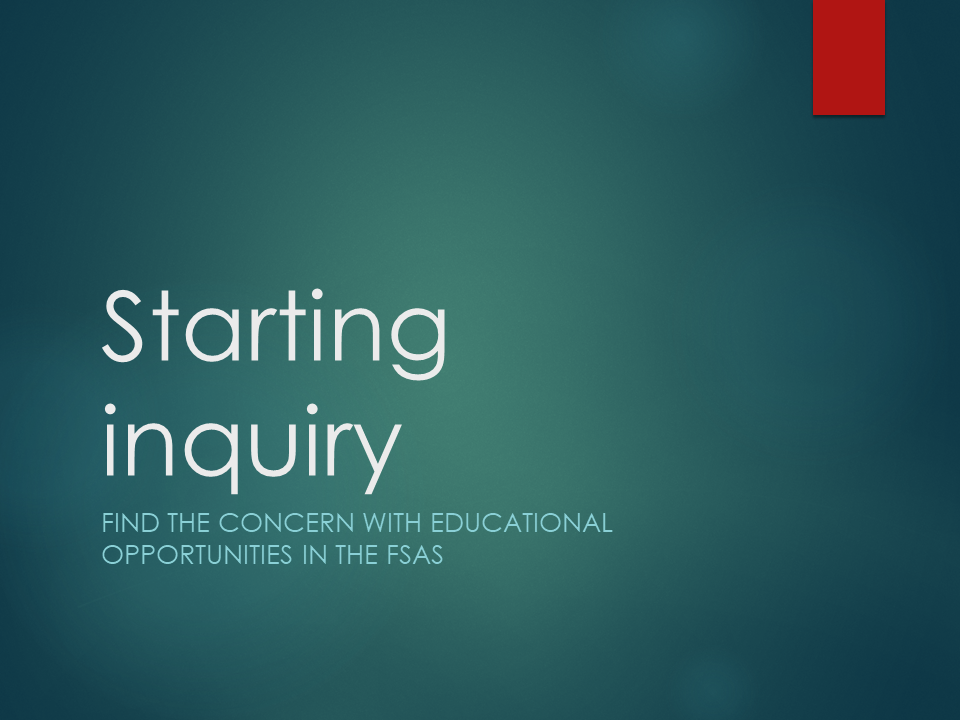}
    \includegraphics[width=0.4\textwidth]{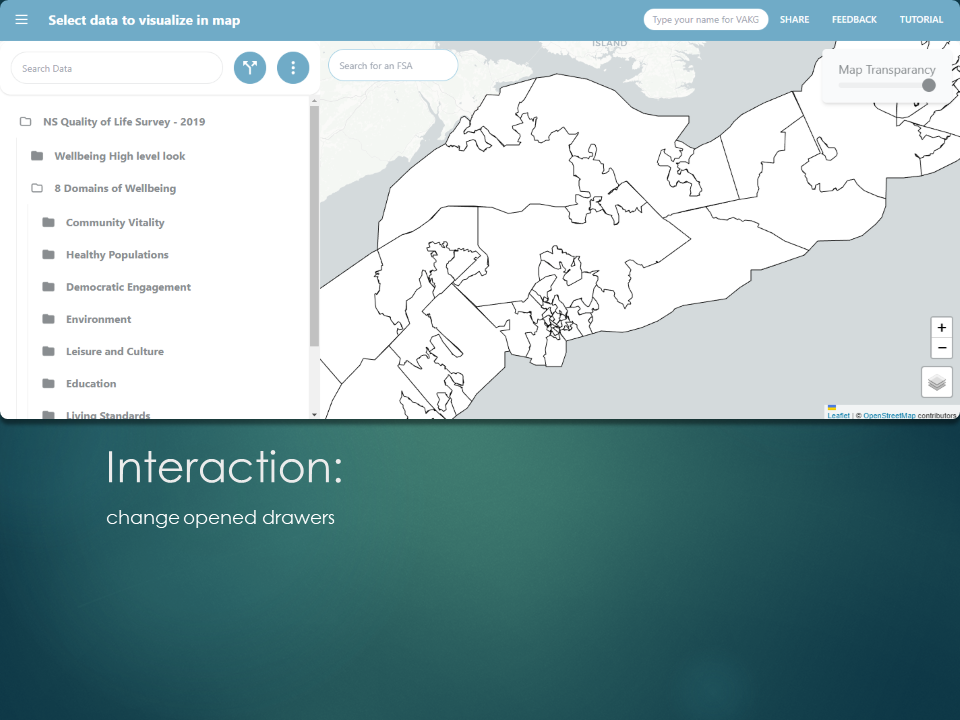}
    \includegraphics[width=0.4\textwidth]{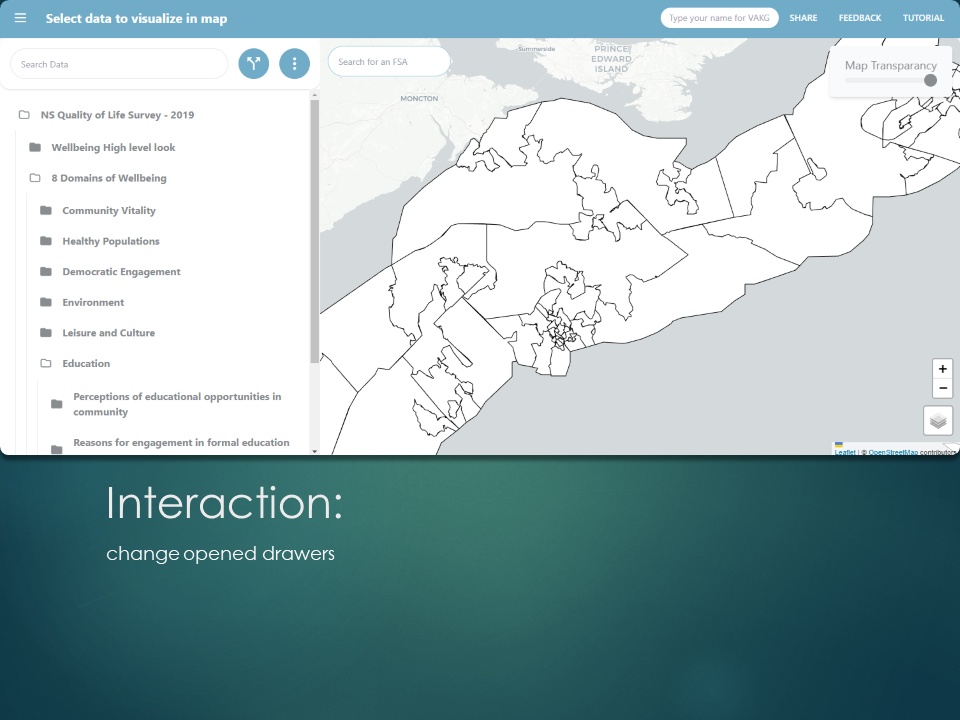}
    \includegraphics[width=0.4\textwidth]{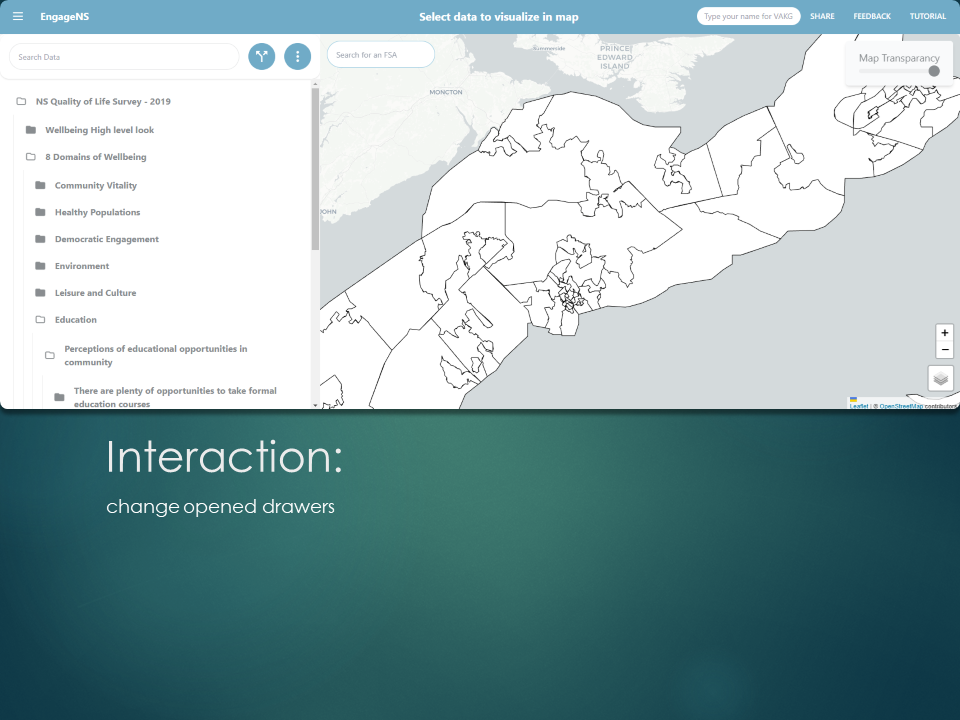}
    \includegraphics[width=0.4\textwidth]{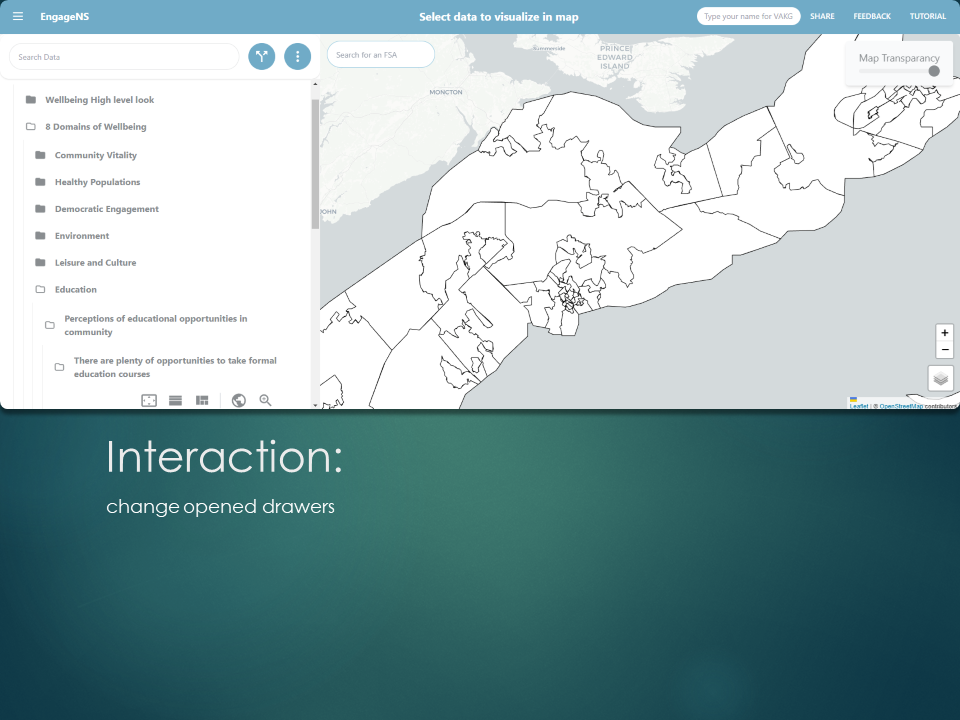}
    \includegraphics[width=0.4\textwidth]{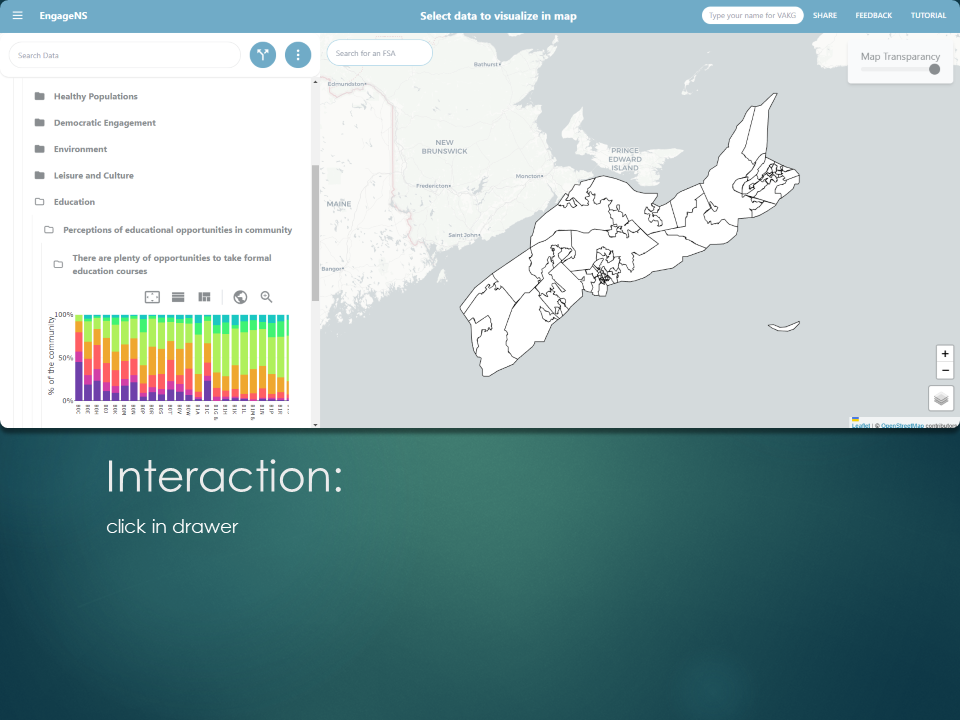}
    \includegraphics[width=0.4\textwidth]{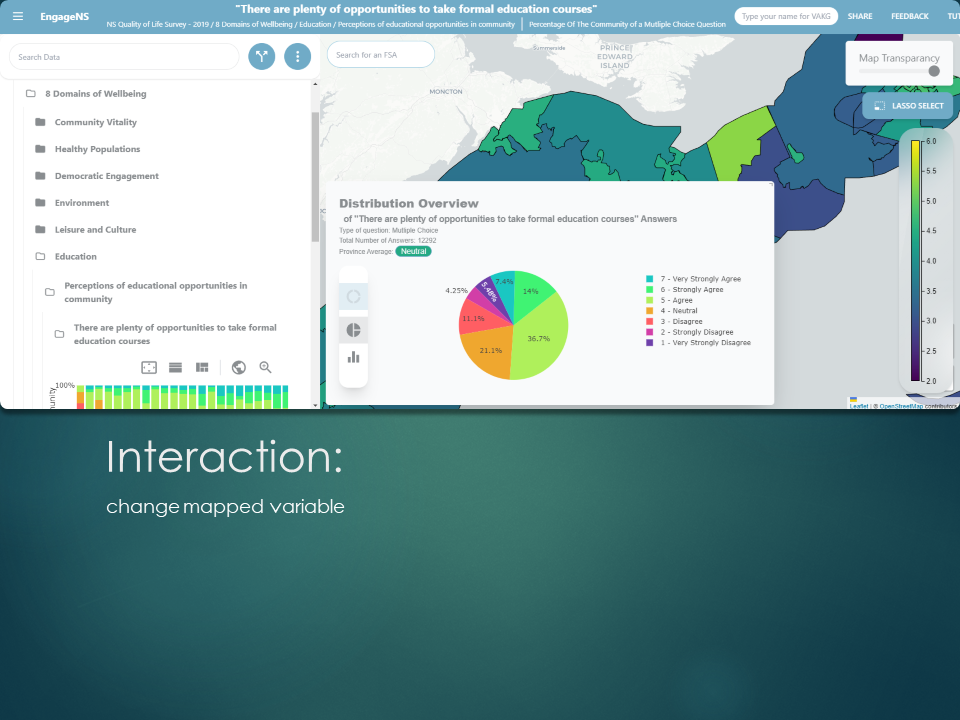}
    \includegraphics[width=0.4\textwidth]{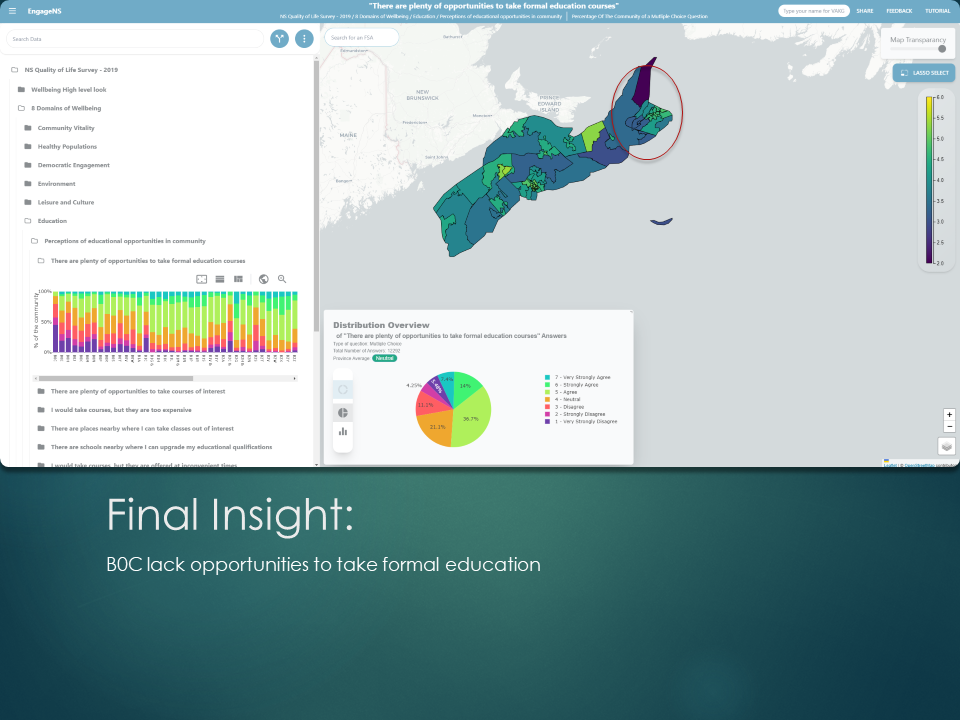}
    \caption{Above is a portion of KG extracted from \textit{Knowledge-Decks} when generating a slide deck, which contains all interactions done in the shortest between an intention and an insight. Visualization created with Neo4J browser~\cite{noauthororeditorneo4j}. The graph's path highlighted in black is used to generate the slide deck below, which is to be read from left to right and top to bottom (part 2). }
    \label{fig:slidedecksampleflow}
\end{figure*}

\bibliographystyle{eurovisDefinitions/eg-alpha-doi}  
\bibliography{mainSup}